%% file: draft_03.tex
\documentclass[twocolumn]{aastex631}
\newcommand\teff{$\mathrm{T_{eff}}$}
\newcommand\logg{$\log g$}

\shorttitle{Young Stars in the APOGEE-2 Survey}
\shortauthors{Rom\'an-Z\'u\~niga et al.}
\graphicspath{{./}}

\begin{document}

\title{Stellar Properties for a Comprehensive Collection of Star Forming Regions in the SDSS APOGEE-2 Survey-PREPRINT version\footnote{Based on SDSS Data Releases 16 and 17}}

\author[0000-0001-8600-4798]{Carlos G. Rom\'an-Z\'u\~niga}
\affiliation{Universidad Nacional Aut\'onoma de M\'exico, Instituto de Astronom\'ia, AP 106,  Ensenada 22800, BC, M\'exico}

\author[0000-0002-5365-1267]{Marina Kounkel}
\affil{Department of Physics and Astronomy, Vanderbilt University, VU Station 1807, Nashville, TN 37235, USA}
\affiliation{Department of Physics and Astronomy, Western Washington University, 516 High St, Bellingham, WA 98225, USA}

\author[0000-0001-9797-5661]{Jes\'us Hern\'andez}
\affiliation{Universidad Nacional Aut\'onoma de M\'exico, Instituto de Astronom\'ia, AP 106,  Ensenada 22800, BC, M\'exico}

\author[0000-0002-5855-401X]{Karla Pe\~na Ram\'irez}
\affiliation{Centro de Astronom\'ia (CITEVA), Universidad de Antofagasta, Av. Angamos 601, Antofagasta, Chile}

\author[0000-0002-7795-0018]{Ricardo L\'opez-Valdivia}
\affiliation{Universidad Nacional Aut\'onoma de M\'exico, Instituto de Astronom\'ia, AP 106,  Ensenada 22800, BC, M\'exico}

\author[0000-0001-6914-7797]{Kevin R. Covey}
\affiliation{Department of Physics and Astronomy, Western Washington University, 516 High St, Bellingham, WA 98225, USA}

\author[0000-0003-2300-8200]{Amelia M.\ Stutz}
\affiliation{Departamento de Astronom\'{i}a, Universidad de Concepci\'{o}n,Casilla 160-C, Concepci\'{o}n, Chile}
\affiliation{Max-Planck-Institute for Astronomy, K\"{o}nigstuhl 17, 69117 Heidelberg, Germany}

\author[0000-0002-1379-4204]{Alexandre Roman-Lopes}
\affiliation{Departamento de Astronom\'ia, Facultad de Ciencias, Universidad de La Serena.  Av. Juan Cisternas 1200, La Serena, Chile}

\author[0000-0001-5436-5388]{Hunter Campbell}
\affiliation{Department of Physics and Astronomy, Western Washington University, 516 High St, Bellingham, WA 98225, USA}

\author[0000-0001-9649-6028]{Elliott Khilfeh}
\affiliation{Department of Physics and Astronomy, Western Washington University, 516 High St, Bellingham, WA 98225, USA}

\author[0000-0002-0506-9854]{Mauricio Tapia}
\affiliation{Universidad Nacional Aut\'onoma de M\'exico, Instituto de Astronom\'ia, AP 106,  Ensenada 22800, BC, M\'exico}

\author[0000-0003-1479-3059]{Guy S. Stringfellow}
\affiliation{Center for Astrophysics and Space Astronomy, Department of Astrophysical and Planetary Sciences, University of Colorado, Boulder,CO, 80309, USA}

\author[0000-0001-6559-2578]{Juan Jos\'e Downes}
\affiliation{Departamento de Astronom\'ia, Facultad de Ciencias, Universidad de la Rep\'ublica, Igu\'a 4225, 14000, Montevideo, Uruguay}

\author[0000-0002-3481-9052]{Keivan G.\ Stassun}
\affil{Department of Physics and Astronomy, Vanderbilt University, VU Station 1807, Nashville, TN 37235, USA}

\author[0000-0002-7064-099X]{Dante Minniti}
\affiliation{Departamento de Ciencias F\'isicas, Facultad de Ciencias Exactas, Universidad Andr\'es Bello, Fern\'andez Concha 700, Las Condes,Santiago, Chile}
\affiliation{Vatican Observatory, V00120 Vatican City State, Italy}

\author[0000-0001-7868-7031]{Amelia Bayo}
\affiliation{Instituto de Física y Astronom\'ia, Universidad de Valparaiso, Gran Breta\~na 1111, Valpara\'iso, Chile}
\affiliation{N\'ucleo Milenio de Formaci\'on Planetaria, NPF, Universidad de Valpara\'iso, Chile}

\author[0000-0001-6072-9344]{Jinyoung Serena Kim}
\affiliation{Steward Observatory, Department of Astronomy, University of Arizona, 933 North Cherry Avenue, Tucson, AZ 85721-0065, USA}

\author[0000-0002-2011-4924]{Genaro Su\'arez}
\affiliation{Department of Physics and Astronomy, The University of Western Ontario, 1151 Richmond St, London, ON N6G 1N9, Canada}
\affiliation{Department of Astrophysics, American Museum of Natural History, 79th Street at Central Park West, New York,
NY 10024, USA}

\author[0000-0002-3576-4508]{Jason E. Ybarra}
\affiliation{Department of Physics, Davidson College, 405 N Main St, Davidson, NC 28035, USA}

\author[0000-0003-3526-5052]{Jos\'e G. Fern\'andez-Trincado}
\affiliation{Instituto de Astronom\'ia, Universidad Cat\'olica del Norte, Av. Angamos 0610, Antofagasta, Chile}

\author[0000-0001-9330-5003]{Pen\'elope Longa-Pe\~na}
\affiliation{Centro de Astronom\'ia (CITEVA), Universidad de Antofagasta, Av. Angamos 601, Antofagasta, Chile}

\author[0000-0002-4013-2716]{Valeria Ram\'irez-Preciado}
\affiliation{Universidad Nacional Aut\'onoma de M\'exico, Instituto de Astronom\'ia, AP 106,  Ensenada 22800, BC, M\'exico}

\author[0000-0001-7351-6540]{Javier Serna}
\affiliation{Universidad Nacional Aut\'onoma de M\'exico, Instituto de Astronom\'ia, AP 106,  Ensenada 22800, BC, M\'exico}

\author[0000-0003-1805-0316]{Richard R. Lane}
\affiliation{Centro de Investigaci\'on en Astronom\'ia, Universidad Bernardo O'Higgins, Avenida Viel 1497, Santiago, Chile}

\author[0000-0002-1693-2721]{D. A. Garc\'ia-Hern\'andez}
\affiliation{Universidad de La Laguna (ULL), Departamento de Astrofísica, E-38206 La Laguna, Tenerife, Spain}

\author[0000-0002-1691-8217]{Rachael L. Beaton}
\affiliation{Department of Astrophysical Sciences, 4 Ivy Lane, Princeton University, Princeton, NJ 08544}
\affiliation{The Observatories of the Carnegie Institution for Science, 813 Santa Barbara St., Pasadena, CA~91101}

\author[0000-0002-3601-133X]{Dmitry Bizyaev}
\affiliation{Apache Point Observatory and New Mexico State
University, P.O. Box 59, Sunspot, NM, 88349-0059, USA}
\affiliation{Sternberg Astronomical Institute, Moscow State
University, Moscow, Russia}

\author[0000-0002-2835-2556]{Kaike Pan}
\affiliation{Apache Point Observatory and New Mexico State
University, P.O. Box 59, Sunspot, NM, 88349-0059, USA}

\accepted{Nov. 15th, 2022}
 
\begin{abstract}

The Sloan Digital Sky Survey IV (SDSS-IV) APOGEE-2 primary science goal was to observe red giant stars throughout the Galaxy to study its dynamics, morphology, and chemical evolution. The APOGEE instrument, a high- resolution 300 fiber H-band (1.55-1.71 micron) spectrograph, is also ideal to study other stellar populations in the Galaxy, among which are a number of star forming regions and young open clusters. We present the results of the determination of six stellar properties (\teff, \logg, [Fe/H], $\mathrm{L/L_\odot, M/M_\odot, and\  age}$) for a sample that is composed of 3360  young stars, of sub-solar to super-solar types, in sixteen Galactic star formation and young open cluster regions. Those sources were selected by using a clustering method that removes most of the field contamination. Samples were also refined by removing targets affected by various systematic effects of the parameter determination. The final samples are presented in a comprehensive catalog that includes all six estimated parameters. This overview study also includes parameter spatial distribution maps for all regions and Hertzprung-Russell ($\mathrm{\log{L/L_\odot}}$\ versus \teff) diagrams. This study serves as a guide for detailed studies on individual regions, and paves the way for the future studies on the global properties of stars in the pre-main sequence phase of stellar evolution using more robust samples.  

\end{abstract}




\section{Introduction} \label{sec:intro}

The complete characterization of the pre-main sequence (PMS) phase of stellar evolution from low to 
intermediate mass stars is an open problem in astrophysics. Various current models \citep[e.g.][]
{Tognelli11, Baraffe15, Dotter16, Bressan12} apply state-of-the-art theory in both the mainly-
convective and the mainly-radiative stages of the PMS. Nevertheless, many processes are still to be considered when those models are compared to spectroscopic observations. For instance, accretion as a function of both mass and age, circumstellar emission, disk dust-veiling of the stellar continuum, rotation or the true chemical content of the youngest stellar populations are still absent from most models. In order to produce better synthetic stellar atmosphere models that compare 
well to observed spectra of young stars, we require of larger and better empirical libraries. Machine-learning automated classification provide the required muscle to process big data samples but their results can only be as good as the input training samples. Thus, both supervised and un-supervised methods to determine PMS parameteres require of robust samples obtained in diverse star forming regions. In turn, large enough samples are provided nowadays by large scale spectroscopic surveys using fiber spectrographs capable of observing hundreds of targets simultaneuously.

In optical wavelengths, the LAMOST survey \citep{LAMOST12} has made important contributions to classify spectra of young stars and to help classifying young stars at other wavelengths \citep[e.g.][]{FangX16,FangX18,FangX20}. The Gaia-ESO survey \citep{GaiaESO12} has provided insight into the chemical content, accretion, dynamics and other properties in young star clusters \citep{Bravi18,Wright19, Bonito20, Binks21, Baratella20, Baratella21, Kos21}. 

The Sloan Digital Sky Survey (SDSS) has hosted several focused studies based on multi-object spectroscopy of young stars and young star clusters in its third \citep[SDSS-III;][]{Eisenstein2011} and fourth phases \citep[SDSS-IV;][]{Blanton17}.  A major advancement came in the infrared with the development of the SDSS-III APOGEE\footnote{Apache Point Galactic Evolution Experiment} \citep{APOGEE17} and SDSS-IV APOGEE-2 programs (Majewski et al. in prep.) In the former, the IN-SYNC program produced a number of studies dedicated to classification of young star spectra
and evolution of young stellar clusters \cite[e.g.][]{Cotaar14, Foster15, Cotaar15, daRio16,daRio17}. In the latter, a more ample APOGEE-2 survey of the Orion Complex \citep{Cottle18}, led to the first 6-D (position-velocity) map of Orion using radial velocities using improved IN-SYNC parameters as well as astrometric parameters from Gaia DR2 \citep{Kounkel18}. APOGEE-2 data has also been used to provide classifications of O and B-type stars in star forming complexes in the 1-3 kpc distance range \citep[e.g.][]{RomanLopes18, Borissova19, Ramirez20, RomanLopes20, RomanLopes20b, Medina21}.

This study collects data from a majority of the star formation and young star cluster observations published in Data Releases 16 and 17 of the SDSS APOGEE-2 program. Our main goals are to overview and compare the extensive data and basic results obtained from the spectral classification and estimation of physical properties in sixteen fields that include star formation complexes in the solar neighborhood $\mathrm{(0.1<d<1\ kpc)}$ and the neighboring spiral arms $\mathrm{(1<d<3\ kpc)}$. 

The common goals in the study of these samples are: a) to identify the young star population; b) to obtain reliable spectral classifications for young stars present; c) to investigate the distribution of spectral properties ($\mathrm{T_{eff}}$, $\log(g)$, $\mathrm{[Fe/H]}$, luminosity, age, mass) among the young star populations; d) to contribute to the characterization of the PMS stage by providing physical HR diagrams; e) to investigate how the spectral and physical properties of young stars relate to the evolution of young clusters, in terms of their circumstellar emission statistics. 

This study relates various datasets with multiple parameters, which are presented in a collection of maps and diagrams. In view of this, we produced a number of Figure Sets that are available in the electronic version of the paper. Examples of such maps and diagrams for selected regions are presented as Figures in the print version. The organization of the paper is as follows: In Section \ref{sec:observations} we describe the generalities of the observed fields and spatial distribution of all targets; in Section \ref{sec:analysis} we describe our methodology to determine stellar parameters from the APOGEE near-infrared spectra, we present a selection of member candidates for each region using a clustering algorithm and we discuss comparisons with other studies. In Section \ref{sec:results}, we present the distribution and basic statistics the parameters determined for each population. In Section \ref{sec:discussion} we present a short discussion on caveats --pertinent to our data-- for the characterization of pre-main sequence populations. Finally, we summarize and discuss our work in Section \ref{sec:summary}.

\begin{figure*}[ht!]
\begin{center}
\includegraphics[scale=0.45]{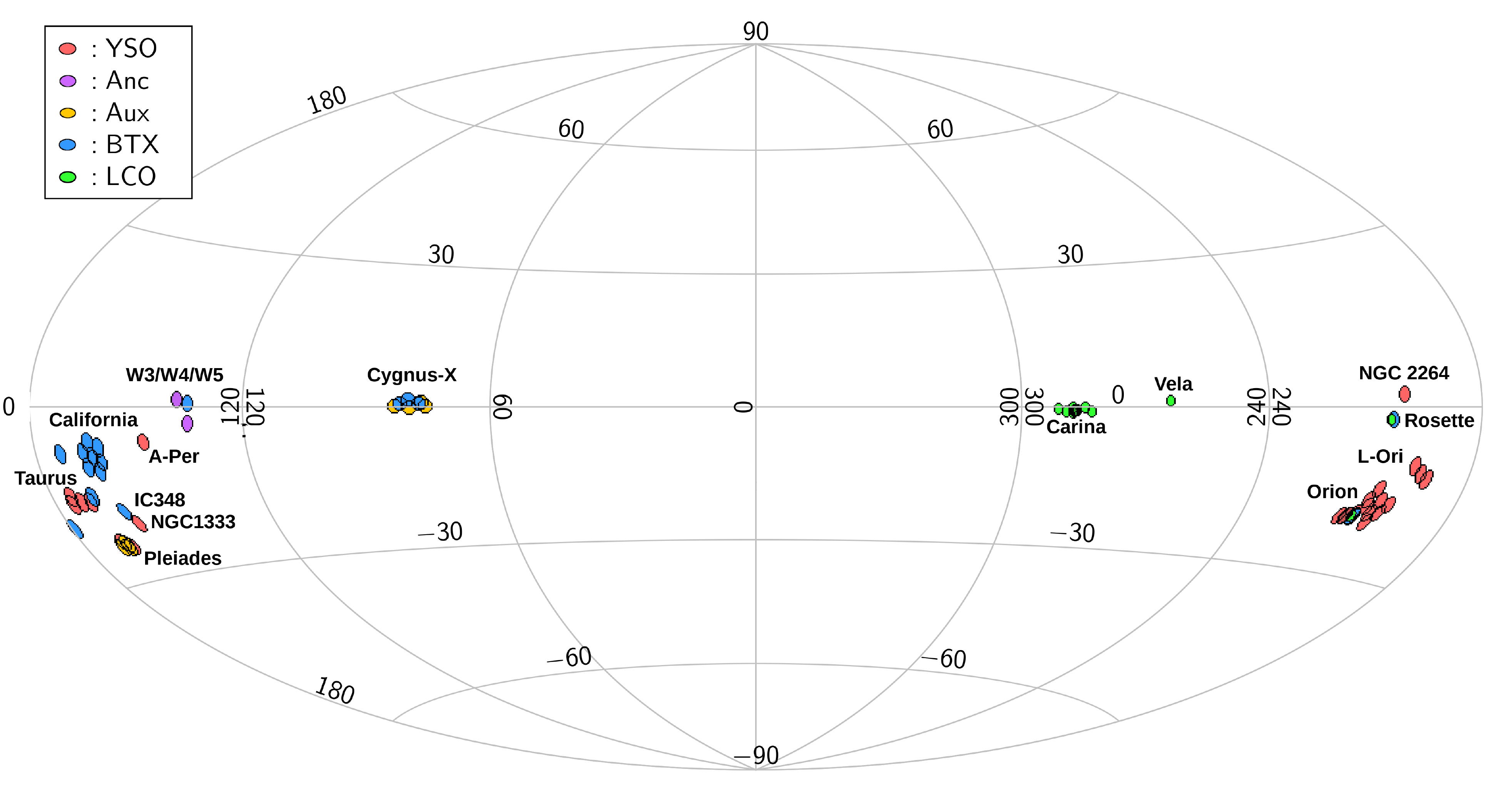}
\caption{A map, in Galactic coordinates, showing the locations of all the fields listed in Table \ref{tab:yso_sample}, along with labels identifying the eighteen main regions reported in this study. Each circular field is projected with a realistic radius of 1.5 or 1.0 deg for regions observed with the North or South APOGEE instruments, respectively. The colors of the fields refer to the field-subprograms described in Section \ref{sec:observations}. \label{fig:bigmap}}
\end{center}
\end{figure*}

\begin{figure*}
\begin{center}
\includegraphics[scale=0.85]{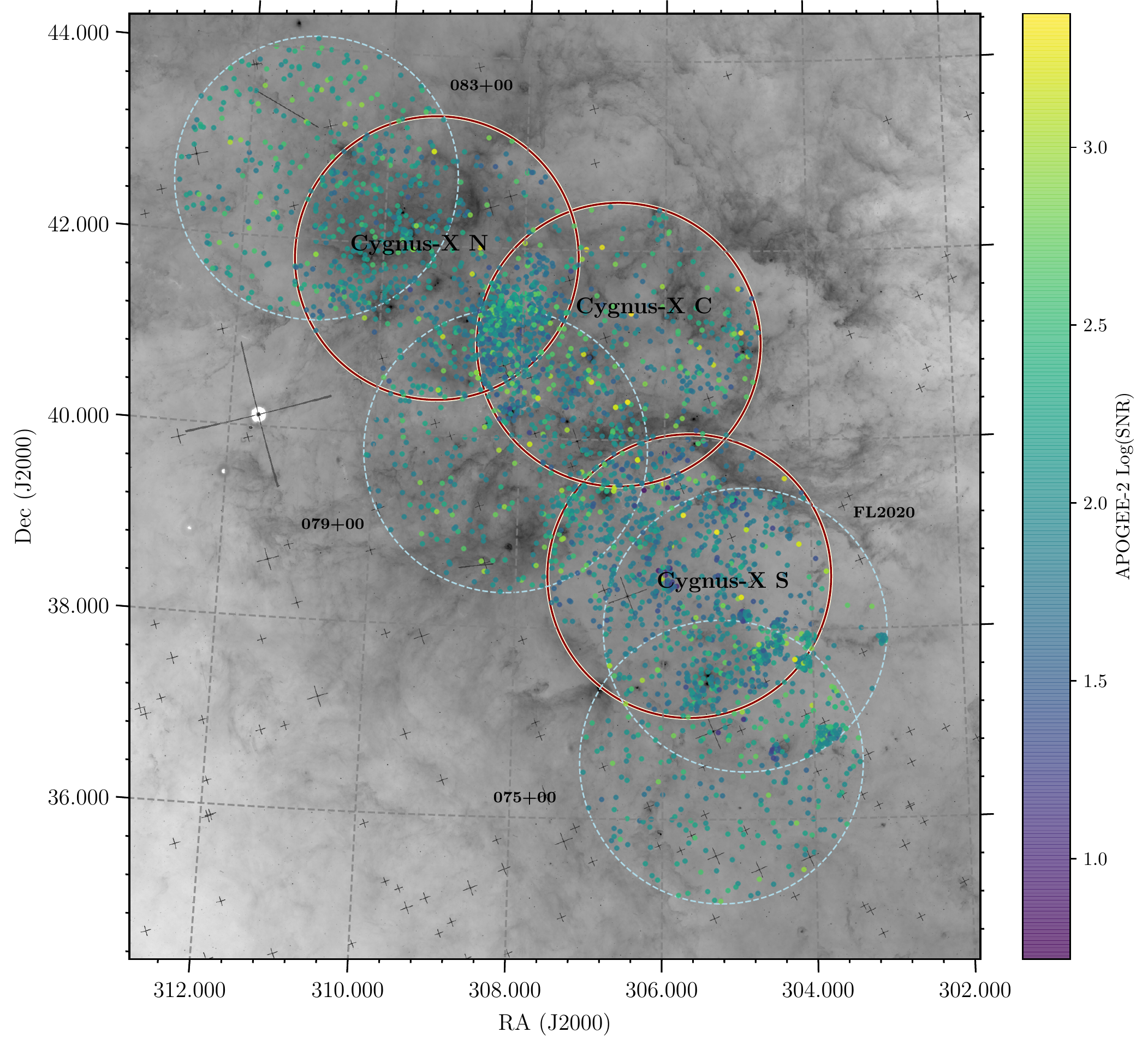}
\caption{APOGEE-2 Cygnus-X Complex Fields are shown, delimited with circles; red circles indicate the main region fields, and the dotted blue circles indicate ``auxiliary" fields (see Appendix \ref{App:sub-programs}) also searched for candidate members. Color dot symbols indicate the positions of all scientific targets. 
The grayscale image in the background shows a 12 $\mu$m dust emission map from the 
WISE WSSA Atlas \citep{WSSA14}. The colorbar indicates values of the signal-to-noise ratio after visit combination. The component figures (51) for all the regions studied in this paper, are available online in the Figure Set \ref{FS1}, which also include maps highlighting H mag brightness and Number of Visits.  
\label{fig:posmap}}
\end{center}
\end{figure*}

\section{Observations} \label{sec:observations}

The APOGEE and APOGEE-2 programs in SDSS used state-of-the-art fiber spectrographs, which use a circular field of view radius and capacity for up to 300 simultaneous fiber observations. In a nutshell, fibers are plugged across a metal plate on the focal plane of the telescope, with target positions pre-drilled on the plate. As described in \cite{Wilson2019}, the fibers are bundled for transmission to the bench spectrographs, where dispersed light is collected onto three near-IR detectors which divide each spectra into three windows: blue $\mathrm{15145-15810\ \AA}$, green $\mathrm{15860-16430\ \AA}$, and red $\mathrm{16480-16950\ \AA}$ \citep[see][for details]{Wilson2019}. One of the spectrographs is installed in the Northern Hemisphere (APOGEE-N), on the Apache Point Observatory 2.5m Sloan telescope \citep{Gunn06}, and the other one is located in the Southern hemisphere (APOGEE-S), on the Las Campanas Observatory 2.5m Du Pont telescope \citep{Bowen1973}.The APOGEE-N and APOGEE-S instruments are nearly identical \citep{Wilson2019}, although some technical aspects at the telescope level are different. For instance, the field of view for the APOGEE-S instrument is smaller: $0\fdg 95$ versus a $1\fdg 5$ maximal FOV radius. At both telescopes, the plate system requires the use of a central post that blocks light in the field center, but while at APO the radius is $1\farcm 6$, the one at LCO is $5\farcm 5$ for the on-axis acquisition camera and another region obscured for the off-axis guider camera. A more detailed description of the SDSS APOGEE instrumentation can be found in \citet{Wilson2019}.  Generalities of how fields and targets were assigned can be consulted in the APOGEE-2 general, APOGEE-2N and APOGEE-2S targeting papers \citep[][]{Zasowski17,Beaton21N,Santana21S}. General spectral data processing is described in \citet{Nidever15} with the APOGEE Stellar Parameters and Chemical Abundance Pipeline described in \citet{GarciaP16}. This paper makes use of reductions from Data Release 17 and the exact data handling is given schematically in \citet{DR17} with the full descriptions in Holtzman et al. (in prep.).

The core program in APOGEE and APOGEE-2 is a survey of all components of the Milky Way and some of the closest galaxies in the Local Group through H band spectra of red giant stars. Young star samples come from a rather diverse series of short, alternative projects in various SDSS-IV APOGEE-2 sub-programs. The targeting papers mentioned above contain a detailed description of all sub-programs of the survey, with some aspects of sub-programs given \citep{Zasowski17,Beaton21N,Santana21S}. In the Appendix \ref{App:sub-programs}, we describe the specific APOGEE-2 sub-programs that are relevant for our study. 

The locations of all the corresponding APOGEE fields for this study are shown in Figure \ref{fig:bigmap}. Then, in Table \ref{tab:yso_sample} in the Appendix \ref{App:RegionsAndTargets}, we list the star formation and young cluster regions considered for the present overview. We list the assigned working names for the projects, the center coordinates of the APOGEE-2 fields observed for each project, as well as the APOGEE-2 sub-program category to which they pertain. We also list the total number of science targets observed in each field, previous to the candidate member selection. It is important to mention that for each field, several 1-hour nominal integrations or ``visits" are listed, using MJD as identifier. Multiple visits were assigned to each field in all cases for several reasons: In the APOGEE program, a nominal 1-hour visit allows the instrument to achieve a signal to noise ratio SNR$>$100 for stars with $\mathrm{H<11.2}$ magnitudes. This is known as the "short" cohort. Medium (3 summed, 1-hour visits) and large (6 summed, 1-hour visits) cohorts, allow to achieve the SNR requirement for stars with brightness limits of 12.1 and 12.5 mag. It is worth noting, however that in several of the star formation and young star cluster fields, the targets in the large cohort were pushed down to H=13 mag because the SNR requirement for some scientific goals could be relaxed\footnote{For example, bright, massive stars at a distance of 2 kpc can be classified down to 1 spectral sub-type using Brackett-11 and Bracket-13 lines which can be well characterized with SNR$>60$ \citep{RomanLopes18}.}. Therefore for the programs described in this paper, the use of multiple visits, is not only justified in terms of nominal brightness cohorts, but also for other reasons, described as follows: a) to reconfigure fibers to assign different targets in each visit. This way it was possible to observe a larger number of targets in crowded regions because the collision radius between fiber connectors limits the minimum separation to approximately 1$\arcmin$ in the Northern instrument and 85$\arcsec$ in the Southern instrument; b) to allow to observe embedded stars in regions of low to moderate extinction, down to H=13 mag; c) to investigate radial velocity (RV) variability, aimed, among other goals, to find and classify spectroscopic binaries.

In Figure \ref{fig:posmap} we show, as an example, a map of the Cygnus-X region where all APOGEE targets are layed out along with the limits for each observed field. We indicate with a color table, the SNR value for each target, showing values above 100 for a majority of sources, although reliable spectral classification and RV determinations can be obtained down to $\mathrm{30<SNR<75}$. In Figure Set FS1 of the Appendix (available in the electronic version of the paper) we show similar maps for each of the main target regions, with color tables indicating SNR, as well as H magnitudes of the targets and the number of times (visits) a target was observed in order to achieve the desired SNR level depending on the brightness of the source.

The selection of target regions allows us to set common goals, in relation to  the classification and characterization of young stars across a diverse set of local environments. In the Appendix \ref{App:selection}, we provide some brief comments on particular regions in terms of individual science goals originally proposed for the various APOGEE internal projects. Several of the targets represent nearby ($0.1<d<0.5$ kpc) complexes or clusters dominated by low-mass star formation (e.g. Taurus, IC 348, Pleiades, $\alpha$-Persei), while another group of regions (Cygnus-X, Carina, W3/W4/W5) can be classified as massive star forming complexes outside the solar neighborhood ($1<d<6$ kpc). In a mid-range group, regions like Orion, NGC 2264 or the Rosette are populations where massive stars are present, but cannot be characterized as massive star forming regions. 

We must stress that conforming equivalent target samples across the diverse star forming environments was not necessarily a goal that framed each APOGEE observational program. The paper by \citet{Cottle18} details a complex selection function for the young stellar objects (YSO) survey in APOGEE-2. For instance the target selection for the Orion Nebula Cluster was focused on the kinematic evolution of the cluster; fields observed toward the Orion A cloud aimed mainly at the characterization of low mass YSOs; then, observations of fields in the Orion OB1a and OB1b regions focus on study of the 5-10 Myr exposed groups that are evolving out from the cloud. The targets for those distinct populations were selected using lists of confirmed members from different authors to define \textit{loci} in color-magnitude diagrams that were used to define candidate members, with additional candidates added using photometric variability as the main criteria. The resultant target lists, then, have to be filtered into brightness groups and run through a prioritization and collision radius avoidance algorithm, producing samples that cannot be statistically complete in terms of spectral type, or mass. 

This last aspect is more notorious for massive star forming region targets, where published lists of known members in the low mass range are less common than lists of OB members, and thus selecting low mass YSO sources from photometric loci was more prone to field star contamination. It is also important to mention that for some regions, particularly those designed during the first two years of the survey, the selection functions could not include distance estimations because Gaia parallaxes were not yet available. One clear example of this is W3/W4/W5, where low mass young star candidate targets were selected using only color-magnitude loci. As will be described in the next section, this unfortunately results in a large fraction of field contaminants. Using only photometric selections for candidate members produce a significant number of contaminants as red giant stars can mimic the colors of embedded young sources. Moreover, even nearby regions like Taurus were significantly affected by the absence of parallax determinations during the selection function observed in the first visits. 

Despite the heterogeneous nature of the sample, our methodology allows us to extract important about diverse young stellar populations in an unprecedented collection collection of spectroscopic samples. This results in one of the largest catalogs of young stellar properties to date.

\section{Methods \label{sec:analysis}}

In this section we describe the processes used to determine six stellar parameters (\teff,\ \logg, [Fe/H] $\log{L/L_\odot}$, mass, age) for the stars in the sixteen regions described in the previous section. We describe a clustering method used to remove field population contaminants from the observed samples, and display the results on maps where we show the distribution of the parameters for each region. Finally, we describe how the selection of members with significant infrared excess was done in order to find and classify, in rough evolutionary stages, those members with circumstellar emission via protoplanetary disks. 

\begin{figure*}[ht]
\includegraphics[scale=0.4]{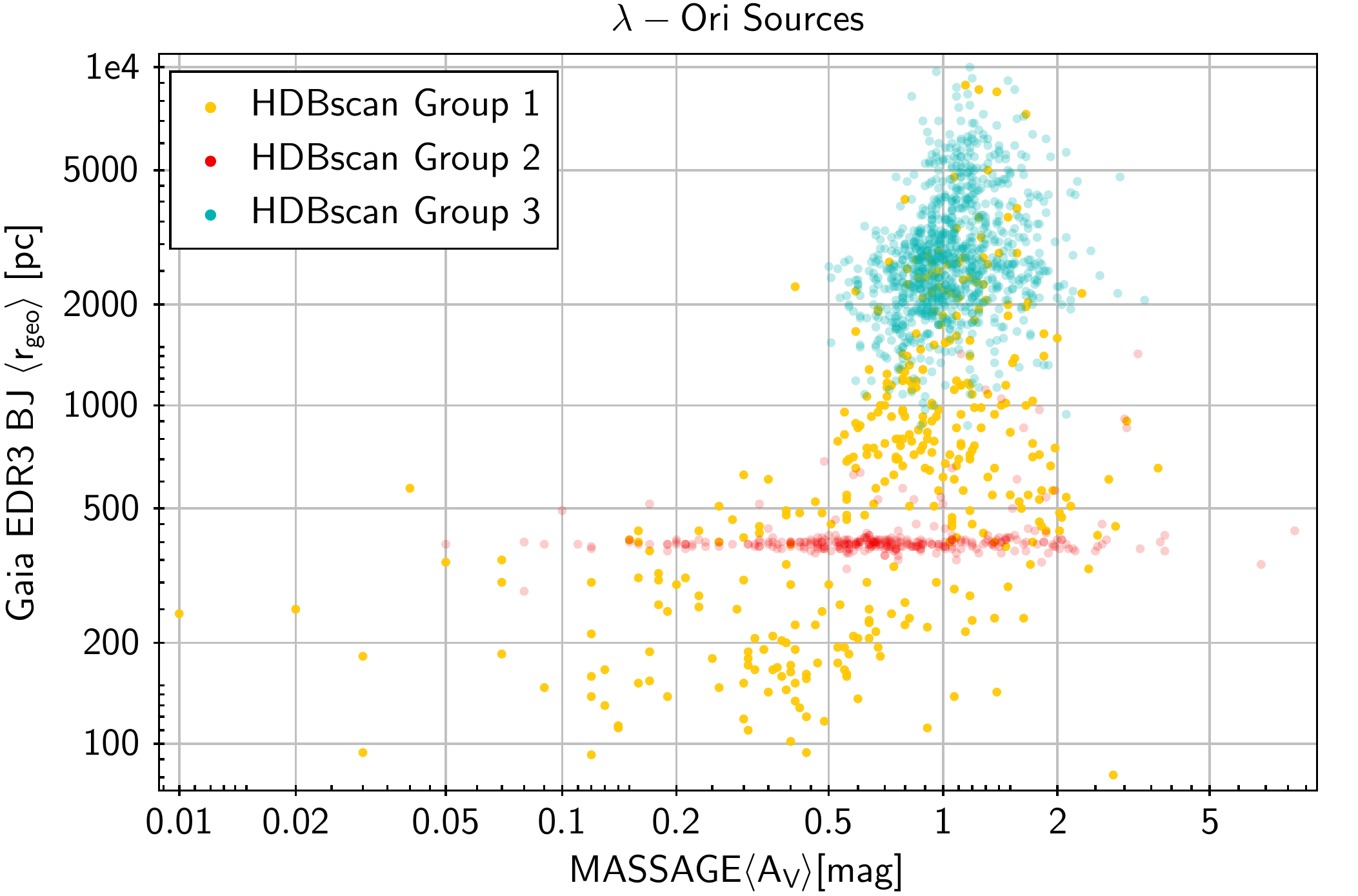} \includegraphics[scale=0.4]{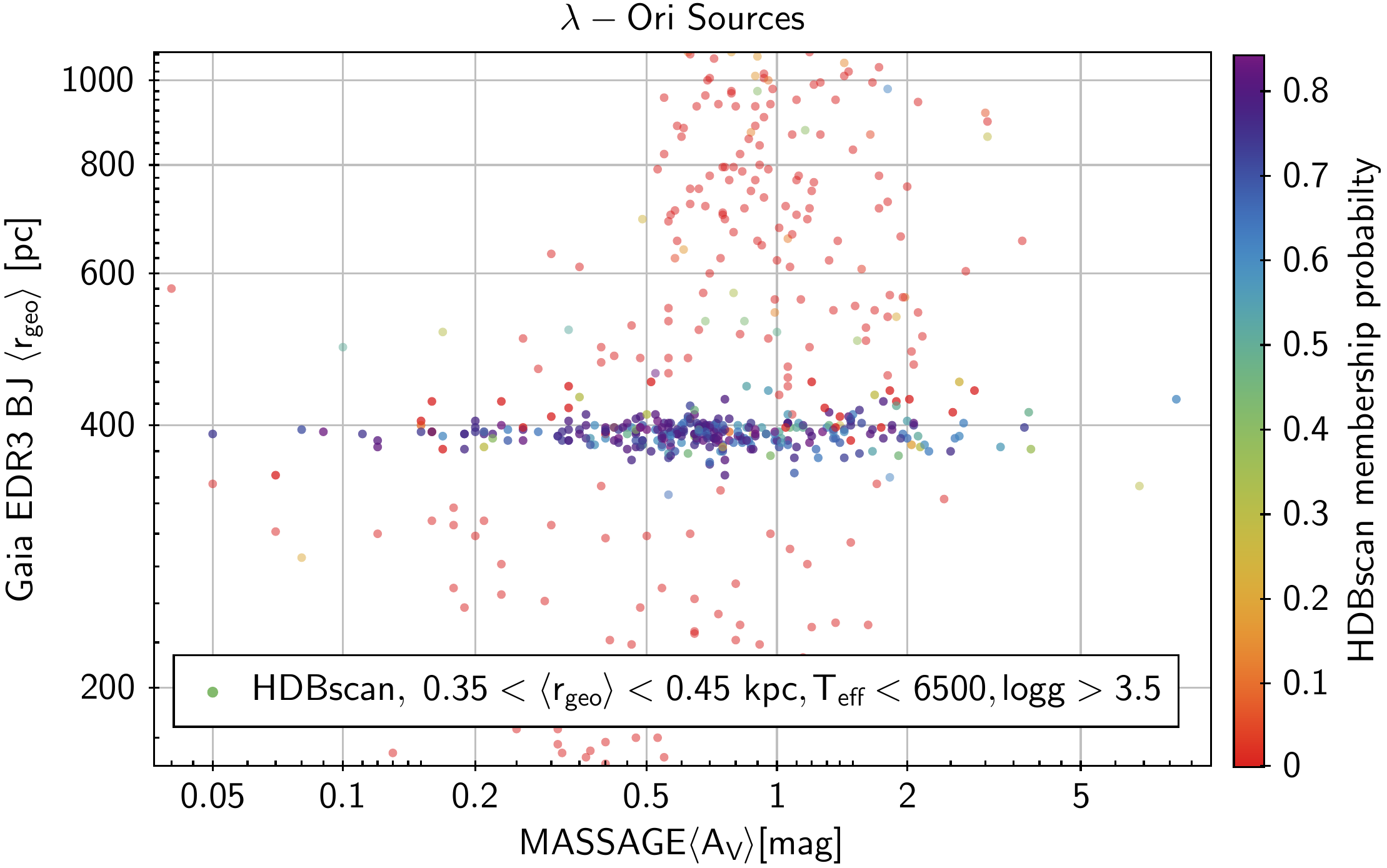}
\caption{Example of region member selection refinements. Both panels show a distribution of the geometric distance estimates from \citet{BJEDR3} versus mean extinction, $\langle A_V\rangle$, from \texttt{MassAge}. The left panel shows the \texttt{HDBSCAN} resultant cluster groups for all targets observed in the $\lambda$-Ori region, where it can be seen how Group 2 stars contain most of the sources coinciding with the expected distance to the cluster (about 380 pc), stars from Group 3 are in most cases background field sources, and group 1 may contain a few sources at the expected distance. In the close-up on the right panel, the colors of the points indicate the cluster membership probability, where the stars with bluer colors pertain to a group that contains stars from Groups 1 and 2 with distances between 350 and 450 pc, \logg$>3.5$ and \teff$>6500$. Those restrictions define the HRMs for this region. \label{fig:avdist}}
\end{figure*}

\subsection{Spectral Parameter determinations with APOGEE Net II \label{sec:analysis:subsec:APOGEENet} }

The standard spectral analysis pipeline for the APOGEE core sample \citep{Holtzman15} is not adequate for the determination of spectral parameters in PMS low mass stars \citep[see][]{Cotaar14}. For this reason, \cite{Olney20} developed APOGEE Net, a data-driven approach to derive \teff, \logg, and [Fe/H] of stars observed with APOGEE. These authors have trained a deep convolutional neural network using previously derived parameters (or ``labels") for red giants and stars with \teff$>4200$\ K from ``The Payne" pipeline \citep{Ting19}, as well as parameters derived from photometric relations for the M- and K-type main sequence dwarfs and PMS stars (for which ``The Payne'' could not derive reliable values). Furthermore, the neural network was trained on the estimates of \teff\ and \logg\ for a sample of PMS stars, derived from interpolation of the photometry of such stars across PARSEC PMS isochrones \citep{Bressan12}. 

As the PMS estimates originally had a lot of scatter, the neural network was trained to replicate the effects of temperature on spectra from main sequence dwarfs, and the effects of surface gravity from giants. This way, APOGEE Net was able to refine the measurements of the parameters of PMS stars. It yielded self-consistent \teff\ measurements (valid for stars with \teff$<$6500 K), eliminating systematic gaps that were present in the earlier extractions, based mainly on cross-matching of the spectra against synthetic templates \citep[e.g.][]{Cotaar14, Kounkel18}. Furthermore, it made possible for the first time to measure \logg\ values calibrated to the isochrones, allowing this parameter to be used as a proxy for age. This has allowed independent confirmation that the spatial distribution of YSOs in the Orion Nebula Cluster (ONC) depends strongly on the age of the stars, as was previously proposed by \citet{Beccari17}, with the older sources more distributed throughout the ONC, and the younger sources being predominantly concentrated near the Trapezium group.

The APOGEE Net was expanded by \citep{Sprague22}, who complemented the training set by adding a comprehensive collection of \teff \ and \logg \ estimates from the literature, allowing a consistent parameter determination across the HR diagram, from late M to OB stars. This way, parameter determination is no longer limited to The Payne coverage and becomes self-consistent across the spectral type and luminosity class space. The \citeauthor{Sprague22} methods were applied to the entire APOGEE-2 target list and the results we present in this study, come from the most recent calibration of the neural network, named APOGEE Net 2, that includes all sources from the SDSS DR16 \citep{DR162020} and DR17 \citep{DR172022}.

\begin{figure*}[ht]
\begin{center}
\includegraphics[scale=1.0]{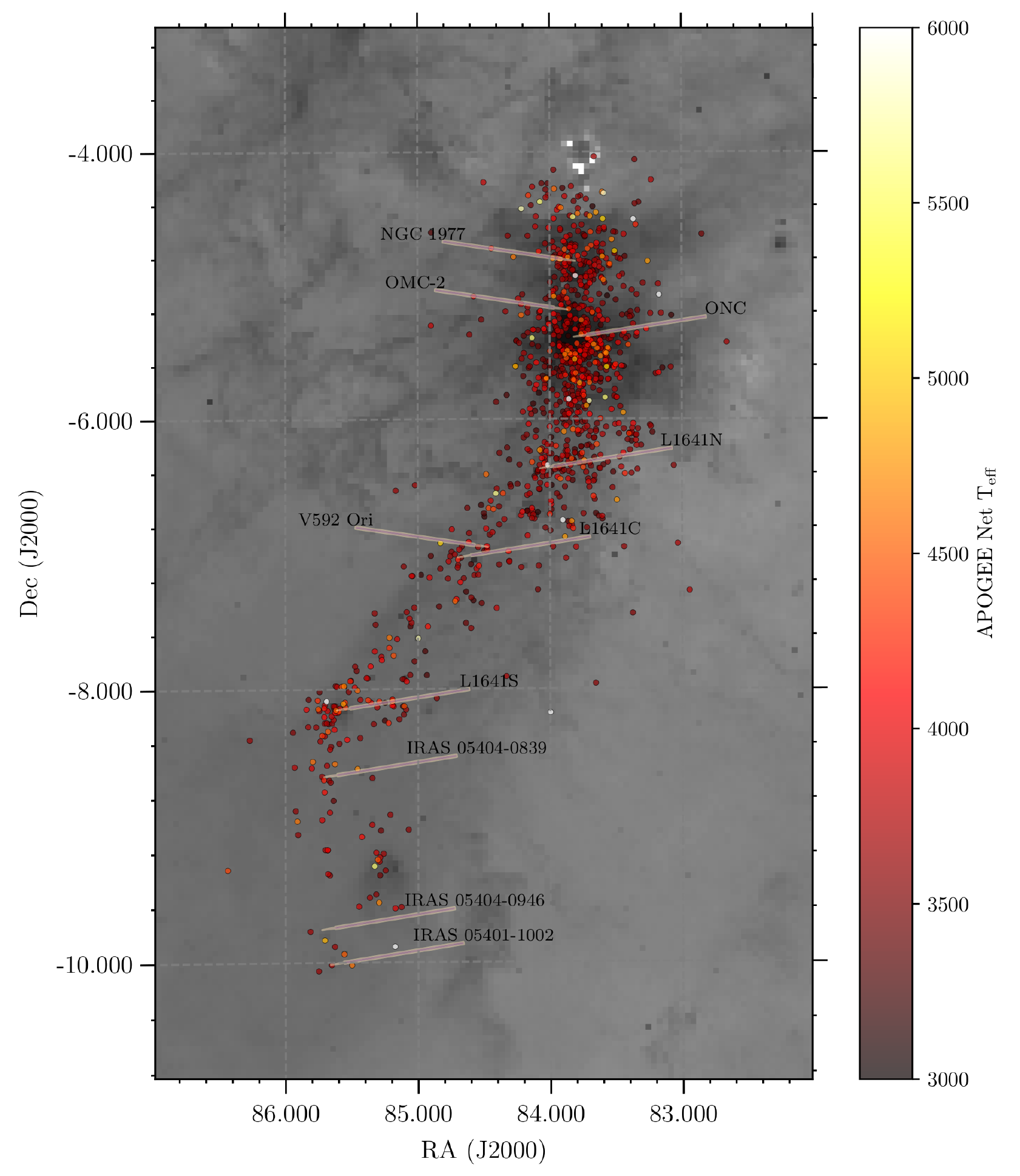}
\caption{A map of the Orion A Region, showing the \teff\ determined for APOGEE-2 targets using APOGEE\ Net. Color dot symbols indicate the positions of all scientific targets. 
The grayscale image in the background shows a 12 $\mu$m dust emission map from the 
WISE WSSA Atlas \citep{WSSA14}. The colorbar indicates \teff. All 
the component figures (102) for the 17 regions studied in this paper, are available online in the Figure Set \ref{FS2}, including maps highlighting\ \logg, [Fe/H], stellar mass, stellar age and luminosities.  
\label{fig:ANMT}}
\end{center}
\end{figure*}		

\subsection{Estimation of Physical Properties \label{sec:analysis:subsec:MASSAGE} }

For each of the regions, we obtained estimations of luminosities, masses, and ages, as well as visual exinction ($\mathrm{A_V}$), using an IDL-based code named \texttt{MassAge} 
(J. Hern\'andez et al. in prep). \texttt{MassAge} uses as input the \teff \ estimates from the APOGEE Net 2 pipeline, geometric distance estimations from \citet{BJEDR3}, and photometry from Gaia EDR3 (Gp, Rp, Bp), and 2MASS (J and H). The uncertainties in the estimated values are obtained using the Monte Carlo method of error propagation \citep{Anderson76}, assuming Gaussian distributions for the uncertainties in each of the input parameters. 

\begin{figure*}[ht]
\includegraphics[scale=0.8]{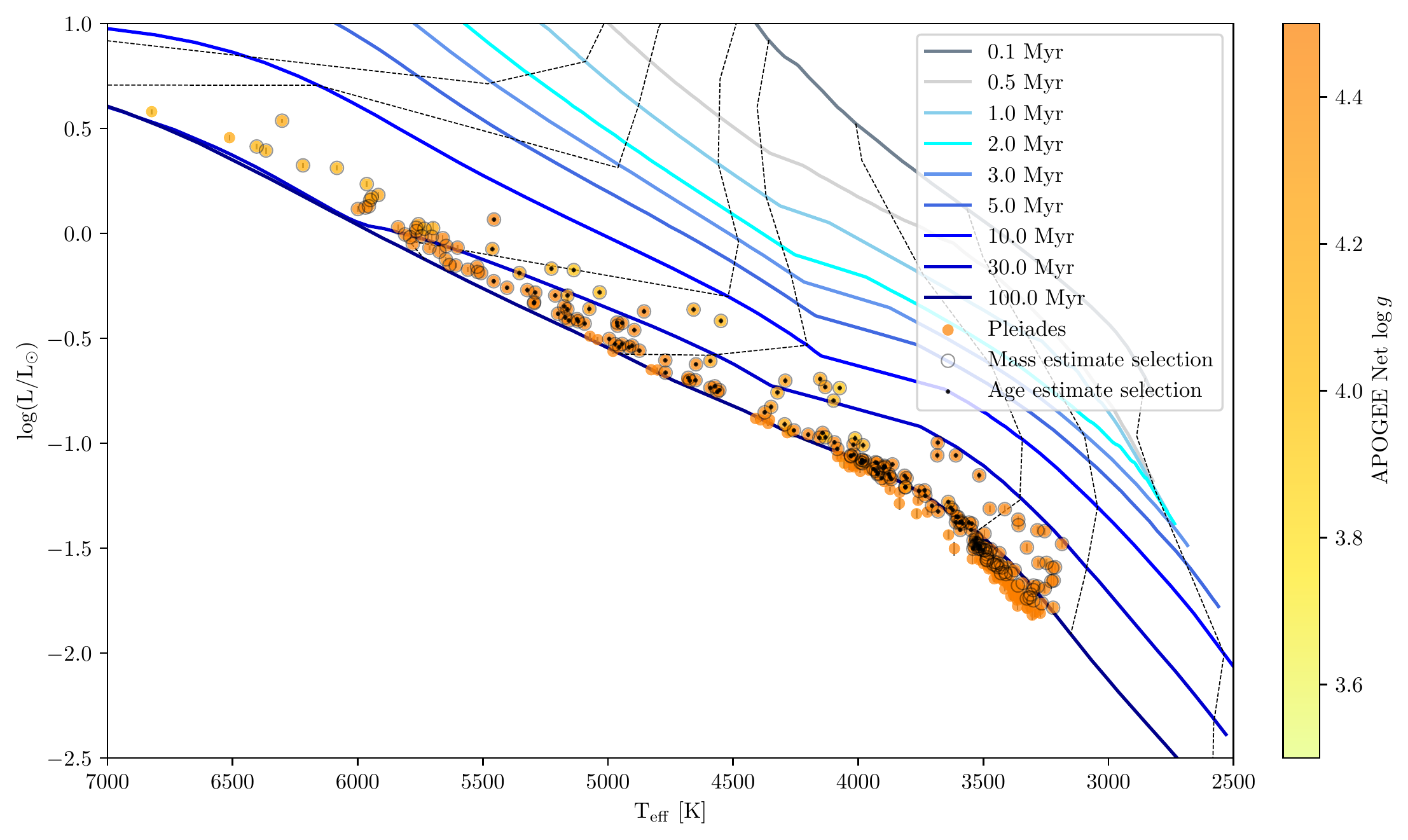}\\ \includegraphics[scale=0.8]{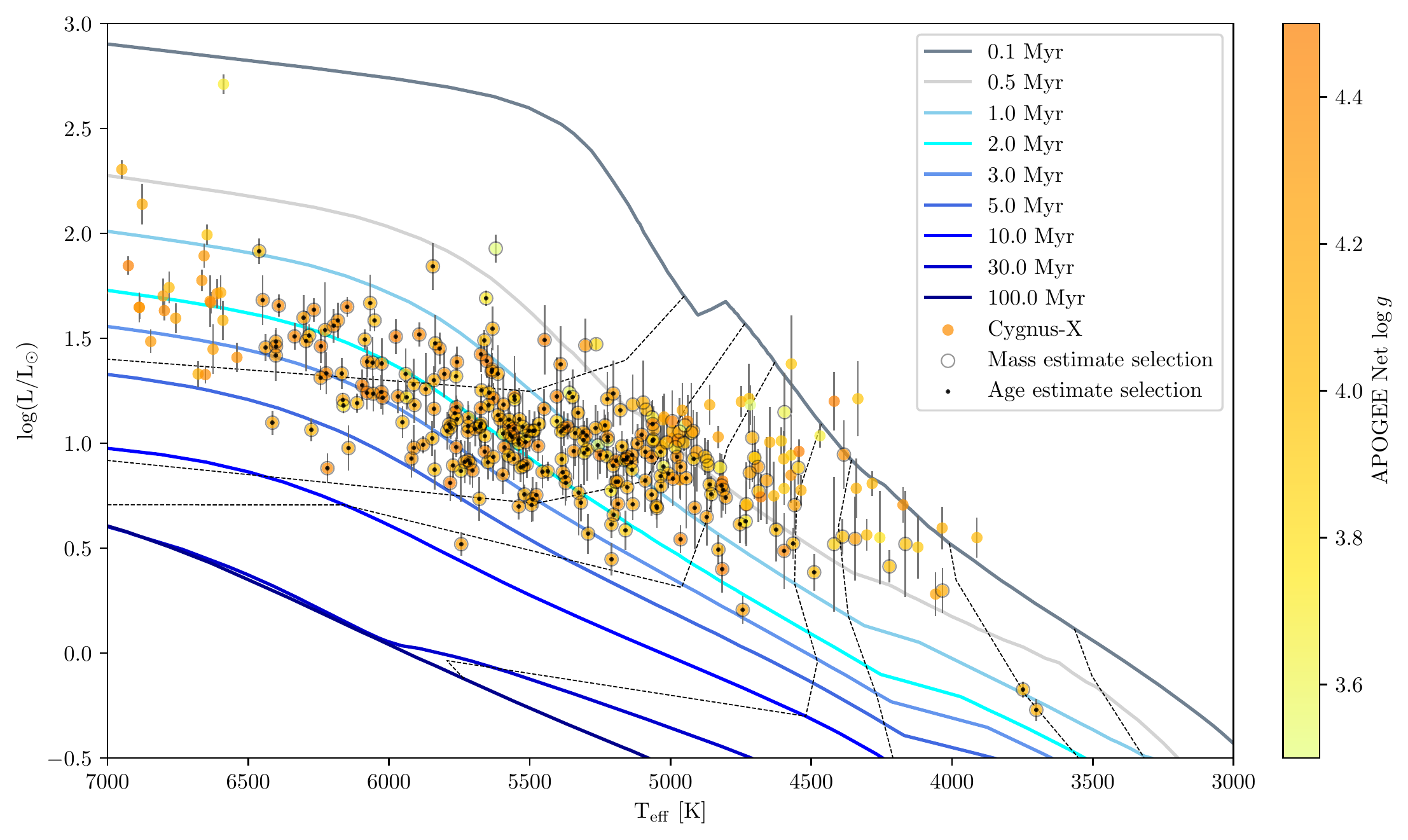}
\caption{Examples HR diagrams for the Pleiades (top) and Cygnus-X regions (bottom). In the diagrams, solid curves are PARSEC-COLIBRI PMS isochrones at 0.1, 0.5, 1.0, 2.0, 3.0, 5.0, 10, 30 and 100 Myr, while the dashed curves indicate constant mass locus for models of 0.1, 0.3, 0.5, 0.8, 1.0, 1.5,  2.0, 3.0 and 5.0 M$_\odot$. All the diagrams for the sixteen regions studied in this paper, are available online in the Figure Set \ref{FS3}.
\label{fig:HRdiags}}
\end{figure*}

The visual extinction, $A_V$, towards each source is estimated, along with a confidence range, by means of a Monte Carlo method that minimizes the differences between the extinction-corrected colors and the expected intrinsic colors from \citet{Luhman20}.  The observed colors are corrected using the extinction law of \citet{CCM89}, assuming a canonical interstellar reddening law (for simplicity we used $R_V$=3.1 for all regions).  Luminosities are derived using the extinction-corrected J magnitude, the bolometric correction for PMS stars from \citet{Pecaut13}, and Gaia EDR3 geometric distance estimations, $\mathrm{r_{geo}}$ from \citet{BJEDR3}. Finally, stellar masses and ages, plus confidence ranges, are obtained by means of Monte-Carlo sampling and interpolation within the PARSEC-COLIBRI evolutionary model grid \citep{Bressan12} on the HR diagram.

\subsection{Field Contamination Removal using Density Based Clustering Methodology \label{sec:analysis:subsec:HDBSCAN} }

As described in Section \ref{sec:observations}, a significant portion of the APOGEE targets are physically unrelated to the corresponding region. This is because the original target selection criteria were rather diverse for the different APOGEE programs. Therefore, it is necessary to apply a method to remove, as uniformly as possible, most field contaminants. Below we describe such methodology. The method combines a clustering algorithm and distance cut refinements in order to narrow down our samples and generate final lists of \textit{probable region members} (hereafter PRMs).

We chose the density-based clustering algorithm \texttt{HDBSCAN} \citep{hdbscan_ref} implemented under the \texttt{R} environment \citep{R_ref} using the \texttt{dbscan} library \citep{fast_hdbscan}. The hierarchical nature of \texttt{HDBSCAN} allows us to detect stellar groups at different densities and discard points in locally non-dense regions as noise. Given the vast diversity of regions explored, we needed a uniform scheme to globally isolate each environmental highest probable member list. The algorithm was applied to four dimensions of our catalog: Gaia proper motions $(\mu_\alpha, \mu_\delta)$, Gaia parallax, and APOGEE Net \logg. We use these parameters as the main discriminators because they combine high-quality astrometric information with the spectroscopically derived parameter that is most sensitive to young sources. \texttt{HDBSCAN} was fed with the uncorrelated principal components that preserved the dataset variance at a 90\% level.

We implemented the nearest-neighbor distances routine of the package under a Euclidean metric, imposing only a minimum number of cluster points of 20 sources (10 sources in the cases of California and NGC\ 2264) which works as a "smoothing" factor of the density estimates implicitly computed from \texttt{HDBSCAN}.   As a result, three output parameters are produced for each source in each region initial catalogs: a data cluster number, membership probability, and outlier score. For most catalogs, the algorithm produced three data clusters, and four in a few cases. 

\input{./Table1}

\subsection{Additional Parameter restrictions \label{sec:analysis:subsec:restrictions} }

In Table \ref{tab:restrictions} we list, for each region, a series of additional restrinctions we applied to the \texttt{HDBSCAN} resultant samples. Restrictions are applied for the astrometric dimensions and at the six stellar parameters

We cover regions within a considerable range of distances (0.1-4.1\ kpc in average). As mentioned by \citet{hunt21}, distant regions are difficult to separate from field stars, as proper motions and parallaxes become decreasingly informative at large distances. The Gaia geometric distance ($\mathrm{r_{geo}}$) errors increase from tens to hundreds of pc as we move from nearby (0.1-0.5 kpc) to distant complexes (1-4 kpc). This way, as distance increases \texttt{HDBSCAN} is more prone to confuse members as field sources, assigning low probability membership or high outlier scores to them. At larger distances, confusion by overlapping sources and multiple components along the line of sight also increase. In order to mitigate this, we removed from the catalogs all sources with Gaia RUWE values larger than 1.4 \citep{Lindegren18}. In addition, we used $A_V$ vs. $\mathrm{r_{geo}}$ plots to determine if the membership probabilities converged well for a specific group. Initially, we considered that PRMs should typically lie on a narrow horizontal band --a co-distant object layer-- that may form above a certain extinction threshold value. In Figure \ref{fig:avdist} we show an example of the derived group distribution for the region $\lambda$-Ori in the  geometrical distance ($\langle \mathrm{r_{geo}} \rangle$) vs. $A_V$ plot. Most sources with \texttt{HDBSCAN} membership probabilities values above 0.6 coincide well with most co-distant sources in the best defined group, but we show also how some objects with low membership scores also coincide with the co-distant object layer. Thus, for each catalog, we define ranges in distance and total proper motion ($\mu=\sqrt{\mu^2_\alpha + \mu^2_\delta}$) that contain the co-distant object layer, and in this way we are able to include a few more sources in each sample that fall within that distance, even if they have low membership scores. This criteria removes additional contaminants from the group of interest while including possible PRMs that could end in other groups due to the nuissances of the clustering algorithm. 

The reliability of the six main parameters depends strongly on both the population environment and the spectral type. It is not trivial to take into account the diversity of conditions the young star groups are subject to, particularly the non-uniform extinction. Spectral classification from APOGEE data is particularly difficult for late M types, where the fit of the continuum is complicated by the overabundance of atomic lines. For early types above solar, model degeneracy from pre- to main-sequence and non-LTE effects rapidly become notorious, resulting in large discrepancies in the spectral-type to \teff \ conversion used for the neural network training samples. In addition, the isochrone/constant mass line grid across the HR diagram is far from regular, which complicates determination of physical parameters, specially for the earlier types and for sources falling close to the edges of the isochrone grid. For all these reasons, in order to provide catalogs with reliable parameters, we decided to further limit our samples according to conservative, yet safer cuts. 

We started by restricting \teff \ to values below $7000$\ K and \logg \ values above 3.0. The first cut prevents us from using underestimated temperatures of intermediate to high mass sources. The second cut removes a majority of field evolved-star contamination.

In the final catalogs, we include only estimates of luminosity and mass for sources with $\mathrm{T_{eff}<6500\ K}$, as we found that the interpolation method of \texttt{MassAge} is optimal below that value. As for age estimations, mass versus age plots for all the samples show that sources for which the interpolation is performed near the edges of the isochrone/constant mass line grid tend to group in a false sequence, while correctly interpolated values lie below that sequence. In addition, for some regions, sources within the lowest \teff \ bins could have overestimated values that, in turn, would bias the sample toward large age values. The last column in Table \ref{tab:restrictions} shows the \teff\ values used in each sample in order to retain only those sources with reliable age estimates. In the IC\ 348, NGC\ 1333 and Taurus, we also removed stars with $\log{L/L_\odot}<-1.0$ from \texttt{MassAge} in order to remove \teff overestimated values from the catalog.

\subsection{Separation of front and back populations in Orion B and Orion OB1 \label{sec:analysis:subsec:frontback} }

In the case of the Orion B and Orion OB1 samples, we noticed that both restricted samples, a plot of extinction vs. distance, showed a clear minimum in the density of sources at $\mathrm{r_{geo}\sim 370\ pc}$. This is consistent with the studies of \citet{Briceno19} and Hern\'andez et al. (subm.). For this reason, we decided to divide those samples into two groups, one with $\mathrm{340<r_{geo}/pc<370}$ (named ``front") and another one with $\mathrm{370<r_{geo}/pc<430}$ (named ``back"). The groups are listed separately in Tables \ref{tab:restrictions} and \ref{tab:nrm_ave}, and they are also shown with distinct symbols in the maps of Figure sets 1 and 2 in Appendix \ref{FS1}.

\section{Results} \label{sec:results}

\par Applying the above selection criteria, a total of 3360 PRM sources were considered for further analysis (this is the sum of the number of sources for which \teff\ and \logg\ were determined for each region, as listed in column 3 of  Table \ref{tab:restrictions}). In Appendix \ref{App:NRM_Tables} we present a master table listing that includes identifications, positions, geometric distance, visual extinction, and the values of the six main parameters determined in this paper: \teff,\ \logg,\ [Fe/H], mass, age, and luminosity. In Table \ref{tab:nrm_ave} we list median values with their corresponding median absolute deviation, for each of these six parameters in each region. 

\input{./Table2}

\subsection{Parameter Maps and Distributions \label{sec:results:subsec:MapsAndHistograms}}

The PRM lists were used to map the distributions of the six main parameters. In Figure \ref{fig:ANMT} we show, as an example, the \teff\ distribution in the Orion A complex. These maps provide an unprecendented overview of the properties of solar and sub-solar type young stars in three kinds of star forming regions: a) nearby ($\mathrm{d<0.4\ kpc)}$ complexes predominantly forming low mass stars, b) intermediate distance ($\mathrm{0.4<d<1.5\ kpc)}$ complexes with a moderate content of massive stars, and large star forming complexes with significant massive star content in the neighbor spiral arms ($\mathrm{d>1.5\ pc)}$, all obtained with the same type of instrument and using the same processing for the resulting datasets. The online figure set contains similar maps for all sixteen regions.

\subsubsection{Landmark Labels \label{sec:results:subsec:MapsAndHistograms:subsubsec:landmarks}}

The target maps and the parameter spatial distribution maps were constructed using as background image cuts from the WSSA Wise 12\ $\mu$m dust emission Atlas \citep{WSSA14}. The  12\ $\mu$m dust emission delineates with great clarity both molecular cloud and photodissociation regions in each map, without adding too much confussion from bright stellar sources or features (some extremely bright stars leave a visible residual mark). However, in those maps it is difficult to assess the location of young star clusters associated with current episodes of stellar birth. Also, most of those episodes in our regions are well characterized in the literature. We added the locations of known embedded clusters and groups listed in several catalogs: for the California, NGC\ 2264, Orion\ A and Orion B regions, we used the catalog of \citet{Porras03}; for the Cygnus-X, IC\ 348, NGC\ 1333 and Taurus regions, we used the catalog of \citet{Bica2003}; for the Carina, $\lambda$-Ori and W3/W4/W5 regions, we used the catalog of \cite{Avedisova02}; in the case of the Rosette Complex, we used the cluster list of \citet{Cambresy13}; for the remaining regions, $\alpha$-Per, Pleiades, Orion OB1 and Vela, we collected locations of known groups directly from the SIMBAD astronomical database \citep{Wenger2000}.

\subsection{Hertzprung-Rusell Diagrams \label{sec:results:subsec:HRdiags}}

We used the derived parameters to construct, for each region, Hertzprung-Russell (HR; Luminosity versus \teff) diagrams. Examples of these diagrams for two regions are shown in Figure \ref{fig:HRdiags}. The \logg\ values were added with a color scale. Symbols are marked according to the parameter restrictions described in the previous section. In all the plots we show isochrone and constant mass curves obtained from the PARSEC-COLIBRI evolutive models \citep{Bressan12}. 

It is clear from these diagrams that APOGEE Net and \texttt{MassAge} provide highly reliable stellar parameters for nearby regions, helped by very precise Gaia parallaxes and low column densities (see the example of the HR diagram for the Pleiades). However, for the most distant regions located outside our local Galactic spiral arm, current Gaia distance uncertainties increase to hundreds of parsecs. In those regions we are also dealing with overlapping clouds and consequently highly variable extinction. For these reasons,  the extinction is likely to be underestimated for many distant sources, and the resultant \texttt{MassAge} luminosities are more uncertain. See Section \ref{sec:discussion} for a more detailed discussion on the HR diagrams for the different regions.

\subsection{Average Metal Abundance}
\label{sec:results:subsec:feh}

In most regions, the average metal abundance, [Fe/H] from the APOGEE Net II tables, appears to have a mostly uniform behavior around solar values for sources with \teff $<4000$ K. Above this value, the parameter dispersion increases significantly. The effect is more notorious for samples in the more distant regions like Carina, W3/W4/W5 and Cygnus-X. \citet{Olney20} discussed how the APOGEE Net algorithm tend to predict lower values of [Fe/H] for hotter stars, but their test samples were smaller and limited to nearby regions, so we confirm this bias with the larger samples used in this this study. 

\section{Discussion} \label{sec:discussion}

One of the main results of this study is the compilation of a significant catalog of stellar properties for young stars in a collection of star forming regions around the Sun, which is presented in Table \ref{tab:nrms}. However, our analysis also helps to evince how the characterization of young star populations poses a number of complications that are not necessarily exclusive of this paper, which are important to consider as similar studies will arise for similar datasets, and similar techniques in big-data treatments.

\textit{Metal abundance}. Methods that compare observed spectra with synthetic models are based on the goodness of fit of groups of metal lines present on the spectral window. Stars of higher \teff\ have less lines, and stars with larger rotation velocities have broader lines, which may propose stars with lower [Fe/H] as better fits for early type sources or rapid rotators. Non-intrinsic variations of the [Fe/H] parameter as a function of temperature are not exclusive of APOGEE data; for instance, this was also discussed by \citet{Kos21} in optical spectra from the GALAH survey. 

By limiting our samples to \teff$<4000$\ K the datasets are reduced but are much more reliable. This way, the APOGEE Net 2 data confirm that star forming region in the solar neighborhood have solar abundance (see Table \ref{tab:nrm_ave}), which in agreement with recent studies like those of \citet{Spina14} or \citet{Kos21}. In Table \ref{tab:nrm_ave} we show how the median value of the APOGEE Net 2 [Fe/H] label for sources with \teff $<4000$ K, are all dead centered on solar for all regions except Carina, Cygnus-X and W3/W4/W5 with no reliable samples in that \teff \ range. It is not expected that star forming regions present significant deviations from solar abundance within a few kpc from the Sun \citep{Spina17}; our most distant remaining samples are in NGC\ 2264, Vela and the Rosette in the 1 kpc range, all showing a solar value as well.

\textit{Age determination}. The APOGEE Net 2 neural network produced stellar parameters across the entire HR diagram and it has been optimized to consider the spectrophotometric properties of young stars \citep{Olney20, Sprague22}. However, as we mentioned in section \ref{sec:analysis:subsec:restrictions}, APOGEE Net 2 \teff\ values in some of the regions may not be precise enough above 7000 K and below 3400 K, affecting the estimation of masses and ages with the \texttt{MassAge} routines. 

\begin{figure}
    \begin{center}
    \includegraphics[scale=0.475]{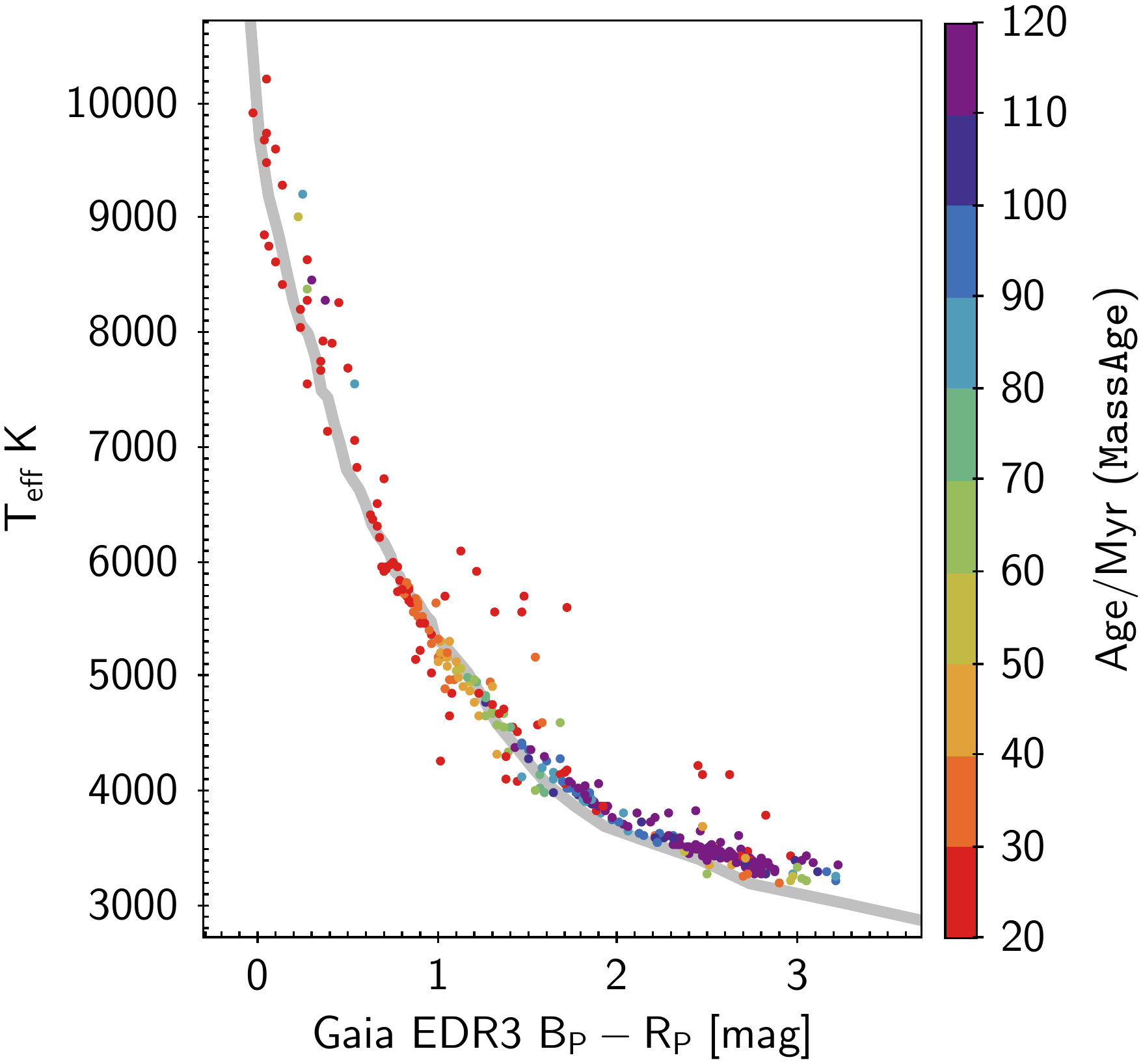} 
    \caption{Effective temperature vs. Gaia EDR3 colors for the APOGEE Pleiades sample. The symbol colors represent age estimates from \texttt{MassAge}. The solid line is the empirical sequence from \citet{Esplin18}. \label{fig:pleidisc}}
    \end{center}
    \end{figure}

In addition, we have to consider that the determination of young cluster ages also includes some model dependent issues. Due to the close distance and minimal extinction towards this cluster, our Pleiades sample presents uniform quality within 0.4 to 1 M$_\odot$ range and forms a tight sequence running slightly above the 100 Myr PARSEC-Colibri isochrone. The expected age for this cluster is 110-160 Myr \citep{Gossage18}, but the median age we obtain is 88$\pm$38 Myr. In Figure \ref{fig:pleidisc} we show a comparison of the Gaia $\mathrm{B_P-R_P}$ colors and APOGEE \teff\ values with those listed in the empirical sequence of \citet{Esplin18}; we can see how the differences between the empirical and observed sequences are very small. However, these small color or \teff\  differences translate as disperson in the individual age estimations with \texttt{Massage}. The PARSEC-Colibri isochrone curves between 30 and 100 Myr have separations barely larger than our bolometric luminosity estimations, and if we compare to other isochrone models, like MIST \citep{Dotter16} we will obtain a different set of values because the isochrone tracks do not run sufficiently close to each other to predict ages with non-significant discrepancies. Age uncertainties of a few tens of Myr are actually typical for clusters of similar age to the Pleiades \citep[e.g.][]{Caiazzo20}. As we go to younger ages the discrepancies may increase, as warned by the MIST group themselves \citep{Choi16}, who expressed the need for better modeling ot the pre-main sequence and low main sequence.

\section{Summary and Future Work \label{sec:summary}}    
    
     The SDSS-III APOGEE and SDSS-IV APOGEE-2 programs represent one of the most important near-infrared large-scale spectroscopic surveys aimed to reconstruct the evolution history of the Milky Way Galaxy. A fraction of the survey telescope time was dedicated to the development of key scientific goals, one of them being the study of young stellar populations in a number of nearby ($\mathrm{d<500\ pc}$) low-mass star forming-complexes and several massive star-forming complexes in the local and neighbor spiral arms ($\mathrm{1<d<3\  kpc}$). This paper compiles and describes data for sixteen star-forming regions surveyed as part of the APOGEE and APOGEE-2 programs.  It provides a common analysis across all datasets, including the determination of spectroscopic parameters (\teff, \logg, [Fe/H]) through the APOGEE Net II neural network \citep{Sprague22} and the estimation of three additional parameters ($\mathrm{L/L_\odot}$, stellar mass, age) through interpolation of PMS evolutionary models \citep{Bressan12} using a code named \texttt{MassAge} (J. Hern\'andez et al., in prep). We constructed HR diagrams for high-probability members of all regions. 

	We developed a simple method to extract, from the original datasets for each region, the sources that are most probable to be physically associated with the very young, sometimes actual star-forming region. This was necessary because the initial lists of targets for each APOGEE program or set of visits were selected for rather diverse scientific goals and following different criteria. We applied a method to remove, as uniformly as possible, most field contaminants or objects that are non members of the respective region. The method combines the application of a clustering algorithm (based on \texttt{HDBSCAN}) followed by distance cut refinements. Among the data considered, are distances and proper motions from Gaia combined with available photometry and for specific regions other ad-hoc conditions. The result was the creation of working samples made of probable members for each region of study, All these constitute a reliable database which allowed us to apply  consistently a common methodology.
     
     Our analysis show that APOGEE Net II reliably provides average abundance estimates for late type stars, but such estimates are systematically less reliable for \teff $>$4500\ K. The Gaia-based distances increases in uncertainty for regions beyond the solar neighborhood, affecting the determination of bolometric luminosities for star forming complexes in the 1-2 kpc. In turn this also affects the estimation of age and mass. We provided a brief discussion of the spatial parameter distributions, HR diagrams, and the circumstellar disk candidates content for each region. The most important contribution of this general analysis is that we are able to put on the table serious complications that are still present in the characterization of the PMS stage of stellar evolution, despite using a very robust sample of excellent quality, and despite using the most precise parallax determinations to date.
     
     APOGEE Net II offers a robust methodology that also has the advantage of being reasonable in terms of computing cost. But it is important that we keep revising the precision of the method, given the complications that arise from the characteristics of stars during the PMS. For instance, the presence of circumstellar disks, strong chromospheric activity, accretion and other aspects, such as deviations from local thermodynamic equilibrium at earlier spectral types, are still to be accounted for in synthetic models. At this point, while we still work in a limited parameter space, we can learn from the comparison with other datasets and methods. Currently, some of those methods are being directly applied to APOGEE data, like the Gradient Tree Boosting heuristic parameter assignment of \citep{Mejia21}, the Asexual Genetic Algorithm (L. Adame et al. in prep) or other machine learning methods like The Payne \citep{Ting19} or The MWM-Payne (I. Straumit et al., submitted). The Milky Way Mapper will provide a much larger sample of young star spectra that will increase the robustness of statistics and will reduce the uncertainties to a level where parameters like accretion, continuum veiling and others associated to the PMS, can reliably be incorporated into models. Samples like  those presented in this study are a solid starting point.

 One dedicated large scale survey of embedded populations in star-forming regions, young star clusters, distributed young stars, and OB associations is still missing from the overall picture. The fifth phase of the SDSS (SDSS-V) is programmed to perform  such survey as part of the goals of the Milky Way Mapper project \citep{Kollmeier17}, promising a much needed large scale perspective and the volume required for robust statistical studies that will serve to narrow down the observational uncertainties and,  finally, constraint  PMS evolution models. Along with the capital contribution of Gaia astrometric parameters \citep{Gaia16b} for large samples of sources in all Galactic environments, and the increasing availability of synoptic data, we are opening a new chapter in the study of young stars, where a precise description of the PMS evolution and a deep understanding of the kinematics of young stellar clusters are clearly feasible.     


\vskip 0.3in

\section{Acknowledgements}

The authors want to acknowledge the help of an anonymous referee, whose comments and suggestions helped us to improve the content of our manuscript. C.R-Z acknowledges support from projects  CONACYT CB2018 A1S-9754, Mexico and UNAM DGAPA PAPIIT IN112620, Mexico.  J.H., M.T. and J.S. acknowledge support from CONACYT project No. 86372 and the UNAM-DGAPA-PAPIIT projects IA102921, IN110422, and IN107519, Mexico. R. L-V acknowledges support from a posdoctoral fellowship from program `Estancias Posdoctorales por M\'exico', CONACYT. D.M. is supported by ANID BASAL project FB210003. K.P.R. acknowledges support from ANID FONDECYT Iniciaci\'on 11201161. DAGH acknowledges support from the State Research Agency (AEI) of the Spanish Ministry of Science, Innovation and Universities (MCIU) and the European Regional Development Fund (FEDER) under grant AYA2017-88254-P. AS gratefully acknowledges funding support through Fondecyt Regular (project code 1180350), from the ANID BASAL project FB210003, and from the Chilean Centro de Excelencia en Astrof\'isica y Tecnolog\'ias Afines (CATA) BASAL grant AFB-170002. J.G.F-T gratefully acknowledges the grant support provided by Proyecto Fondecyt Iniciaci\'on No. 11220340, and also from ANID Concurso de Fomento a la Vinculaci\'on Internacional para Instituciones de Investigaci\'on Regionales (Modalidad corta duraci\'on) Proyecto No. FOVI210020, and from the Joint Committee ESO-Government of Chile 2021 (ORP 023/2021), and from Becas Santander Movilidad Internacional Profesores 2022, Banco Santander Chile.

 Funding for the Sloan Digital Sky Survey IV has been provided by the Alfred P. Sloan Foundation, the U.S. Department of Energy Office of Science, and the Participating Institutions. SDSS-IV acknowledges support and resources from the Center for High Performance Computing  at the University of Utah. The SDSS website is www.sdss.org.

 SDSS-IV is managed by the Astrophysical Research Consortium for the Participating Institutions of the SDSS Collaboration including the Brazilian Participation Group, the Carnegie Institution for Science, Carnegie Mellon University, Center for Astrophysics | Harvard \& Smithsonian, the Chilean Participation Group, the French Participation Group, Instituto de Astrof\'isica de Canarias, The Johns Hopkins University, Kavli Institute for the Physics and Mathematics of the Universe (IPMU) / University of Tokyo, the Korean Participation Group, Lawrence Berkeley National Laboratory, Leibniz Institut f\"ur Astrophysik Potsdam (AIP),  Max-Planck-Institut f\"ur Astronomie (MPIA Heidelberg), Max-Planck-Institut f\"ur Astrophysik (MPA Garching), Max-Planck-Institut f\"ur Extraterrestrische Physik (MPE), National Astronomical Observatories of China, New Mexico State University, NewYork University, University of Notre Dame, Observat\'ario Nacional / MCTI, The Ohio State University, Pennsylvania State University, Shanghai Astronomical Observatory, United Kingdom Participation Group, Universidad Nacional Aut\'onoma de M\'exico, University of Arizona, University of Colorado Boulder, University of Oxford, University of Portsmouth, University of Utah, University of Virginia, University of Washington, University of Wisconsin, Vanderbilt University, and Yale University.

  This work has made use of data from the European Space Agency (ESA) mission {\it Gaia} (\url{https://www.cosmos.esa.int/gaia}), processed by the {\it Gaia} Data Processing and Analysis Consortium (DPAC,
\url{https://www.cosmos.esa.int/web/gaia/dpac/consortium}). Funding for the DPAC has been provided by national institutions, in particular the institutions participating in the {\it Gaia} Multilateral Agreement.

This research has made use of the SIMBAD database, operated at CDS, Strasbourg, France.


%

\facilities{Du Pont (APOGEE), Sloan (APOGEE), 2MASS, Gaia, PanStars1}

\software{astropy\citep{Astropy13,Astropy18}, TopCat \citep{TOPCAT05}, R \citep{R_ref}. Additional data analyses were done using IDL version 7.0 (Exelis Visual Information Solutions, Boulder, Colorado).}



\bibliography{draft_03}{}
\bibliographystyle{aasjournal}

\appendix

\section{Target Summary for all Regions \label{App:RegionsAndTargets}}

\input{./TableA1}

\section{APOGEE-2 sub-programs relevant to this study \label{App:sub-programs}}

\begin{itemize}
 
\item \textit{Goal Science Program} (YSO): APOGEE-2 Goal Science Programs were defined to fill spaces in the LST distribution after scheduling of the core program, without driving survey requirements. Several young stellar clusters were targeted to characterize embedded populations. These included an extension of the SDSS III IN-SYNC program \citep{Cotaar14}.
      
\item \textit{Ancillary Science Program} (ASP): two calls for ancillary science projects were made during the 5 year span of the main survey, assigning time to 23 projects, among which was the pilot survey of the W3/W4/W5 Complex, aimed to investigate the massive star forming region.
      
\item \textit{Bright Time Extension Program} (BTX): This program was a 1.5 year extension to the main survey that resulted from an unexpected bright time observing efficiency gain. Several additional star forming regions were proposed in this program.
	  
\item \textit{After Sloan IV Program} (AS4): As part of the BTX, a number of fields were assigned to science projects aimed to enable the transition towards the SDSS-V survey. The survey of the California Complex is in this category.
	  
\item \textit{APOGEE-2S Chilean Time Allocation} (LOC): A number of contributed\footnote{``Contributed" in this case means that the data obtained for such programs end up forming part of the main survey after proprietary time. This is in contrast with the so called ``External" programs which would remain proprietary to the principal investigators after the end of the survey.} programs were proposed by Chilean scientists through the Chilean National Time Allocation Committee (CNTAC) or through the TAC of the Carnegie Institution for Science (CIS). Several star forming regions in the Southern sky were observed through these programs. 

\end{itemize}

\section{Program description and selection of individual regions  \label{App:selection} }

 \textbf{$\alpha$-Persei Cluster}. The $\alpha$-Persei cluster was proposed for the YSO program among a list of flagship clusters containing young stars in low extinction environments. This nearby stellar group with a significant content of intermediate mass stars and, with an spectroscopic age of about 50 Myr \citep{Basri99}, provides a good comparison field for the Pleiades and other young, no longer embedded populations.  
			
\textbf{California Cloud}. The seven field survey of the California Molecular Cloud for the AS4 sub-program, includes observations of a sample of embedded stars in the young cluster LKH$\alpha$-101 but also a significant number of solar type (G, F) targets selected to be at the expected distance of the cloud \citep[about 400 pc, see][]{Lada09}, to compare the kinematics of the local population with that of the molecular cloud gas. 


 \textbf{Carina Arm}. The Carina Arm data presented in this paper are composed of observations proposed by several authors, mostly through CNTAC/CIS assignations for researchers based in Chilean universities: (a) Programs 283-01-C, 290-01-C and 291-00-C were proposed as a survey of massive stars towards the young complexes Westerlund 2, NGC 3496, NGC 3576 and NGC 3603 in the Carina Arm including surrounding regions, as described in \citet{RomanLopes20}. (b) Fields 287-01-C, 288-01-C, 288-00-C and 288-01-C, along with the fields 160-60-C, 160-61-C and 162-60-C comprise two proposals with the common goal of exploring the kinematics of the young population in the NGC 3372 Complex at the Carina Molecular Cloud and to compare the physical and dynamical properties of the young population with that of other regions like Orion.
	
\textbf{Cygnus-X Complex}. The three main regions proposed for the Cygnus-X region as part of the BTX extension of the young star program, aim to study the population of the Cygnus OB2 association, targets in the embedded cluster population at the north and south portions of the Cygnus-X molecular complex, and candidate YSO to intermediate and massive stars towards the southern M29/Cygnus OB1/OB9 regions.

 \textbf{IC 348 and NGC 1333 clusters}. These datasets had the main goal to reinforce the SDSS III IN-SYNC samples obtained for the embedded clusters IC 348 and NGC 1333 in the Perseus molecular cloud \citep{Cotaar15,Yao18}. Targets were selected in order to obtain precise radial velocities to explore the presence of spectroscopic binaries. Visits were divided among the YSO and the BTX programs, with the second group being specifically designed to measure RV variations aimed to find spectroscopic binaries. 

  \textbf{NGC 2264} The NGC 2264 cluster in Monoceros was also studied as part of the IN-SYNC survey during SDSS III with the main goal of characterizing spectroscopic binaries through RV analysis. Partial results that combine targets in NGC 2264, the two Perseus clusters and Orion A were published by \citet{Fernandez17}. The YSO and BTX program for NGC 2264 in APOGEE-2 was aimed to explore the properties of young stars and to expand the study of spectroscopic binaries in the embedded population.

 \textbf{Orion Complex}. We are including a new dataset for the Orion Complex survey as part of the uniform exploration of the stellar parameters for young stars across the DR16 and DR17 datasets. A detailed description of the Orion Complex survey using data from the main APOGEE-2 YSO goal science program, can already be found in \citet{Cottle18} and \citet{Kounkel18}. In this paper, we are also including targets from the program 209-20-C which had as main goal to study the kinematics of members in and around the Orion Nebula Cluster (ONC) to test cluster formation scenarios, particularly how they relate to the integral shaped filament that runs across the region \citep{Stutz16,Stutz18}. 
	
	Orion is a very extended region with a complex morphology and kinematics, and contains groups which vary in evolutionary state from deeply-embedded to fully-exposed. We made a simple separation of the survey into four sub-regions: Orion A, covering the ONC and the active molecular complex to the south; Orion B covering the embededed clusters NGC 2071 and NGC 2024 in the northern filament, as well as the cluster $\sigma-$Ori which expands from the southern-most tip west of the Horsehead Nebula; Orion OB1 encompassing all the APOGEE-2 fields to the northwest of $\sigma$-Ori;  $\lambda$-Ori, which spans three fields covering the Collinder 69 cluster and the flanking groups B30 and B35, expanding along with a large molecular ring \citep[e.g.][]{Dolan01,Dolan02}. 
	
	In order to make this division, the fields labeled as ORIONA, ORIONC,D,E and ONCB were assigned to Orion A. The fields labeled as ORIONB (minus those corresponding to ONC), and those belonging to fields ORIONOB1A-A located East and South of $(\alpha,\delta)\mathrm{=(83.94,-0.96)}$ were assigned to Orion B. The rest of the fields labeled as ORIONOB1 were assigned to the region Orion OB1 (see Table \ref{tab:yso_sample} for details).
	
 \textbf{Pleiades Cluster}. The Pleiades cluster is a prototypical young open cluster located at a nearby distance of 134 pc \citep[e.g.][]{Percival05}. It stands as a very important calibrator for the determination of the spectral parameters of young stars \citep[SDSS III IN-SYNC,][]{Cotaar14} and its study is capital for the understanding of how young clusters in the solar neighborhood evolve, expand and eventually integrate to the disk population. The Pleiades fields proposed in the YSO program aim to expand the pilot set explored in SDSS III, eventually contributing to a robust collection of Pleiades members with precise spectral characterization.
		
 \textbf{Rosette Complex}. The Rosette was proposed as a BTX target for the YSO program by teams in both hemispheres. A first proposal for LCO aimed to obtain, for the first time, a significantly large sample of embedded star targets in the Rosette Molecular Cloud, and compare against targets in the NGC 2244 OB association at the Rosette Nebula. The other team also proposed to observe a sample combining embedded and emerged targets in the cluster population. In the end, scientific targets were merged for a joint project with 6 visits using BTX time, using as common goals to trace the young star population and to explore the evolution of the family of young clusters across the complex. 
				
	\textbf{Taurus-Auriga Complex}. This region, one of the closest low mass star formation complexes, has been the target of numerous spectroscopic studies in the past. The main goals are to characterize the properties, evolution and kinematics of groups of young low-mass stars across the filamentary structure of the Taurus Molecular Cloud. A first approach to the 6-D distribution of the young star groups in Taurus combining precise parallax information from both Gaia DR2 and VLA observations, along with radial velocities information from APOGEE Net \citep{Olney20} can be found in \citet{Galli19}. This and subsequent studies aim for the characterization of the young stellar population using APOGEE-2 data.

 \textbf{Vela C Cloud}. One APOGEE field, was dedicated to explore the Vela C molecular cloud complex, emcompassing the RCW34 and RCW36 (Gum19 and Gum20) active star forming regions. The Vela C complex an interesting region that combines both low mass and massive star formation \citep{Massi2019}, and the APOGEE observations  aim at similar goals as those in Carina, Cygnus-X and W3/W4/W5: the exploration of young cluster sources in the presence of massive members. 

 \textbf{W3/W4/W5 Complex}. An APOGEE-2 survey of the W3, W4 and W5 massive star forming regions in the Perseus Arm, was proposed as an ASP program with two main goals: The first goal was to test APOGEE for the study of hot, massive sources. Until then, a small number of blue sources were added to each core survey plate for the purposes of removing telluric atmospheric lines during the reduction process \citep{Nidever15}, but it was not clear if the scarcity of features in the H band spectral window were enough to provide a reliable spectral classification for O and B sources. This way a list of 240 targets containing both confirmed types and photometric candidates was provided, resulting in a successful methodology for the identification and classification of O and B stars down to 1 spectral sub-type \citep[][]{RomanLopes18,Ramirez20}, which appears to extend nicely to A stars (V. Ramirez-Preciado in prep) and represents to date the most reliable classification of intermediate to high mass stars in the absence of a set of synthetic atmosphere models adequate for the APOGEE near-IR spectral range. The second goal was to study the young stellar population in the FGK type regime in a massive star forming region like W3/W4/W5, to investigate the limitations of our methodologies in the presence of high extinction and less reliable parallax data. Two APOGEE fields were used for this program, covering nicely the areas of the W3-4 and W5 regions respectively.  Two small extensions to add visits to these fields were added during the second ancillary call and the BTX programs.

\section{Stellar Properties \label{App:NRM_Tables}}

\par This version of the manuscript shows an abridged version of the table containing the fist 10 entries of each region in the catalog, in order to facilitate commenting and edition of the manuscript. It is expected that the electronic version of the paper would contain a full version of the table in electronic format. 

\input{./TableA2_short}

\section{PROVISIONAL LOCATION FOR FIGURE SETS}

\par It is expected that figure sets will be available for figures 2, 4 and 5
in the electronic version of the paper. A preview of the figures can
be accessed by personal communication with corresponding author.

\input{./figset1}
\input{./figset2}
\input{./figset3}




\end{document}

%% file: Table1.tex
\begin{longrotatetable}
\begin{deluxetable*}{lLcccccccc}
\tablecaption{Sample Restrictions \label{tab:restrictions}}
\tabletypesize{\footnotesize}
\tablehead{
\colhead{Main Target Region} &
\colhead{Restrictions} &
\multicolumn{6}{c}{No. of Sources in cut for parameter determination} & \colhead{\teff \ limits, Age cut}\\
\colhead{}& \colhead{} &  \colhead{\teff,\ \logg} &
\colhead{[Fe/H]} & \colhead{$\mathrm{\log{L/L_\odot}}$} &
\colhead{$\mathrm{M/M_\odot}$} & \colhead{Age/Myr} & 
\colhead{$\mathrm{RUWE >1.4}$} & \colhead{} }
\startdata
$\alpha$-Per & \mathrm{ \   33<\mu/mas\cdot yr^{-1}<36  \land\ 160<r_{geo}/pc<190}    &  122  & 69  & 98   & 98   & 95  & 28  & 3000,6000\\ 
California   & \mathrm{ \    5<\mu/mas\cdot yr^{-1}<10  \land\ 400<r_{geo}/pc<600}    &  29   & 16  & 29   & 29   & 25  & 9   & 3000,5500\\
Carina       & \mathrm{ \  5.8<\mu/mas\cdot yr^{-1}<7.8 \land\ 2.0<r_{geo}/kpc<3.0}   &  58   & 5   & 37   & 37   & 18  & 13  & 3500,6000\\
Cygnus-X     & \mathrm{ \  5.8<\mu/mas\cdot yr^{-1}<7.8 \land\ 0.9<r_{geo}/kpc<2.0}   &  340  & 3   & 269  & 269  & 152 & 46  & 3400,6500\\
IC\ 348      & \mathrm{ \  6.5<\mu/mas\cdot yr^{-1}<9.5 \land\ 280<r_{geo}/pc<340}    &  73   & 43  & 72   & 73   & 65  & 27  & 3600,5500\\
NGC\ 1333    & \mathrm{ \ 10.5<\mu/mas\cdot yr^{-1}<13  \land\ 240<r_{geo}/pc<360}    &  69   & 59  & 67   & 67   & 40  & 50  & 3400,5500\\
NGC\ 2264    & \mathrm{ \  3.8<\mu/mas\cdot yr^{-1}<4.8 \land\ 600<r_{geo}/pc<800}    &  125  & 101 & 124  & 124  & 121 & 19  & 3400,5500\\
$\lambda$-Ori & \mathrm{ \    1<\mu/mas\cdot yr^{-1}<5   \land\ 360<r_{geo}/pc<430}   &  223  & 172 & 222  & 222  & 212 & 68  & 3100,5500\\
Orion\ A     & \mathrm{ \    1<\mu/mas\cdot yr^{-1}<5   \land\ 360<r_{geo}/pc<430}    &  1095 & 884 & 1069 & 1069 & 994 & 307 & 3000,5500\\
Orion\ B, front & \mathrm{ 0.0<\mu/mas\cdot yr^{-1}<3.5 \land\ 340<r_{geo}/pc<370}    &  51   & 36  & 51   & 51   & 50  & 18  & 3000,5500\\
Orion\ B, back & \mathrm{ 0.0<\mu/mas\cdot yr^{-1}<3.5 \land\ 370<r_{geo}/pc<430}     &  227  & 192 & 224  & 224  & 209 & 61  & 3000,5500\\
Orion\ OB1\, front & \mathrm{ 0.0<\mu/mas\cdot yr^{-1}<3.5 \land\ 320<r_{geo}/pc<370} &  237  & 120 & 236  & 236  & 220 & 120 & 3000,5500\\
Orion\ OB1\, back & \mathrm{ 0.0<\mu/mas\cdot yr^{-1}<3.5 \land\ 370<r_{geo}/pc<430}  &   72  &  28 &  72  &  72  &  66 &  82 & 3000,5500\\
Pleiades     & \mathrm{       47<\mu/mas\cdot yr^{-1}<52 \land\ 125<r_{geo}/pc<145}   &  293  & 168 & 212  & 212  & 177 & 0   & 3500,5500\\
Rosette      & \mathrm{   1.5<\mu/mas\cdot yr^{-1}<2.5 \land\ 1.0<r_{geo}/kpc<2.0}    &  163  & 30  & 136  & 136  & 90  & 26  & 3400,5500\\ 
Taurus       & \mathrm{    20<\mu/mas\cdot yr^{-1}<30 \land\ 120<r_{geo}/pc<170}      &  71   & 58  & 71   & 71   & 67  & 109 & 3400,5500\\
Vela Ridge   & \mathrm{   7.5<\mu/mas\cdot yr^{-1}<12.5 \land\ 0.8<r_{geo}/kpc<1.1}   &  194  & 95  & 178  & 1178 & 141 & 16  & 3400,5500\\
W3/W4/W5     & \mathrm{   0.0<\mu/mas\cdot yr^{-1}<2.5 \land\ 1.8<r_{geo}/kpc<2.2}    &  118  & 0   & 107  & 107  & 32  & 17  & 3500,5700\\
\enddata
\end{deluxetable*}
\end{longrotatetable}

%% file: Table2.tex
\begin{deluxetable*}{lcccccc}
\tablecaption{HRM Median Values per Region\tablenotemark{1} \label{tab:nrm_ave}}
\tabletypesize{\scriptsize}
\tablehead{
\colhead{Region} & \colhead{\teff/K} & \colhead{\logg} & \colhead{[Fe/H]} &
\colhead{$\mathrm{\log{L/L_\odot}}$} & \colhead{$\mathrm{M/M_\odot}$} & \colhead{Age/Myr} } 
\startdata
$\alpha$-Per      & 3602$\pm$371 & 4.5$\pm$0.13 &  0.0$\pm$0.03 & -1.3$\pm$0.42 & 0.5$\pm$0.15 & 51.8$\pm$14.1 \\
California        & 3917$\pm$387 & 4.2$\pm$0.18 & -0.1$\pm$0.03 &  0.0$\pm$0.37 & 0.6$\pm$0.20 & 2.7$\pm$1.7 \\
Carina            & 4903$\pm$545 & 3.9$\pm$0.18 & ***           &  0.8$\pm$0.25 & 1.7$\pm$0.86 & 1.5$\pm$1.1 \\
Cygnus-X          & 5420$\pm$426 & 4.2$\pm$0.14 & ***           &  1.0$\pm$0.17 & 2.3$\pm$0.34 & 1.4$\pm$0.8 \\
IC\ 348           & 3878$\pm$233 & 4.0$\pm$0.11 &  0.0$\pm$0.02 & -0.3$\pm$0.28 & 0.7$\pm$0.08 & 3.5$\pm$1.9 \\
NGC\ 1333         & 3536$\pm$219 & 3.8$\pm$0.13 &  0.0$\pm$0.02 & -0.7$\pm$0.39 & 0.5$\pm$0.10 & 3.2$\pm$2.3 \\
NGC\ 2264         & 3723$\pm$187 & 4.3$\pm$0.15 &  0.0$\pm$0.03 & -0.5$\pm$0.19 & 0.7$\pm$0.08 & 3.5$\pm$1.8 \\
$\lambda$-Ori     & 3727$\pm$186 & 4.1$\pm$0.14 &  0.0$\pm$0.02 & -0.4$\pm$0.23 & 0.6$\pm$0.09 & 3.0$\pm$1.4 \\
Orion\ A          & 3642$\pm$177 & 4.0$\pm$0.16 &  0.0$\pm$0.02 & -0.4$\pm$0.23 & 0.6$\pm$0.10 & 2.6$\pm$1.4 \\
Orion\ B, front   & 3669$\pm$151 & 4.1$\pm$0.09 &  0.0$\pm$0.02 & -0.5$\pm$0.22 & 0.7$\pm$0.06 & 4.2$\pm$2.6 \\
Orion\ B, back    & 3667$\pm$178 & 3.9$\pm$0.14 &  0.0$\pm$0.02 & -0.3$\pm$0.24 & 0.5$\pm$0.09 & 1.5$\pm$0.7 \\
Orion\ OB1, front & 3973$\pm$329 & 4.3$\pm$0.15 &  0.0$\pm$0.02 & -0.4$\pm$0.22 & 0.7$\pm$0.11 & 4.7$\pm$1.9 \\
Orion\ OB1, back  & 4243$\pm$446 & 4.2$\pm$0.15 &  0.0$\pm$0.03 & -0.2$\pm$0.25 & 0.8$\pm$0.21 & 4.1$\pm$1.8 \\
Pleiades          & 3873$\pm$481 & 4.6$\pm$0.08 &  0.0$\pm$0.02 & -1.1$\pm$0.47 & 0.6$\pm$0.15 & 87.6$\pm$37.6 \\
Rosette           & 4418$\pm$352 & 4.0$\pm$0.17 & -0.1$\pm$0.04 &  0.4$\pm$0.26 & 0.9$\pm$0.30 & 1.1$\pm$0.4 \\
Taurus            & 3552$\pm$241 & 4.1$\pm$0.14 &  0.0$\pm$0.03 & -0.6$\pm$0.23 & 0.5$\pm$0.13 & 2.9$\pm$1.4 \\
Vela              & 4010$\pm$353 & 4.1$\pm$0.16 &  0.0$\pm$0.03 & -0.1$\pm$0.30 & 0.7$\pm$0.19 & 2.5$\pm$1.5 \\
W3/W4/W5          & 5265$\pm$475 & 4.0$\pm$0.16 &  ***          &  1.3$\pm$0.27 & 2.6$\pm$0.51 & 1.3$\pm$0.7 \\
\enddata
\tablenotetext{1}{Targets with multiple observations were averaged for analysis. Values listed correspond to median and median absolute deviation, except for age values in Cygnus-X, Carina and W3/W4/W5, where we used simple mean values.}
\end{deluxetable*}

%% file: TableA1.tex
\begin{longrotatetable}
\begin{deluxetable*}{lccccrc}
\tablecaption{Summary of APOGEE-2 Young Cluster Targets \label{tab:yso_sample}}
\tabletypesize{\footnotesize}
\tablehead{
\colhead{Main Target Region} & \colhead{RA} & 
\colhead{Dec} & \colhead{Plate\ ID} & 
\colhead{MJDs} & \colhead{Subprogram} & \colhead{No. of targets\tablenotemark{1} \tablenotemark{2}} }
\startdata
\multicolumn{7}{c}{\textit{Single Field Regions}}\\
\hline
\hline
$\alpha$-Persei & 51.62 & 49.0 & 9662 & 57821, 58097, 58148, 58179, 58419, 58446 & YSO & 255\\ 
IC348-RV & 55.90 & 32.00 & 10100 & 58058, 58061, 58084, 58085, 58087 & YSO, BTX & 968\\
          & & & 12706 & 59063, 59066, 59071, 59072, 59146 & & \\
NGC\ 1333 & 52.60 & 31.25 & 6224 & 56561 & YSO & 645 \\
          & & & 6225 & 56236 & & \\
          & & & 6226 & 56315 & & \\
          & & & 7070 & 56671 & & \\
          & & & 7071 & 56674 & & \\
          & & & 7072 & 56563, 56607 & & \\
NGC\ 1333-RV & & & 11425 & 58447, 58527, 58744, 58733, 58775, 58791, & YSO, BTX & 264\\
             & & &       & 58796, 58803, 58820, 58828, 58831, 58850, 59077 & & \\
NGC\ 2264 & 100.27 & 9.68 & 10302 & 58061, 58067, 58068, 58086, 58089, 58096 & YSO & 252\\
Vela Ridge (265+01-C) & 134.43072 & -43.54588 & 12276 & 58882, 58883 & LCO & 252\\
\hline
\hline
\multicolumn{7}{c}{\textit{Multiple Field Regions}}\\
\hline
\hline
\textbf{California Cloud} & & & & & & \\
California (160-09) & 63.06 & 38.40 & 11432 & 58383, 58384 & AS4 & 1763\\
California (161-06) & 65.71 & 39.93 & 11433 & 58443 & &  \\
California (162-11) & 63.39 & 35.80 & 11434 & 58418 & & \\
California (163-05) & 68.70 & 38.74 & 11435 & 58441 & & \\
California (163-08) & 66.16 & 37.25 & 11436 & 58444 & & \\
California (165-07) & 68.80 & 35.97 & 11437 & 58442 & & \\
California (164-10) & 66.38 & 34.59 & 11438 & 58448 & & \\
\textbf{Carina Arm} & & & & & & \\
Carina (160-60-C) & 160.86041 & -59.62363 & 11620 & 58560, 59180 & LCO & 1406\\
Carina (161-60-C) & 161.15234 & -59.82986 & 11621 & 58560, 58626 & & \\
Carina (162-60-C) & 161.63592 & -59.83055 & 11622 & 58654, 58883 & & \\
Carina (287-01-C) & 160.54833 & -59.5016 & 12357 & 58914, 59187 &  & \\
Carina (288-00-C) & 162.4802 & -59.54877 & 12356 & 58914, 59187, 59223 & & \\
Carina (288-01-C-b) & 161.00154 & -59.87431 & 9752 & 59202 &  & \\
Carina (288-01-C-m) & 161.33551 & -60.43674 & 12358 & 58914 &  & \\
\textbf{Cygnus-X Complex} & & & & & & \\
Cygnus-X\ N & 309.25 & 41.87 & 11271 & 58389 & BTX & 3372 \\
           &  &  & 11272 & 58390 &  & \\
           &  &  & 11273 & 58418 &  & \\
           &  &  & 11274 & 58643, 58646 &  & \\
           &  &  & 11275 & 58650 &  & \\
           &  &  & 11276 & 58653 &  & \\
Cygnus-X\ C & 306.65 & 41.01 & 11409 & 58625, 58626, 58627 & BTX  & \\
           &  &  & 11410 & 58629 &  & \\
           &  &  & 11411 & 58647, 58649 &  & \\
           &  &  & 11412 & 58654, 58655 &  & \\
           &  &  & 11413 & 58656 &  & \\
           &  &  & 11414 & 58659 &  & \\
Cygnus-X\ S & 305.66 & 38.57 & 11415 & 58637, 58642, 58732 & BTX  & \\
           &  &  & 11416 & 58653 &  & \\
           &  &  & 11417 & 58654 &  & \\
           &  &  & 11418 & 58656 &  & \\
           &  &  & 11419 & 58657, 58658 &  & \\
           &  &  & 11420 & 58659 &  & \\
\textbf{$\lambda$-Ori} & & & & & & \\           
$\lambda$-Ori\ B & 82.34 & 11.73 & 8885 & 57409 & YSO & 1742\\
            & & & 8886 & 57411, 57413 &  & \\
            & & & 8887 & 57678 &  & \\
$\lambda$-Ori\ C & 86.6 & 8.99 & 9482 & 57685 & YSO & \\
            & & & 9537 & 58115, 58116 &  & \\
            & & & 9538 & 58067 &  & \\  
$\lambda$-Ori\ A & 84.14 & 10.34 & 8879 & 57406, 57411 & YSO & \\
            & & & 8880 & 57413, 57648, 57649 &  & \\ 
            & & & 8881 & 57650 &  & \\
            & & & 8882 & 57651, 57655 &  & \\
            & & & 8883 & 57656, 57675 &  & \\
            & & & 8884 & 57676, 57677 &  & \\
\textbf{Orion AB Complex} & & & & & & \\
Orion\ A-A & 84.1001 & -5.1003 & 9481 & 58391, 58417, 58420, 58424, 58441 & YSO & 9538\\
Orion\ A-B & 83.5496 & -5.2996 & 9533 & 57737, 57762, 57794, 58085, 58060, 58063 & & \\
Orion\ A-C & 84.2507 & -6.8994 & 9659 & 57790, 57792, 57793 &  & \\
Orion\ A-D & 84.4999 & -7.2005 & 9660 & 57790, 57792, 57793 &  & \\
Orion\ A-E & 85.2001 & -8.6996 & 9661 & 57795 & & \\
Orion\ B-A & 86.654 & 0.134 & 8890 & 57433 & YSO & \\
 & & & 8891 & 57443 &  & \\
 & & & 8892 & 57675, 57677 &  & \\
 & & & 8893 & 57677, 57678 &  & \\
Orion\ B-B & 85.416 & -2.12 & 8894 & 57433 & & \\
 & & & 8895 & 57434 &  & \\
 & & & 8896 & 57435 &  & \\
 & & & 8897 & 57436 &  & \\
 & & & 8898 & 57654 &  & \\
 & & & 8899 & 57680 &  & \\
Orion\ OB1ab-A & 84.1 & -2.2 & 9468 & 57794, 57795 & YSO & \\
 & & & 9469 & 58033, 58037 &  & \\
 & & & 9470 & 58038 &  & \\
Orion\ OB1ab-B & 84.0 & 0.7 & 9471 & 57683 &  & \\
               & & & 9472 & 57684 &  & \\
               & & & 9473 & 57707 &  & \\
               & & & 9474 & 57708 &  & \\
Orion\ OB1ab-C & 82.5 & -1.5 & 9475 & 57732,57734 &  & \\
               & & & 9476 & 57764 &  & \\
               & & & 9477 & 57796 &  & \\
Orion\ OB1ab-D & 80.7 & -1.8 & 9478 & 57708 &  & \\
               & & & 9479 & 57713 &  & \\
               & & & 9480 & 58384 &  & \\
Orion\ OB1ab-E & 81.5 & 1.0 & 8900 & 57648 &  & \\
               & & & 8901 & 57649, 57650 &  & \\
               & & & 8902 & 57652, 57653 &  & \\
               & & & 8903 & 57675 &  & \\
Orion\ OB1ab-F & 82.0 & 3.0 & 8904 & 57676 &  & \\
               & & & 8905 & 57410 &  & \\
               & & & 8906 & 57411, 57412 &  & \\
Orion\ Nebula\ Cluster (209-20-C) & 83.71766 & -5.55237 & 12273 & 58882,58883 & LCO & (278) \\
Orion\ A-RV & 83.5 & -5.4 & 11593 & 58477, 58496, 58500, 58532, 58767, 58769 & BTX & (263) \\
            & & & & 58772, 58775, 58778, 58797, 58802, 58805 & & \\
Orion\ B-RV & 84.4 & -7.1 & 11594 & 58497, 58501, 58504, 58773, 58779, 58801, 58804 & BTX & (265)\\
            & & & & 58822, 58828, 58831, 58835, 58850, 58853, 58861 &  & \\
\textbf{Pleiades Cluster} & & & & & & \\            
Pleiades   & 56.712917 & 24.175889 & 5534 & 55847, 55851, 55854 & Core\ Survey & 786 \\
Pleiades-E & 57.5 & 24.1 & 8888 & 57408, 57469, 57652 & YSO & (523) \\
Pleiades-W & 55.45 & 24.7 & 9257 & 57684, 57764, 58037 & & \\
\textbf{Rosette Complex} & & & & & & \\
Rosette (206-02) & 98.20038 & 4.27645 & 11440 & 58474 & LCO & 1031\\
                 & & & 11441 & 58474, 59201, 59224 &  & \\
Rosette & 98.01 & 4.8 & 12266 & 58828,58829,58830,58831,58834,58857,58860 & YSO,BTX & \\
\textbf{Taurus-Auriga Complex} & & & & & & \\
Taurus\ L1495 & 64.75 & 27.65 & 9258 & 57682, 58007, 58009, 58010, 58036, 58039 & YSO & 2833\\
Taurus\ L1495 & 64.5 & 29.2 & 11592 & 58499, 58503, 58744, 58854 & YSO, BTX & \\
Taurus\ L1517 & 74.0 & 30.75 & 11426 & 58414, 58417 & YSO,BTX & \\
  			   &  &  & 11426 & 58417 &  & \\
   			   &  &  & 11427 & 58438, 58439 &  & \\
   			   &  &  & 11428 & 58444 &  & \\
Taurus\ L1521 & 67.0  & 25.9  & 9287 & 57710, 57737, 57761 & YSO & \\
Taurus\ L1527 & 69.45 & 24.75 & 9288 & 57711, 57738, 57765 & & \\
Taurus\ L1536 & 68.48 & 23.65 & 9259 & 57683, 57712, 58010 & & \\
Taurus\ L1551 & 68.25 & 18.125 & 11429 & 58385 & YSO, BTX & \\
   			   &  &  & 11430 & 58386 &  & \\
   		       &  &  & 11431 & 58392,58421 &  & \\
\textbf{W3/W4/W5 Complex} & & & & & & \\   		       
W3/W4       & 37.1726 & 61.4056 & 9245 & 57710, 57764 & ASP & 1802 \\
            & & & 9246 & 57650 &  & \\
            & & & 9247 & 57677 &  & \\
            & & & 9248 & 57762 &  & \\
            & & & 9249 & 57766, 58008 &  & \\
            & & & 9250 & 58037 &  & \\
W3/W4\ (btx) & 37.17 & 61.4 & 11600 & 58467 & YSO, BTX & \\
             & & & 11601 & 58501 &  & \\
             & & & 11602 & 58733, 58738 &  & \\ 
  W5        & 43.37 & 60.8982 & 9251 & 58007 & ASP & (1061)\\
  W5        & & & 9252 & 58008 &  & \\
  W5        & & & 9253 & 58009 &  & \\
  W5        & & & 9255 & 58029 &  & \\
  W5        & & & 9256 & 58039 &  & \\             
W5-a       & 43.5 & 60.65 & 9542 & 57733 & ASP & \\
\hline
\multicolumn{7}{c}{Auxiliary Fields}\\
\hline
Carina (285+00) & 157.78603 & -58.04926 & 10174 & 58114, 58117, 58145 & Core\ Survey & (261)\\
Cygnus-X (075+00) & 305.22904 & 36.605611 & 5525 & 56466, 56472, 56541, 56560 & Core\ Survey & (259)\\
Cygnus-X (079+00) & 308.1882 & 39.8618 & 7534 & 56934, 56937, 56962 & Core\ Survey & (390)\\
                  & & & 7789 & 57297, 57579, 57643 & & \\
Cygnus-X (083+00) & 311.0139 & 42.6438 & 6082 & 56231, 56439, 56440, 56449 & Core\ Survey & (263)\\
Pleiades (169-24) & 58.161499 & 22.248699 & 10362 & 58121, 58146, 58151, 58386 & BTX & (462) \\
                  & 58.1615 & 22.2487 & 11220 & 58441, 58447, 58477 & & \\
Pleiades (167-24) & 56.3324 & 23.6689 & 12129 & 58777, 58881 & BTX & (264)\\
\enddata
\tablenotetext{1}{We list the total number of unique targets in the first entry of each region. For some regions the count for specific sub-programs are indicated between parentheses}
\end{deluxetable*}
\end{longrotatetable}

%% file: TableA2_short.tex
\begin{longrotatetable}
\begin{deluxetable*}{lccrrrrrrrr}
\tablecaption{Probable Region Members}
\label{tab:nrms}
\tablehead{
\colhead{APOGEE-ID} & \colhead{RA} & \colhead{Dec} & 
\colhead{A$_V$} & \colhead{$\mathrm{r_{geo}}$}  & 
\colhead{$\mathrm{T_{eff}}$} & \colhead{$\log{g}$} & 
\colhead{[Fe/H]} & \colhead{$\mathrm{\log{L/L_\odot}}$} & 
\colhead{$\mathrm{M/M_\odot}$} & \colhead{Age}\\
\colhead{} & \multicolumn{2}{c}{J2000} & 
\colhead{[mag]} & \colhead{[pc]} & \colhead{[K]} & 
\colhead{} & \colhead{} & \colhead{} & \colhead{} & \colhead{[Myr]}\\} 
\startdata
\hline
\multicolumn{11}{c}{$\alpha$-Persei Cluster}\\
\hline
2M03190241+4933375 &    49.759998 &    49.560398 &    0.7$_{- 0.11}^{+ 0.13}$ &   175.37$_{-   0.55}^{+   0.81}$ &     3.60$\pm$   0.006 &     4.29$\pm$   0.030 &    -0.09$\pm$   0.019 &    -1.07$\pm$   0.020 &    0.6$_{- 0.00}^{+ 0.01}$ &   87.4$_{- 7.32}^{+ 6.25}$ \\
2M03192494+4859402 &    49.853901 &    48.994499 &    0.1$_{- 0.05}^{+ 0.07}$ &   164.89$_{-   2.27}^{+   2.16}$ &     3.51$\pm$   0.003 &     4.63$\pm$   0.052 &     0.04$\pm$   0.010 &    -1.72$\pm$   0.030 &    0.4$_{- 0.01}^{+ 0.01}$ &   82.4$_{- 3.92}^{+ 5.44}$ \\
2M03193117+4941171 &    49.879902 &    49.688099 &    0.8$_{- 0.24}^{+ 0.23}$ &   176.31$_{-   1.90}^{+   1.90}$ &     3.53$\pm$   0.011 &     4.26$\pm$   0.086 &    -0.08$\pm$   0.036 &    -1.59$\pm$   0.027 &    0.4$_{- 0.03}^{+ 0.01}$ &  140.1$_{-39.12}^{+90.49}$ \\
2M03195729+4904215 &    49.988701 &    49.072701 &    0.1$_{- 0.04}^{+ 0.04}$ &   163.91$_{-   1.10}^{+   0.98}$ &     3.54$\pm$   0.001 &     4.74$\pm$   0.021 &     0.04$\pm$   0.006 &    -1.37$\pm$   0.014 &    0.5$_{- 0.00}^{+ 0.00}$ &   60.7$_{- 3.63}^{+ 3.91}$ \\
2M03201212+4856412 &    50.050499 &    48.944801 &    0.1$_{- 0.07}^{+ 0.05}$ &   169.36$_{-   0.94}^{+   1.11}$ &     3.53$\pm$   0.002 &     4.76$\pm$   0.028 &     0.06$\pm$   0.009 &    -1.38$\pm$   0.020 &    0.5$_{- 0.01}^{+ 0.01}$ &   47.7$_{- 2.38}^{+ 2.75}$ \\
2M03204185+4824375 &    50.174400 &    48.410400 &    0.1$_{- 0.04}^{+ 0.05}$ &   169.20$_{-   0.54}^{+   0.57}$ &     3.71$\pm$   0.001 &     4.50$\pm$   0.009 & \nodata               &    -0.41$\pm$   0.014 &    0.8$_{- 0.01}^{+ 0.01}$ &   40.0$_{- 1.07}^{+ 2.42}$ \\
2M03205158+4927574 &    50.214901 &    49.466000 &    0.2$_{- 0.03}^{+ 0.04}$ &   172.75$_{-   0.35}^{+   0.32}$ &     3.67$\pm$   0.001 &     4.39$\pm$   0.013 & \nodata               &    -0.57$\pm$   0.013 &    0.8$_{- 0.01}^{+ 0.01}$ &   36.1$_{- 1.74}^{+ 1.99}$ \\
2M03210651+4826127 &    50.277100 &    48.436901 &    0.2$_{- 0.04}^{+ 0.04}$ &   171.98$_{-   0.34}^{+   0.30}$ &     3.68$\pm$   0.001 &     4.59$\pm$   0.011 & \nodata               &    -0.60$\pm$   0.013 &    0.8$_{- 0.01}^{+ 0.01}$ &   47.7$_{- 2.19}^{+ 3.69}$ \\
2M03212020+4845270 &    50.334202 &    48.757500 &    0.2$_{- 0.05}^{+ 0.04}$ &   172.87$_{-   0.44}^{+   0.42}$ &     3.71$\pm$   0.001 &     4.52$\pm$   0.009 & \nodata               &    -0.37$\pm$   0.012 &    0.8$_{- 0.00}^{+ 0.01}$ &   39.8$_{- 1.44}^{+ 1.85}$ \\
2M03212047+4753152 &    50.335300 &    47.887600 &    1.1$_{- 0.29}^{+ 0.25}$ &   170.03$_{-   0.87}^{+   0.83}$ &     3.60$\pm$   0.012 &     4.21$\pm$   0.031 & \nodata               &    -1.11$\pm$   0.030 &    0.6$_{- 0.02}^{+ 0.01}$ &  121.7$_{-11.29}^{+ 7.31}$ \\
\hline
\multicolumn{11}{c}{California Cloud}\\
\hline
2M04212799+3345498 &    65.366699 &    33.763901 &    0.6$_{- 0.05}^{+ 0.06}$ &   496.91$_{-   4.92}^{+   4.73}$ &     3.71$\pm$   0.004 &     4.43$\pm$   0.037 & \nodata               &    -0.05$\pm$   0.023 &    1.1$_{- 0.01}^{+ 0.02}$ &   14.9$_{- 0.83}^{+ 0.80}$ \\
2M04222320+3820491 &    65.596703 &    38.347000 &    2.3$_{- 0.06}^{+ 0.09}$ &   522.27$_{-   8.13}^{+   6.99}$ &     3.67$\pm$   0.004 &     4.34$\pm$   0.061 & \nodata               &     0.24$\pm$   0.022 &    1.2$_{- 0.05}^{+ 0.09}$ &    2.0$_{- 0.20}^{+ 0.25}$ \\
2M04232189+3352172 &    65.841202 &    33.871498 &    0.5$_{- 0.23}^{+ 0.23}$ &   525.67$_{-   5.10}^{+   4.66}$ &     3.73$\pm$   0.017 &     3.75$\pm$   0.232 & \nodata               &     0.22$\pm$   0.060 &    1.3$_{- 0.02}^{+ 0.02}$ &    9.2$_{- 1.37}^{+ 1.96}$ \\
2M04254131+3537189 &    66.422096 &    35.621899 &    1.5$_{- 0.05}^{+ 0.06}$ &   538.79$_{-   4.43}^{+   4.95}$ &     3.73$\pm$   0.003 &     4.22$\pm$   0.028 & \nodata               &     0.59$\pm$   0.018 &    1.8$_{- 0.02}^{+ 0.02}$ &    3.6$_{- 0.24}^{+ 0.21}$ \\
2M04254864+3900413 &    66.452698 &    39.011501 &    1.0$_{- 0.33}^{+ 0.24}$ &   513.42$_{-   3.89}^{+   4.52}$ &     3.74$\pm$   0.020 &     3.34$\pm$   0.266 & \nodata               &     0.27$\pm$   0.070 &    1.3$_{- 0.04}^{+ 0.03}$ &    9.6$_{- 2.30}^{+ 2.46}$ \\
2M04260844+3335174 &    66.535202 &    33.588200 &    0.9$_{- 0.04}^{+ 0.03}$ &   533.22$_{-   3.51}^{+   5.28}$ &     3.71$\pm$   0.002 &     3.92$\pm$   0.042 & \nodata               &     0.55$\pm$   0.014 &    1.8$_{- 0.02}^{+ 0.02}$ &    2.2$_{- 0.09}^{+ 0.10}$ \\
2M04271110+3521514 &    66.796303 &    35.364300 &    1.6$_{- 0.04}^{+ 0.05}$ &   539.05$_{-  10.10}^{+   6.09}$ &     3.68$\pm$   0.002 &     4.30$\pm$   0.034 & \nodata               &     0.20$\pm$   0.018 &    1.3$_{- 0.02}^{+ 0.02}$ &    3.3$_{- 0.20}^{+ 0.24}$ \\
2M04275080+3631263 &    66.961700 &    36.523998 &    4.5$_{- 0.37}^{+ 0.39}$ &   535.75$_{-  37.69}^{+  81.01}$ &     3.57$\pm$   0.012 &     3.98$\pm$   0.134 &    -0.11$\pm$   0.066 &    -0.56$\pm$   0.140 &    0.6$_{- 0.09}^{+ 0.07}$ &    4.3$_{- 2.40}^{+ 2.99}$ \\
2M04291155+3504495 &    67.298103 &    35.080399 &    1.6$_{- 0.07}^{+ 0.07}$ &   528.00$_{-  12.11}^{+  19.11}$ &     3.58$\pm$   0.003 &     4.17$\pm$   0.048 &    -0.01$\pm$   0.013 &    -0.30$\pm$   0.027 &    0.6$_{- 0.02}^{+ 0.01}$ &    2.0$_{- 0.19}^{+ 0.23}$ \\
2M04291206+3734394 &    67.300301 &    37.577599 &    1.8$_{- 0.08}^{+ 0.11}$ &   515.17$_{-   4.95}^{+   5.02}$ &     3.73$\pm$   0.008 &     3.85$\pm$   0.084 & \nodata               &     0.38$\pm$   0.033 &    1.5$_{- 0.02}^{+ 0.03}$ &    6.4$_{- 0.65}^{+ 0.64}$ \\
\hline
\multicolumn{11}{c}{Carina Complex}\\
\hline
2M10442327-5928115 &   161.097000 &   -59.469898 &    1.5$_{- 0.09}^{+ 0.08}$ &  2415.07$_{- 261.81}^{+ 295.17}$ &     3.68$\pm$   0.011 &     3.73$\pm$   0.239 & \nodata               &     0.50$\pm$   0.104 &    1.3$_{- 0.15}^{+ 0.20}$ &    1.1$_{- 0.38}^{+ 0.49}$ \\
2M10445950-5926040 &   161.248001 &   -59.434498 &    2.6$_{- 0.16}^{+ 0.19}$ &  2619.04$_{- 347.70}^{+ 532.31}$ &     3.69$\pm$   0.018 &     3.87$\pm$   0.247 & \nodata               &     1.08$\pm$   0.151 &    1.7$_{- 0.43}^{+ 0.52}$ &\nodata                     \\
2M10431426-5912128 &   160.809402 &   -59.203602 &    1.2$_{- 0.10}^{+ 0.13}$ &  2351.07$_{- 206.38}^{+ 224.81}$ &     3.69$\pm$   0.012 &     3.71$\pm$   0.242 & \nodata               &     0.63$\pm$   0.083 &    1.6$_{- 0.20}^{+ 0.19}$ &    1.1$_{- 0.26}^{+ 0.41}$ \\
2M10431400-5923538 &   160.808395 &   -59.398300 &    1.6$_{- 0.08}^{+ 0.07}$ &  2925.19$_{- 334.78}^{+ 496.76}$ &     3.69$\pm$   0.007 &     4.16$\pm$   0.087 & \nodata               &     0.76$\pm$   0.111 &    1.6$_{- 0.15}^{+ 0.19}$ &    0.7$_{- 0.24}^{+ 0.26}$ \\
2M10444199-6007385 &   161.175003 &   -60.127399 &    1.0$_{- 0.17}^{+ 0.22}$ &  2154.54$_{- 125.79}^{+ 145.01}$ &     3.69$\pm$   0.022 &     4.11$\pm$   0.137 & \nodata               &     0.71$\pm$   0.072 &    1.6$_{- 0.35}^{+ 0.40}$ &    0.8$_{- 0.32}^{+ 0.53}$ \\
2M10441794-5929263 &   161.074799 &   -59.490601 &    1.5$_{- 0.09}^{+ 0.10}$ &  2494.46$_{- 163.00}^{+ 230.46}$ &     3.71$\pm$   0.011 &     3.52$\pm$   0.275 & \nodata               &     0.99$\pm$   0.081 &    2.3$_{- 0.26}^{+ 0.28}$ &    0.8$_{- 0.16}^{+ 0.19}$ \\
2M10421485-5923140 &   160.561905 &   -59.387199 &    3.2$_{- 0.15}^{+ 0.21}$ &  2658.67$_{- 283.10}^{+ 290.96}$ &     3.71$\pm$   0.018 &     4.20$\pm$   0.093 & \nodata               &     1.47$\pm$   0.103 &    3.2$_{- 0.74}^{+ 0.39}$ &\nodata                     \\
2M10451232-5949490 &   161.301300 &   -59.830299 &    4.6$_{- 0.24}^{+ 0.28}$ &  2370.16$_{- 699.37}^{+ 947.61}$ &     3.72$\pm$   0.014 &     4.25$\pm$   0.118 & \nodata               &     0.64$\pm$   0.263 &    1.9$_{- 0.32}^{+ 0.45}$ &    2.4$_{- 1.32}^{+ 2.11}$ \\
2M10440449-5958274 &   161.018707 &   -59.974300 &    1.6$_{- 0.13}^{+ 0.12}$ &  2226.21$_{- 153.00}^{+ 179.77}$ &     3.72$\pm$   0.013 &     3.52$\pm$   0.248 & \nodata               &     0.79$\pm$   0.071 &    2.1$_{- 0.13}^{+ 0.12}$ &    1.8$_{- 0.54}^{+ 0.63}$ \\
2M10443931-5923550 &   161.163803 &   -59.398602 &    0.0$_{- 0.00}^{+ 0.00}$ &  2469.66$_{-  68.37}^{+  85.38}$ &     3.72$\pm$   0.016 &     3.99$\pm$   0.122 & \nodata               &     0.88$\pm$   0.033 &    2.3$_{- 0.11}^{+ 0.06}$ &    1.6$_{- 0.40}^{+ 0.63}$ \\
\hline
\multicolumn{11}{c}{Cygnus-X Complex}\\
\hline
2M20261538+3909148 &   306.564087 &    39.154099 &    4.9$_{- 0.13}^{+ 0.11}$ &  1911.58$_{- 319.51}^{+ 442.69}$ &     3.70$\pm$   0.012 &     4.28$\pm$   0.078 & \nodata               &     0.89$\pm$   0.157 &    1.7$_{- 0.21}^{+ 0.30}$ &    0.6$_{- 0.19}^{+ 0.43}$ \\
2M20261791+3855247 &   306.574707 &    38.923500 &    2.7$_{- 0.09}^{+ 0.10}$ &  1653.57$_{-  61.35}^{+  64.02}$ &     3.74$\pm$   0.013 &     4.07$\pm$   0.087 & \nodata               &     0.87$\pm$   0.046 &    2.3$_{- 0.09}^{+ 0.08}$ &    2.0$_{- 0.46}^{+ 0.41}$ \\
2M20262157+3957387 &   306.589905 &    39.960800 &    5.4$_{- 0.09}^{+ 0.16}$ &  1652.08$_{- 101.28}^{+ 103.69}$ &     3.78$\pm$   0.006 &     4.09$\pm$   0.048 & \nodata               &     1.49$\pm$   0.049 &    2.9$_{- 0.09}^{+ 0.09}$ &    1.8$_{- 0.15}^{+ 0.18}$ \\
2M20262249+3848142 &   306.593689 &    38.804001 &    2.6$_{- 0.30}^{+ 0.15}$ &  1650.49$_{-  36.82}^{+  39.02}$ &     3.80$\pm$   0.021 &     4.42$\pm$   0.061 & \nodata               &     1.51$\pm$   0.063 &    2.7$_{- 0.08}^{+ 0.06}$ &    2.2$_{- 0.29}^{+ 0.27}$ \\
2M20262334+4047281 &   306.597290 &    40.791199 &    3.6$_{- 0.19}^{+ 0.17}$ &  1660.74$_{- 119.48}^{+ 138.13}$ &     3.71$\pm$   0.019 &     4.32$\pm$   0.064 & \nodata               &     0.71$\pm$   0.082 &    2.0$_{- 0.23}^{+ 0.15}$ &    1.9$_{- 0.60}^{+ 0.71}$ \\
2M20262833+4104084 &   306.618103 &    41.069000 &    3.2$_{- 0.19}^{+ 0.21}$ &  1738.39$_{- 106.80}^{+ 155.15}$ &     3.69$\pm$   0.020 &     3.98$\pm$   0.150 & \nodata               &     0.76$\pm$   0.084 &    1.5$_{- 0.42}^{+ 0.35}$ &    0.6$_{- 0.26}^{+ 0.28}$ \\
2M20263035+4009464 &   306.626495 &    40.162899 &    4.7$_{- 0.17}^{+ 0.18}$ &  1490.33$_{- 123.91}^{+ 157.92}$ &     3.70$\pm$   0.016 &     4.19$\pm$   0.129 & \nodata               &     1.07$\pm$   0.087 &    2.0$_{- 0.42}^{+ 0.53}$ &\nodata                     \\
2M20263850+3933473 &   306.660400 &    39.563099 &    4.1$_{- 0.07}^{+ 0.08}$ &  1712.00$_{- 105.44}^{+ 105.40}$ &     3.80$\pm$   0.011 &     4.05$\pm$   0.083 & \nodata               &     1.06$\pm$   0.055 &    1.9$_{- 0.08}^{+ 0.07}$ &    5.7$_{- 0.62}^{+ 0.62}$ \\
2M20264171+3837422 &   306.673798 &    38.628399 &    4.0$_{- 0.22}^{+ 0.10}$ &  1616.41$_{-  47.99}^{+  59.67}$ &     3.83$\pm$   0.021 &     4.23$\pm$   0.075 & \nodata               &     1.60$\pm$   0.071 &    2.7$_{- 0.08}^{+ 0.07}$ &\nodata                     \\
2M20265498+3937459 &   306.729095 &    39.629398 &    4.0$_{- 0.43}^{+ 0.13}$ &  1721.35$_{-  73.27}^{+  62.54}$ &     3.79$\pm$   0.020 &     4.32$\pm$   0.095 & \nodata               &     1.59$\pm$   0.086 &    3.0$_{- 0.07}^{+ 0.12}$ &    1.7$_{- 0.26}^{+ 0.19}$ \\
\hline
\multicolumn{11}{c}{IC 348}\\
\hline
2M03441361+3215542 &    56.056702 &    32.265099 &    4.5$_{- 0.14}^{+ 0.17}$ &   330.33$_{-  20.74}^{+  19.47}$ &     3.56$\pm$   0.001 &     3.79$\pm$   0.035 &     0.02$\pm$   0.005 &    -0.56$\pm$   0.060 &    0.6$_{- 0.03}^{+ 0.03}$ &    3.8$_{- 1.03}^{+ 1.31}$ \\
2M03441642+3209552 &    56.068401 &    32.165401 &    1.2$_{- 0.04}^{+ 0.03}$ &   311.22$_{-   3.85}^{+   3.01}$ &     3.65$\pm$   0.001 &     4.13$\pm$   0.018 & \nodata               &    -0.13$\pm$   0.016 &    1.0$_{- 0.01}^{+ 0.01}$ &    4.6$_{- 0.30}^{+ 0.28}$ \\
2M03441816+3204570 &    56.075699 &    32.082500 &    6.9$_{- 0.08}^{+ 0.11}$ &   326.02$_{-  16.59}^{+  13.91}$ &     3.69$\pm$   0.002 &     4.29$\pm$   0.021 & \nodata               &     0.31$\pm$   0.043 &    1.5$_{- 0.03}^{+ 0.04}$ &    2.9$_{- 0.32}^{+ 0.60}$ \\
2M03441857+3212530 &    56.077400 &    32.214699 &    5.4$_{- 0.26}^{+ 0.26}$ &   293.19$_{-  25.74}^{+  29.66}$ &     3.62$\pm$   0.020 &     4.08$\pm$   0.139 & \nodata               &    -0.67$\pm$   0.084 &    0.8$_{- 0.02}^{+ 0.05}$ &   20.4$_{- 7.43}^{+ 8.92}$ \\
2M03442129+3211563 &    56.088699 &    32.199001 &    2.2$_{- 0.06}^{+ 0.04}$ &   314.77$_{-   7.11}^{+   9.04}$ &     3.56$\pm$   0.001 &     4.03$\pm$   0.033 &     0.04$\pm$   0.005 &    -0.65$\pm$   0.025 &    0.6$_{- 0.01}^{+ 0.01}$ &    5.8$_{- 0.61}^{+ 0.85}$ \\
2M03442155+3210174 &    56.089802 &    32.171501 &    2.0$_{- 0.06}^{+ 0.07}$ &   319.95$_{-   7.94}^{+  11.80}$ &     3.56$\pm$   0.002 &     3.72$\pm$   0.044 &    -0.01$\pm$   0.007 &    -0.65$\pm$   0.028 &    0.6$_{- 0.01}^{+ 0.01}$ &    5.4$_{- 0.73}^{+ 0.70}$ \\
2M03442161+3210376 &    56.090099 &    32.177101 &    3.1$_{- 0.07}^{+ 0.07}$ &   321.73$_{-   9.89}^{+   8.40}$ &     3.57$\pm$   0.001 &     3.96$\pm$   0.037 &    -0.05$\pm$   0.008 &    -0.43$\pm$   0.029 &    0.6$_{- 0.02}^{+ 0.02}$ &    2.9$_{- 0.37}^{+ 0.46}$ \\
2M03442166+3206248 &    56.090302 &    32.106899 &    2.0$_{- 0.05}^{+ 0.07}$ &   315.11$_{-   7.00}^{+   7.67}$ &     3.56$\pm$   0.001 &     3.95$\pm$   0.033 &     0.03$\pm$   0.007 &    -0.62$\pm$   0.025 &    0.6$_{- 0.01}^{+ 0.01}$ &    4.8$_{- 0.52}^{+ 0.53}$ \\
2M03442228+3205427 &    56.092899 &    32.095200 &    3.3$_{- 0.06}^{+ 0.07}$ &   305.26$_{-  10.72}^{+  10.90}$ &     3.59$\pm$   0.003 &     3.80$\pm$   0.077 &    -0.05$\pm$   0.009 &    -0.43$\pm$   0.033 &    0.7$_{- 0.01}^{+ 0.00}$ &    4.0$_{- 0.45}^{+ 0.52}$ \\
2M03442398+3211000 &    56.099998 &    32.183300 &    2.1$_{- 0.03}^{+ 0.04}$ &   315.29$_{-   2.21}^{+   2.12}$ &     3.76$\pm$   0.002 &     4.22$\pm$   0.020 & \nodata               &     0.42$\pm$   0.013 &    1.4$_{- 0.02}^{+ 0.01}$ &\nodata                     \\
\hline
\multicolumn{11}{c}{NGC 1333}\\
\hline
2M03280010+3008469 &    52.000401 &    30.146400 &    3.0$_{- 0.06}^{+ 0.06}$ &   268.30$_{-   5.87}^{+   7.51}$ &     3.53$\pm$   0.001 &     3.79$\pm$   0.006 &    -0.07$\pm$   0.002 &    -0.55$\pm$   0.023 &    0.4$_{- 0.01}^{+ 0.01}$ &    1.7$_{- 0.17}^{+ 0.22}$ \\
2M03283173+3059158 &    52.132198 &    30.987700 &    6.6$_{- 0.14}^{+ 0.10}$ &   290.35$_{-  12.73}^{+  10.94}$ &     3.65$\pm$   0.006 &     4.06$\pm$   0.043 & \nodata               &     0.09$\pm$   0.041 &    0.9$_{- 0.07}^{+ 0.07}$ &    1.7$_{- 0.25}^{+ 0.30}$ \\
2M03285101+3118184 &    52.212601 &    31.305099 &    2.4$_{- 0.11}^{+ 0.12}$ &   303.16$_{-   5.27}^{+   4.43}$ &     3.55$\pm$   0.001 &     3.92$\pm$   0.023 &    -0.13$\pm$   0.005 &    -0.19$\pm$   0.022 &    0.3$_{- 0.01}^{+ 0.01}$ &    0.6$_{- 0.07}^{+ 0.04}$ \\
2M03285105+3116324 &    52.212700 &    31.275700 &    2.6$_{- 0.28}^{+ 0.24}$ &   282.88$_{-   7.70}^{+   8.94}$ &     3.55$\pm$   0.009 &     3.78$\pm$   0.129 &    -0.04$\pm$   0.028 &    -0.99$\pm$   0.073 &    0.6$_{- 0.05}^{+ 0.03}$ &   21.0$_{- 2.27}^{+ 2.65}$ \\
2M03285119+3119548 &    52.213299 &    31.331900 &    3.5$_{- 0.03}^{+ 0.05}$ &   295.54$_{-   3.52}^{+   3.77}$ &     3.61$\pm$   0.000 &     3.78$\pm$   0.010 & \nodata               &    -0.08$\pm$   0.015 &    0.7$_{- 0.00}^{+ 0.00}$ &    1.5$_{- 0.08}^{+ 0.08}$ \\
2M03285213+3115471 &    52.217201 &    31.263100 &    3.8$_{- 0.23}^{+ 0.20}$ &   298.97$_{-  11.92}^{+  10.40}$ &     3.57$\pm$   0.013 &     3.37$\pm$   0.106 &    -0.26$\pm$   0.063 &    -0.68$\pm$   0.055 &    0.7$_{- 0.04}^{+ 0.01}$ &    7.9$_{- 0.97}^{+ 1.25}$ \\
2M03285216+3122453 &    52.217300 &    31.379299 &    1.8$_{- 0.04}^{+ 0.05}$ &   286.53$_{-   4.04}^{+   3.23}$ &     3.56$\pm$   0.000 &     3.66$\pm$   0.013 &    -0.01$\pm$   0.003 &    -0.52$\pm$   0.018 &    0.6$_{- 0.01}^{+ 0.01}$ &    2.9$_{- 0.28}^{+ 0.26}$ \\
2M03285461+3116512 &    52.227501 &    31.280899 &    4.7$_{- 0.17}^{+ 0.14}$ &   311.51$_{-  10.78}^{+  11.23}$ &     3.56$\pm$   0.002 &     3.66$\pm$   0.043 &    -0.03$\pm$   0.007 &    -0.45$\pm$   0.036 &    0.5$_{- 0.02}^{+ 0.02}$ &    2.3$_{- 0.34}^{+ 0.40}$ \\
2M03285514+3116247 &    52.229801 &    31.273500 &    3.4$_{- 0.17}^{+ 0.19}$ &   290.48$_{-   9.86}^{+  10.34}$ &     3.55$\pm$   0.005 &     3.96$\pm$   0.092 &     0.01$\pm$   0.022 &    -0.75$\pm$   0.056 &    0.6$_{- 0.03}^{+ 0.02}$ &    7.0$_{- 1.02}^{+ 1.74}$ \\
2M03285622+3117457 &    52.234299 &    31.296000 &    3.5$_{- 0.36}^{+ 0.39}$ &   313.03$_{-  12.13}^{+  13.24}$ &     3.54$\pm$   0.014 &     3.47$\pm$   0.195 &    -0.21$\pm$   0.068 &    -0.94$\pm$   0.117 &    0.6$_{- 0.08}^{+ 0.06}$ &   12.2$_{- 2.32}^{+ 2.43}$ \\
\hline
\multicolumn{11}{c}{NGC 2264}\\
\hline
2M06390996+1005127 &    99.791496 &    10.086900 &    1.8$_{- 0.13}^{+ 0.14}$ &   727.35$_{-  33.19}^{+  35.10}$ &     3.56$\pm$   0.002 &     4.31$\pm$   0.066 &    -0.05$\pm$   0.009 &    -0.34$\pm$   0.048 &    0.5$_{- 0.03}^{+ 0.03}$ &    1.5$_{- 0.27}^{+ 0.37}$ \\
2M06392200+1006233 &    99.841698 &    10.106500 &    0.2$_{- 0.06}^{+ 0.05}$ &   671.48$_{-  14.01}^{+  15.13}$ &     3.60$\pm$   0.002 &     4.59$\pm$   0.020 &    -0.06$\pm$   0.010 &    -0.46$\pm$   0.024 &    0.7$_{- 0.00}^{+ 0.00}$ &    4.8$_{- 0.40}^{+ 0.47}$ \\
2M06392550+0931394 &    99.856300 &     9.527600 &    1.3$_{- 0.10}^{+ 0.10}$ &   656.11$_{-  31.26}^{+  31.19}$ &     3.59$\pm$   0.004 &     4.46$\pm$   0.071 &    -0.08$\pm$   0.020 &    -0.49$\pm$   0.048 &    0.7$_{- 0.01}^{+ 0.01}$ &    4.9$_{- 0.65}^{+ 0.89}$ \\
2M06393398+0949208 &    99.891602 &     9.822500 &    0.3$_{- 0.04}^{+ 0.05}$ &   690.43$_{-  14.32}^{+  17.30}$ &     3.59$\pm$   0.001 &     4.32$\pm$   0.037 &    -0.01$\pm$   0.006 &    -0.27$\pm$   0.022 &    0.6$_{- 0.01}^{+ 0.02}$ &    1.9$_{- 0.15}^{+ 0.22}$ \\
2M06393931+0955596 &    99.913803 &     9.933200 &    0.2$_{- 0.07}^{+ 0.07}$ &   676.12$_{-  22.85}^{+  17.63}$ &     3.58$\pm$   0.002 &     4.20$\pm$   0.038 &    -0.07$\pm$   0.009 &    -0.23$\pm$   0.030 &    0.6$_{- 0.02}^{+ 0.02}$ &    1.3$_{- 0.13}^{+ 0.24}$ \\
2M06394147+0946196 &    99.922798 &     9.772100 &    0.3$_{- 0.05}^{+ 0.05}$ &   710.17$_{-  10.31}^{+  12.10}$ &     3.66$\pm$   0.002 &     4.40$\pm$   0.025 & \nodata               &     0.00$\pm$   0.021 &    1.1$_{- 0.02}^{+ 0.02}$ &    4.0$_{- 0.41}^{+ 0.48}$ \\
2M06394355+0936039 &    99.931503 &     9.601100 &    4.4$_{- 0.11}^{+ 0.11}$ &   674.50$_{-  64.17}^{+  76.53}$ &     3.66$\pm$   0.002 &     4.47$\pm$   0.037 & \nodata               &     0.05$\pm$   0.079 &    1.0$_{- 0.05}^{+ 0.03}$ &    2.5$_{- 0.81}^{+ 0.82}$ \\
2M06395109+0936328 &    99.962898 &     9.609100 &    1.1$_{- 0.11}^{+ 0.14}$ &   711.65$_{-  42.72}^{+  44.03}$ &     3.56$\pm$   0.005 &     3.98$\pm$   0.081 &    -0.06$\pm$   0.022 &    -0.62$\pm$   0.059 &    0.7$_{- 0.03}^{+ 0.03}$ &    5.6$_{- 1.08}^{+ 1.91}$ \\
2M06395957+0956243 &    99.998199 &     9.940100 &    0.0$_{- 0.00}^{+ 0.00}$ &   720.34$_{-  14.42}^{+  14.28}$ &     3.60$\pm$   0.003 &     4.04$\pm$   0.033 &    -0.12$\pm$   0.009 &    -0.30$\pm$   0.024 &    0.7$_{- 0.00}^{+ 0.00}$ &    2.6$_{- 0.19}^{+ 0.22}$ \\
2M06395984+0933416 &    99.999397 &     9.561600 &    0.0$_{- 0.06}^{+ 0.07}$ &   685.16$_{-  19.71}^{+  18.91}$ &     3.60$\pm$   0.004 &     4.35$\pm$   0.081 &    -0.11$\pm$   0.016 &    -0.59$\pm$   0.029 &    0.7$_{- 0.00}^{+ 0.00}$ &    8.1$_{- 0.81}^{+ 1.07}$ \\
\hline
\multicolumn{11}{c}{$\lambda$-Ori Complex}\\
\hline
2M05301704+1215410 &    82.570999 &    12.261400 &    0.5$_{- 0.04}^{+ 0.05}$ &   399.88$_{-   6.75}^{+   5.70}$ &     3.56$\pm$   0.001 &     4.04$\pm$   0.037 &     0.05$\pm$   0.007 &    -0.61$\pm$   0.019 &    0.6$_{- 0.01}^{+ 0.01}$ &    4.5$_{- 0.35}^{+ 0.48}$ \\
2M05303607+1206406 &    82.650299 &    12.111300 &    0.3$_{- 0.06}^{+ 0.06}$ &   380.25$_{-   5.96}^{+   6.36}$ &     3.52$\pm$   0.002 &     4.06$\pm$   0.032 &    -0.00$\pm$   0.007 &    -0.81$\pm$   0.021 &    0.4$_{- 0.02}^{+ 0.01}$ &    3.6$_{- 0.42}^{+ 0.29}$ \\
2M05303614+1030543 &    82.650597 &    10.515100 &    0.3$_{- 0.05}^{+ 0.05}$ &   390.44$_{-   2.31}^{+   2.17}$ &     3.66$\pm$   0.002 &     4.21$\pm$   0.039 & \nodata               &     0.00$\pm$   0.018 &    1.1$_{- 0.02}^{+ 0.01}$ &    3.3$_{- 0.27}^{+ 0.30}$ \\
2M05303619+1236524 &    82.650803 &    12.614600 &    0.8$_{- 0.09}^{+ 0.08}$ &   385.16$_{-   3.54}^{+   3.73}$ &     3.56$\pm$   0.003 &     3.78$\pm$   0.076 &    -0.03$\pm$   0.008 &    -0.17$\pm$   0.022 &    0.4$_{- 0.01}^{+ 0.02}$ &    0.7$_{- 0.06}^{+ 0.05}$ \\
2M05303638+1236566 &    82.651604 &    12.615700 &    0.9$_{- 0.08}^{+ 0.09}$ &   385.72$_{-   7.66}^{+   8.48}$ &     3.53$\pm$   0.002 &     3.95$\pm$   0.048 &    -0.01$\pm$   0.010 &    -0.70$\pm$   0.024 &    0.5$_{- 0.02}^{+ 0.01}$ &    3.3$_{- 0.37}^{+ 0.37}$ \\
2M05304096+1244161 &    82.670700 &    12.737800 &    3.9$_{- 0.14}^{+ 0.19}$ &   415.80$_{-  16.13}^{+  18.94}$ &     3.55$\pm$   0.003 &     3.74$\pm$   0.056 &    -0.05$\pm$   0.009 &    -0.26$\pm$   0.045 &    0.4$_{- 0.02}^{+ 0.02}$ &    0.8$_{- 0.09}^{+ 0.11}$ \\
2M05304136+1030325 &    82.672401 &    10.509000 &    0.3$_{- 0.08}^{+ 0.08}$ &   396.12$_{-   2.43}^{+   2.34}$ &     3.72$\pm$   0.006 &     4.03$\pm$   0.046 & \nodata               &     0.51$\pm$   0.025 &    1.8$_{- 0.03}^{+ 0.04}$ &    3.2$_{- 0.39}^{+ 0.43}$ \\
2M05304878+1005218 &    82.703300 &    10.089400 &    0.6$_{- 0.06}^{+ 0.09}$ &   391.78$_{-   5.54}^{+   5.88}$ &     3.56$\pm$   0.002 &     4.17$\pm$   0.054 &     0.01$\pm$   0.009 &    -0.61$\pm$   0.022 &    0.7$_{- 0.01}^{+ 0.02}$ &    5.5$_{- 0.53}^{+ 0.57}$ \\
2M05305801+1001153 &    82.741699 &    10.020900 &    0.6$_{- 0.08}^{+ 0.04}$ &   400.45$_{-   5.44}^{+   4.74}$ &     3.56$\pm$   0.002 &     3.96$\pm$   0.071 &     0.02$\pm$   0.009 &    -0.58$\pm$   0.017 &    0.6$_{- 0.01}^{+ 0.01}$ &    3.7$_{- 0.26}^{+ 0.39}$ \\
2M05310101+1204142 &    82.754204 &    12.070600 &    0.7$_{- 0.09}^{+ 0.07}$ &   383.79$_{-   5.04}^{+   4.38}$ &     3.56$\pm$   0.003 &     3.88$\pm$   0.053 &    -0.02$\pm$   0.007 &    -0.65$\pm$   0.019 &    0.6$_{- 0.01}^{+ 0.02}$ &    5.4$_{- 0.40}^{+ 0.56}$ \\\hline
\hline
\multicolumn{11}{c}{Orion A Complex}\\
\hline
2M05324266-0459313 &    83.177803 &    -4.992000 &    0.2$_{- 0.10}^{+ 0.15}$ &   372.21$_{-   2.71}^{+   3.22}$ &     3.57$\pm$   0.005 &     4.01$\pm$   0.067 &     0.01$\pm$   0.014 &    -0.42$\pm$   0.028 &    2.3$_{- 1.76}^{+-1.71}$ &    2.3$_{- 0.20}^{+ 0.25}$ \\
2M05324731-0539426 &    83.197098 &    -5.661900 &    0.7$_{- 0.06}^{+ 0.04}$ &   372.94$_{-   3.32}^{+   3.39}$ &     3.55$\pm$   0.001 &     4.07$\pm$   0.030 &     0.06$\pm$   0.006 &    -0.53$\pm$   0.015 &    2.6$_{- 2.09}^{+-2.07}$ &    2.6$_{- 0.20}^{+ 0.16}$ \\
2M05325352-0606011 &    83.223000 &    -6.100300 &    0.8$_{- 0.10}^{+ 0.13}$ &   372.94$_{-   7.03}^{+   8.01}$ &     3.53$\pm$   0.003 &     4.15$\pm$   0.097 &     0.03$\pm$   0.015 &    -0.86$\pm$   0.027 &    6.0$_{- 5.57}^{+-5.53}$ &    6.0$_{- 0.59}^{+ 0.88}$ \\
2M05325531-0412390 &    83.230499 &    -4.210800 &    0.4$_{- 0.04}^{+ 0.05}$ &   356.97$_{-   3.24}^{+   2.91}$ &     3.57$\pm$   0.001 &     4.28$\pm$   0.039 &    -0.02$\pm$   0.006 &    -0.53$\pm$   0.017 &    4.9$_{- 4.19}^{+-4.17}$ &    4.9$_{- 0.29}^{+ 0.45}$ \\
2M05325630-0603189 &    83.234596 &    -6.055300 &    0.7$_{- 0.14}^{+ 0.08}$ &   348.48$_{-   4.57}^{+   5.58}$ &     3.56$\pm$   0.004 &     4.04$\pm$   0.077 &    -0.01$\pm$   0.010 &    -0.84$\pm$   0.023 &   14.7$_{-14.02}^{-14.01}$ &   14.7$_{- 1.41}^{+ 1.75}$ \\
2M05325690-0512476 &    83.237099 &    -5.213200 &    1.2$_{- 0.07}^{+ 0.07}$ &   393.12$_{-   4.90}^{+   4.94}$ &     3.57$\pm$   0.001 &     4.24$\pm$   0.030 &     0.01$\pm$   0.003 &    -0.50$\pm$   0.018 &    3.2$_{- 2.64}^{+-2.62}$ &    3.2$_{- 0.26}^{+ 0.24}$ \\
2M05325791-0602429 &    83.241302 &    -6.045300 &    0.6$_{- 0.12}^{+ 0.11}$ &   380.64$_{-   4.96}^{+   5.93}$ &     3.55$\pm$   0.004 &     4.32$\pm$   0.058 &     0.04$\pm$   0.015 &    -0.76$\pm$   0.027 &    7.0$_{- 6.46}^{+-6.42}$ &    7.0$_{- 0.81}^{+ 0.75}$ \\
2M05330175-0449184 &    83.257301 &    -4.821800 &    1.3$_{- 0.20}^{+ 0.14}$ &   377.56$_{-   4.44}^{+   3.86}$ &     3.68$\pm$   0.024 &     3.28$\pm$   0.208 & \nodata               &    -0.70$\pm$   0.046 &   67.0$_{-66.27}^{-66.17}$ &   67.0$_{- 8.39}^{+ 4.94}$ \\
2M05330432-0519410 &    83.267998 &    -5.328100 &    0.2$_{- 0.12}^{+ 0.09}$ &   375.15$_{-   2.35}^{+   2.69}$ &     3.61$\pm$   0.006 &     4.03$\pm$   0.092 & \nodata               &    -0.27$\pm$   0.020 &    3.0$_{- 2.30}^{+-2.29}$ &    3.0$_{- 0.13}^{+ 0.14}$ \\
2M05331059-0459078 &    83.294098 &    -4.985500 &    0.8$_{- 0.15}^{+ 0.13}$ &   373.95$_{-   5.00}^{+   5.46}$ &     3.55$\pm$   0.005 &     3.99$\pm$   0.077 &    -0.04$\pm$   0.019 &    -0.74$\pm$   0.031 &    7.8$_{- 7.16}^{+-7.11}$ &    7.8$_{- 0.82}^{+ 1.08}$ \\
\hline
\multicolumn{11}{c}{Orion B Complex}\\
\hline
2M05361923-0123588 &    84.080101 &    -1.399700 &    0.7$_{- 0.05}^{+ 0.08}$ &   397.79$_{-   5.06}^{+   4.72}$ &     3.59$\pm$   0.002 &     4.22$\pm$   0.043 &    -0.09$\pm$   0.006 &    -0.49$\pm$   0.019 &    0.7$_{- 0.00}^{+ 0.00}$ &    5.1$_{- 0.40}^{+ 0.32}$ \\
2M05362908-0235482 &    84.121201 &    -2.596700 &    0.8$_{- 0.07}^{+ 0.05}$ &   359.18$_{-   3.72}^{+   3.40}$ &     3.56$\pm$   0.001 &     3.98$\pm$   0.029 &     0.03$\pm$   0.005 &    -0.57$\pm$   0.017 &    0.6$_{- 0.01}^{+ 0.01}$ &    4.1$_{- 0.30}^{+ 0.32}$ \\
2M05363302-0123334 &    84.137604 &    -1.392600 &    0.5$_{- 0.03}^{+ 0.03}$ &   340.99$_{-   1.61}^{+   1.66}$ &     3.64$\pm$   0.001 &     4.38$\pm$   0.032 & \nodata               &     0.10$\pm$   0.013 &    0.8$_{- 0.01}^{+ 0.01}$ &    1.4$_{- 0.06}^{+ 0.09}$ \\
2M05363878-0254196 &    84.161598 &    -2.905500 &    0.7$_{- 0.07}^{+ 0.06}$ &   361.40$_{-   4.48}^{+   4.21}$ &     3.55$\pm$   0.002 &     4.25$\pm$   0.049 &     0.02$\pm$   0.007 &    -0.71$\pm$   0.018 &    0.6$_{- 0.01}^{+ 0.01}$ &    6.9$_{- 0.40}^{+ 0.59}$ \\
2M05365019-0247099 &    84.209198 &    -2.786100 &    0.7$_{- 0.08}^{+ 0.09}$ &   365.75$_{-   3.79}^{+   3.60}$ &     3.56$\pm$   0.001 &     4.24$\pm$   0.045 &     0.03$\pm$   0.008 &    -0.72$\pm$   0.019 &    0.7$_{- 0.01}^{+ 0.01}$ &    8.4$_{- 0.67}^{+ 0.79}$ \\
2M05365170-0210104 &    84.215401 &    -2.169600 &    3.5$_{- 0.20}^{+ 0.22}$ &   362.20$_{-   9.52}^{+   7.56}$ &     3.69$\pm$   0.026 &     4.29$\pm$   0.094 & \nodata               &    -0.17$\pm$   0.062 &    1.1$_{- 0.03}^{+ 0.03}$ &   12.3$_{- 3.35}^{+ 6.54}$ \\
2M05365231-0121246 &    84.218002 &    -1.356900 &    0.7$_{- 0.03}^{+ 0.03}$ &   348.45$_{-   1.95}^{+   1.51}$ &     3.62$\pm$   0.001 &     4.14$\pm$   0.046 & \nodata               &    -0.18$\pm$   0.012 &    0.7$_{- 0.00}^{+ 0.00}$ &    2.4$_{- 0.09}^{+ 0.09}$ \\
2M05365409-0253155 &    84.225403 &    -2.887700 &    0.4$_{- 0.05}^{+ 0.04}$ &   380.19$_{-   2.30}^{+   2.06}$ &     3.64$\pm$   0.002 &     4.24$\pm$   0.032 & \nodata               &     0.06$\pm$   0.016 &    0.9$_{- 0.03}^{+ 0.03}$ &    1.6$_{- 0.12}^{+ 0.17}$ \\
2M05365500-0135508 &    84.229202 &    -1.597500 &    0.4$_{- 0.04}^{+ 0.03}$ &   419.06$_{-   5.09}^{+   5.39}$ &     3.59$\pm$   0.002 &     4.26$\pm$   0.053 &    -0.01$\pm$   0.007 &    -0.59$\pm$   0.016 &    0.7$_{- 0.00}^{+ 0.00}$ &    7.6$_{- 0.35}^{+ 0.63}$ \\
2M05365983-0120245 &    84.249298 &    -1.340100 &    0.9$_{- 0.10}^{+ 0.16}$ &   417.41$_{-   9.01}^{+  11.73}$ &     3.52$\pm$   0.004 &     4.48$\pm$   0.061 &    -0.02$\pm$   0.014 &    -0.67$\pm$   0.032 &    0.4$_{- 0.03}^{+ 0.03}$ &    2.2$_{- 0.32}^{+ 0.39}$ \\
\hline
\multicolumn{11}{c}{Orion OB1 Complex}\\
\hline
2M05191549-0204529 &    79.814598 &    -2.081400 &    1.6$_{- 0.07}^{+ 0.05}$ &   359.50$_{-   7.43}^{+   8.15}$ &     3.55$\pm$   0.002 &     4.17$\pm$   0.078 &     0.01$\pm$   0.009 &    -0.63$\pm$   0.026 &    0.6$_{- 0.02}^{+ 0.02}$ &    4.4$_{- 0.46}^{+ 0.58}$ \\
2M05192175-0217193 &    79.840698 &    -2.288700 &    0.6$_{- 0.03}^{+ 0.02}$ &   349.42$_{-   2.47}^{+   2.09}$ &     3.63$\pm$   0.002 &     4.44$\pm$   0.036 & \nodata               &    -0.29$\pm$   0.014 &    0.9$_{- 0.02}^{+ 0.02}$ &    5.2$_{- 0.38}^{+ 0.45}$ \\
2M05194349-0116397 &    79.931198 &    -1.277700 &    0.4$_{- 0.03}^{+ 0.04}$ &   364.45$_{-   3.73}^{+   3.49}$ &     3.58$\pm$   0.001 &     4.43$\pm$   0.027 &     0.02$\pm$   0.006 &    -0.53$\pm$   0.015 &    0.7$_{- 0.00}^{+ 0.00}$ &    5.3$_{- 0.21}^{+ 0.33}$ \\
2M05195766-0301262 &    79.990303 &    -3.024000 &    0.8$_{- 0.04}^{+ 0.02}$ &   351.16$_{-   2.11}^{+   2.00}$ &     3.65$\pm$   0.001 &     4.50$\pm$   0.025 & \nodata               &    -0.18$\pm$   0.013 &    1.0$_{- 0.01}^{+ 0.01}$ &    6.2$_{- 0.28}^{+ 0.35}$ \\
2M05201480-0225456 &    80.061699 &    -2.429300 &    0.5$_{- 0.03}^{+ 0.05}$ &   335.54$_{-   1.61}^{+   1.47}$ &     3.67$\pm$   0.001 &     4.43$\pm$   0.034 & \nodata               &    -0.33$\pm$   0.015 &    1.0$_{- 0.01}^{+ 0.02}$ &   14.3$_{- 0.95}^{+ 0.72}$ \\
2M05201885-0149010 &    80.078598 &    -1.817000 &    0.8$_{- 0.04}^{+ 0.03}$ &   350.42$_{-   1.78}^{+   2.08}$ &     3.65$\pm$   0.002 &     4.33$\pm$   0.026 & \nodata               &    -0.17$\pm$   0.014 &    1.0$_{- 0.01}^{+ 0.02}$ &    4.6$_{- 0.29}^{+ 0.30}$ \\
2M05202650-0300325 &    80.110397 &    -3.009100 &    1.0$_{- 0.08}^{+ 0.07}$ &   339.28$_{-   4.73}^{+   4.65}$ &     3.54$\pm$   0.002 &     4.31$\pm$   0.078 &     0.03$\pm$   0.015 &    -0.78$\pm$   0.023 &    0.6$_{- 0.01}^{+ 0.01}$ &    6.3$_{- 0.55}^{+ 0.68}$ \\
2M05203382-0155237 &    80.140900 &    -1.923300 &    1.0$_{- 0.13}^{+ 0.16}$ &   350.53$_{-   3.98}^{+   3.57}$ &     3.57$\pm$   0.007 &     4.17$\pm$   0.066 &    -0.02$\pm$   0.015 &    -0.52$\pm$   0.030 &    0.7$_{- 0.04}^{+ 0.05}$ &    4.1$_{- 0.42}^{+ 0.41}$ \\
2M05204202-0128109 &    80.175102 &    -1.469700 &    0.3$_{- 0.06}^{+ 0.06}$ &   342.74$_{-   1.91}^{+   2.42}$ &     3.63$\pm$   0.004 &     4.01$\pm$   0.039 & \nodata               &    -0.05$\pm$   0.020 &    0.7$_{- 0.00}^{+ 0.01}$ &    1.6$_{- 0.11}^{+ 0.10}$ \\
2M05205489+0102022 &    80.228699 &     1.034000 &    0.4$_{- 0.03}^{+ 0.04}$ &   354.09$_{-   1.53}^{+   1.79}$ &     3.65$\pm$   0.003 &     4.55$\pm$   0.028 & \nodata               &    -0.16$\pm$   0.016 &    1.0$_{- 0.02}^{+ 0.02}$ &    4.4$_{- 0.32}^{+ 0.33}$ \\
\hline
\multicolumn{11}{c}{Pleiades Cluster}\\
\hline
2M03465940+2431124 &    56.747475 &    24.520124 &    0.1$_{- 0.05}^{+ 0.06}$ &   136.89$_{-   0.59}^{+   0.63}$ &     3.99$\pm$   0.006 &     3.96$\pm$   0.017 & \nodata               &     1.57$\pm$   0.027 &    2.3$_{- 0.05}^{+ 0.07}$ &\nodata                     \\
2M03470141+2329419 &    56.755917 &    23.494942 &    0.0$_{- 0.00}^{+ 0.00}$ &   134.65$_{-   0.32}^{+   0.32}$ &     3.71$\pm$   0.001 &     4.65$\pm$   0.011 & \nodata               &    -0.42$\pm$   0.012 &    0.8$_{- 0.00}^{+ 0.00}$ &   43.3$_{- 1.60}^{+ 1.81}$ \\
2M03470358+2409349 &    56.764927 &    24.159687 &    0.3$_{- 0.10}^{+ 0.12}$ &   136.61$_{-   0.37}^{+   0.43}$ &     3.61$\pm$   0.006 &     4.39$\pm$   0.035 & \nodata               &    -1.10$\pm$   0.023 &    0.6$_{- 0.01}^{+ 0.01}$ &  121.8$_{- 6.32}^{+ 6.41}$ \\
2M03470376+2336588 &    56.765732 &    23.616312 &    0.6$_{- 0.07}^{+ 0.08}$ &   137.70$_{-   0.48}^{+   0.31}$ &     3.57$\pm$   0.002 &     4.54$\pm$   0.038 &    -0.03$\pm$   0.006 &    -1.06$\pm$   0.018 &    0.6$_{- 0.01}^{+ 0.01}$ &   43.2$_{- 3.51}^{+ 2.73}$ \\
2M03470421+2359426 &    56.767555 &    23.995214 &    0.4$_{- 0.04}^{+ 0.04}$ &   134.46$_{-   0.70}^{+   0.45}$ &     3.90$\pm$   0.004 &     4.05$\pm$   0.023 & \nodata               &     0.92$\pm$   0.018 &    1.7$_{- 0.01}^{+ 0.02}$ &\nodata                     \\
2M03470678+2342546 &    56.778248 &    23.715149 &    0.2$_{- 0.03}^{+ 0.03}$ &   135.60$_{-   0.25}^{+   0.26}$ &     3.70$\pm$   0.002 &     4.70$\pm$   0.018 & \nodata               &    -0.54$\pm$   0.013 &    0.8$_{- 0.01}^{+ 0.01}$ &   72.0$_{-16.88}^{+20.85}$ \\
2M03470734+2313349 &    56.780598 &    23.226358 &    0.2$_{- 0.08}^{+ 0.07}$ &   135.98$_{-   0.34}^{+   0.33}$ &     3.63$\pm$   0.004 &     4.19$\pm$   0.031 & \nodata               &    -0.91$\pm$   0.017 &    0.7$_{- 0.01}^{+ 0.01}$ &  118.8$_{-30.83}^{+95.79}$ \\
2M03470813+2418246 &    56.783913 &    24.306841 &    0.3$_{- 0.06}^{+ 0.08}$ &   134.27$_{-   0.91}^{+   1.11}$ &     3.53$\pm$   0.004 &     4.69$\pm$   0.045 &     0.02$\pm$   0.011 &    -1.67$\pm$   0.017 &    0.4$_{- 0.01}^{+ 0.01}$ &\nodata                     \\
2M03470918+2403078 &    56.788261 &    24.052174 &    0.4$_{- 0.13}^{+ 0.11}$ &   135.96$_{-   0.95}^{+   0.83}$ &     3.55$\pm$   0.008 &     4.60$\pm$   0.076 &    -0.03$\pm$   0.020 &    -1.55$\pm$   0.024 &    0.5$_{- 0.01}^{+ 0.01}$ &  217.6$_{-10.05}^{+ 7.94}$ \\
2M03471352+2342515 &    56.806385 &    23.714289 &    0.1$_{- 0.03}^{+ 0.03}$ &   134.55$_{-   0.29}^{+   0.22}$ &     3.68$\pm$   0.001 &     4.64$\pm$   0.015 & \nodata               &    -0.66$\pm$   0.012 &    0.8$_{- 0.01}^{+ 0.01}$ &  100.8$_{-35.05}^{+25.61}$ \\
2M03471365+2349535 &    56.806923 &    23.831497 &    0.0$_{- 0.08}^{+ 0.06}$ &   134.42$_{-   0.63}^{+   0.64}$ &     3.53$\pm$   0.003 &     4.51$\pm$   0.046 &     0.01$\pm$   0.014 &    -1.36$\pm$   0.017 &    0.5$_{- 0.02}^{+ 0.01}$ &\nodata                     \\
\hline
\multicolumn{11}{c}{Rosette Complex}\\
\hline
2M06322147+0450274 &    98.089500 &     4.841000 &    1.6$_{- 0.05}^{+ 0.05}$ &  1505.39$_{-  70.94}^{+  77.93}$ &     3.68$\pm$   0.004 &     3.93$\pm$   0.115 & \nodata               &     0.63$\pm$   0.046 &    1.3$_{- 0.07}^{+ 0.07}$ &    0.7$_{- 0.10}^{+ 0.14}$ \\
2M06322421+0358061 &    98.100899 &     3.968400 &    3.4$_{- 0.13}^{+ 0.14}$ &  1632.39$_{- 119.09}^{+ 110.22}$ &     3.72$\pm$   0.016 &     4.34$\pm$   0.091 & \nodata               &     1.03$\pm$   0.070 &    2.4$_{- 0.36}^{+ 0.25}$ &    0.9$_{- 0.40}^{+ 0.33}$ \\
2M06322511+0447396 &    98.104698 &     4.794300 &    1.5$_{- 0.04}^{+ 0.04}$ &  1441.40$_{-  69.04}^{+  77.22}$ &     3.69$\pm$   0.004 &     4.32$\pm$   0.046 & \nodata               &     0.50$\pm$   0.045 &    1.4$_{- 0.06}^{+ 0.06}$ &    1.3$_{- 0.22}^{+ 0.21}$ \\
2M06322658+0452383 &    98.110802 &     4.877300 &    2.6$_{- 0.13}^{+ 0.16}$ &  1819.46$_{- 420.47}^{+ 655.46}$ &     3.64$\pm$   0.015 &     3.10$\pm$   0.233 & \nodata               &     0.14$\pm$   0.234 &    0.8$_{- 0.06}^{+ 0.15}$ &    1.1$_{- 0.59}^{+ 1.64}$ \\
2M06322766+0445035 &    98.115303 &     4.751000 &    1.3$_{- 0.05}^{+ 0.05}$ &  1332.67$_{-  58.01}^{+  60.13}$ &     3.67$\pm$   0.004 &     4.19$\pm$   0.070 & \nodata               &     0.48$\pm$   0.044 &    1.2$_{- 0.06}^{+ 0.06}$ &    0.9$_{- 0.11}^{+ 0.18}$ \\
2M06322807+0454037 &    98.116997 &     4.901000 &    1.2$_{- 0.10}^{+ 0.07}$ &  1409.32$_{-  38.42}^{+  30.00}$ &     3.73$\pm$   0.009 &     3.66$\pm$   0.251 & \nodata               &     0.92$\pm$   0.035 &    2.4$_{- 0.14}^{+ 0.07}$ &    1.4$_{- 0.27}^{+ 0.30}$ \\
2M06322894+0449308 &    98.120598 &     4.825200 &    1.8$_{- 0.18}^{+ 0.17}$ &  1195.05$_{- 217.10}^{+ 229.08}$ &     3.57$\pm$   0.008 &     4.06$\pm$   0.148 &     0.00$\pm$   0.032 &    -0.27$\pm$   0.133 &    0.5$_{- 0.08}^{+ 0.08}$ &    1.3$_{- 0.49}^{+ 1.22}$ \\
2M06322969+0501344 &    98.123703 &     5.026200 &    1.5$_{- 0.10}^{+ 0.10}$ &  1326.31$_{- 108.70}^{+ 132.70}$ &     3.66$\pm$   0.002 &     4.07$\pm$   0.067 & \nodata               &     0.34$\pm$   0.077 &    1.1$_{- 0.02}^{+ 0.03}$ &    1.1$_{- 0.26}^{+ 0.29}$ \\
2M06323100+0450059 &    98.129204 &     4.835000 &    1.8$_{- 0.11}^{+ 0.20}$ &  1371.56$_{-  34.32}^{+  48.69}$ &     3.78$\pm$   0.007 &     4.15$\pm$   0.072 & \nodata               &     1.15$\pm$   0.042 &\nodata                     &\nodata                     \\
2M06323336+0334525 &    98.139000 &     3.581300 &    1.9$_{- 0.23}^{+ 0.18}$ &  1476.47$_{- 133.81}^{+ 178.67}$ &     3.73$\pm$   0.029 &     4.22$\pm$   0.105 & \nodata               &     0.43$\pm$   0.115 &    1.6$_{- 0.13}^{+ 0.13}$ &    5.5$_{- 1.92}^{+ 3.43}$ \\
\hline
\multicolumn{11}{c}{Taurus Complex}\\
\hline
2M04331003+2433433 &    68.291801 &    24.562000 &    1.1$_{- 0.03}^{+ 0.03}$ &   129.97$_{-   0.24}^{+   0.28}$ &     3.62$\pm$   0.001 &     3.86$\pm$   0.014 & \nodata               &    -0.10$\pm$   0.011 &    0.7$_{- 0.00}^{+ 0.00}$ &    1.8$_{- 0.06}^{+ 0.07}$ \\
2M04332621+2245293 &    68.359200 &    22.758200 &    5.5$_{- 0.06}^{+ 0.07}$ &   157.24$_{-   2.68}^{+   2.73}$ &     3.56$\pm$   0.001 &     3.78$\pm$   0.025 &    -0.02$\pm$   0.004 &    -0.49$\pm$   0.020 &    0.5$_{- 0.01}^{+ 0.01}$ &    2.5$_{- 0.21}^{+ 0.27}$ \\
2M04333278+1800436 &    68.386597 &    18.012100 &    1.2$_{- 0.05}^{+ 0.06}$ &   145.08$_{-   0.46}^{+   0.55}$ &     3.56$\pm$   0.001 &     4.15$\pm$   0.018 &     0.02$\pm$   0.004 &    -0.56$\pm$   0.017 &    0.6$_{- 0.01}^{+ 0.01}$ &    3.4$_{- 0.26}^{+ 0.28}$ \\
2M04333405+2421170 &    68.391899 &    24.354700 &    2.2$_{- 0.08}^{+ 0.12}$ &   129.32$_{-   0.42}^{+   0.35}$ &     3.59$\pm$   0.001 &     3.77$\pm$   0.016 &    -0.02$\pm$   0.003 &    -0.02$\pm$   0.016 &    0.5$_{- 0.01}^{+ 0.01}$ &    0.7$_{- 0.04}^{+ 0.04}$ \\
2M04334171+1750402 &    68.423798 &    17.844500 &    1.4$_{- 0.08}^{+ 0.08}$ &   143.62$_{-   0.68}^{+   0.63}$ &     3.53$\pm$   0.002 &     4.29$\pm$   0.031 &    -0.05$\pm$   0.005 &    -0.86$\pm$   0.017 &    0.5$_{- 0.01}^{+ 0.01}$ &    7.5$_{- 0.54}^{+ 0.66}$ \\
2M04334298+2235566 &    68.429100 &    22.599100 &    2.1$_{- 0.04}^{+ 0.04}$ &   163.29$_{-   3.29}^{+   3.14}$ &     3.54$\pm$   0.001 &     4.54$\pm$   0.022 &    -0.01$\pm$   0.003 &    -0.67$\pm$   0.020 &    0.5$_{- 0.01}^{+ 0.01}$ &    3.4$_{- 0.33}^{+ 0.32}$ \\
2M04335200+2250301 &    68.466698 &    22.841700 &    1.7$_{- 0.05}^{+ 0.06}$ &   160.20$_{-   0.44}^{+   0.47}$ &     3.61$\pm$   0.001 &     3.65$\pm$   0.013 & \nodata               &     0.08$\pm$   0.012 &    0.7$_{- 0.01}^{+ 0.01}$ &    0.8$_{- 0.04}^{+ 0.04}$ \\
2M04335283+1803166 &    68.470100 &    18.054600 &    1.4$_{- 0.08}^{+ 0.10}$ &   144.24$_{-   0.76}^{+   0.67}$ &     3.53$\pm$   0.001 &     4.32$\pm$   0.022 &    -0.07$\pm$   0.004 &    -0.70$\pm$   0.021 &    0.4$_{- 0.01}^{+ 0.01}$ &\nodata                     \\
2M04341099+2251445 &    68.545799 &    22.862400 &    2.4$_{- 0.04}^{+ 0.06}$ &   160.87$_{-   0.48}^{+   0.57}$ &     3.58$\pm$   0.001 &     4.05$\pm$   0.018 &     0.06$\pm$   0.003 &    -0.32$\pm$   0.014 &    0.6$_{- 0.01}^{+ 0.01}$ &    2.0$_{- 0.11}^{+ 0.13}$ \\
2M04341527+2250309 &    68.563599 &    22.841900 &    4.7$_{- 0.11}^{+ 0.19}$ &   164.89$_{-   9.73}^{+  11.83}$ &     3.52$\pm$   0.004 &     3.83$\pm$   0.091 & \nodata               &    -1.35$\pm$   0.060 &\nodata                     &\nodata                     \\
\hline
\multicolumn{11}{c}{Vela C Complex}\\
\hline
2M08572257-4353178 &   134.343994 &   -43.888302 &    1.2$_{- 0.08}^{+ 0.11}$ &   801.77$_{-  27.51}^{+  29.63}$ &     3.73$\pm$   0.008 &     4.02$\pm$   0.144 & \nodata               &     0.06$\pm$   0.040 &    1.2$_{- 0.05}^{+ 0.04}$ &   13.7$_{- 1.53}^{+ 2.77}$ \\
2M08572811-4257594 &   134.367203 &   -42.966499 &    0.7$_{- 0.05}^{+ 0.05}$ &   880.66$_{-  17.35}^{+  22.17}$ &     3.62$\pm$   0.003 &     4.10$\pm$   0.085 & \nodata               &     0.00$\pm$   0.022 &    0.7$_{- 0.01}^{+ 0.01}$ &    1.2$_{- 0.06}^{+ 0.08}$ \\
2M08572851-4243479 &   134.368805 &   -42.730000 &    3.5$_{- 0.10}^{+ 0.10}$ &   916.20$_{-  44.85}^{+  54.95}$ &     3.72$\pm$   0.012 &     3.84$\pm$   0.211 & \nodata               &     0.05$\pm$   0.056 &    1.2$_{- 0.05}^{+ 0.06}$ &   13.4$_{- 2.54}^{+ 2.64}$ \\
2M08573006-4323570 &   134.375305 &   -43.399200 &    2.0$_{- 0.07}^{+ 0.06}$ &   907.24$_{-  23.54}^{+  30.43}$ &     3.70$\pm$   0.003 &     4.07$\pm$   0.086 & \nodata               &     0.01$\pm$   0.028 &    1.2$_{- 0.03}^{+ 0.02}$ &    8.8$_{- 1.01}^{+ 0.75}$ \\
2M08573446-4314463 &   134.393600 &   -43.246201 &    0.7$_{- 0.10}^{+ 0.07}$ &   899.93$_{-  45.96}^{+  35.97}$ &     3.58$\pm$   0.004 &     4.45$\pm$   0.092 &     0.07$\pm$   0.017 &    -0.61$\pm$   0.039 &    0.7$_{- 0.00}^{+ 0.00}$ &    6.8$_{- 0.82}^{+ 1.43}$ \\
2M08573501-4308355 &   134.395905 &   -43.143200 &    0.9$_{- 0.19}^{+ 0.16}$ &   886.26$_{-  35.78}^{+  43.05}$ &     3.60$\pm$   0.006 &     4.52$\pm$   0.075 &     0.03$\pm$   0.019 &    -0.46$\pm$   0.053 &    0.7$_{- 0.00}^{+ 0.01}$ &    5.1$_{- 0.65}^{+ 1.13}$ \\
2M08573557-4259590 &   134.398193 &   -42.999699 &    1.1$_{- 0.15}^{+ 0.16}$ &   849.14$_{-  37.00}^{+  54.82}$ &     3.59$\pm$   0.009 &     4.11$\pm$   0.114 &    -0.02$\pm$   0.040 &    -0.61$\pm$   0.059 &    0.7$_{- 0.01}^{+ 0.01}$ &    7.6$_{- 1.36}^{+ 1.82}$ \\
2M08573649-4310134 &   134.402100 &   -43.170399 &    1.1$_{- 0.04}^{+ 0.04}$ &   891.54$_{-  21.59}^{+  24.08}$ &     3.63$\pm$   0.003 &     4.46$\pm$   0.050 & \nodata               &    -0.10$\pm$   0.024 &    0.7$_{- 0.00}^{+ 0.01}$ &    2.0$_{- 0.12}^{+ 0.20}$ \\
2M08573704-4239514 &   134.404404 &   -42.664299 &    3.3$_{- 0.15}^{+ 0.14}$ &  1043.98$_{- 126.72}^{+ 155.27}$ &     3.59$\pm$   0.008 &     4.08$\pm$   0.154 &    -0.10$\pm$   0.044 &    -0.12$\pm$   0.090 &    0.6$_{- 0.09}^{+ 0.06}$ &    1.0$_{- 0.34}^{+ 0.51}$ \\
2M08573823-4302379 &   134.409302 &   -43.043900 &    1.0$_{- 0.16}^{+ 0.15}$ &   845.97$_{-  39.12}^{+  32.95}$ &     3.56$\pm$   0.005 &     4.06$\pm$   0.196 &     0.01$\pm$   0.028 &    -0.54$\pm$   0.044 &    0.6$_{- 0.03}^{+ 0.04}$ &    3.2$_{- 0.51}^{+ 0.73}$ \\
\hline
\multicolumn{11}{c}{W3/W4/W5 Complex}\\
\hline
2M02341416+6109306 &    38.558998 &    61.158501 &    1.9$_{- 0.13}^{+ 0.26}$ &  2010.31$_{-  70.57}^{+  62.78}$ &     3.78$\pm$   0.011 &     4.06$\pm$   0.060 & \nodata               &     1.41$\pm$   0.051 &    2.7$_{- 0.08}^{+ 0.09}$ &    2.0$_{- 0.30}^{+ 0.30}$ \\
2M02341820+6147375 &    38.575901 &    61.793800 &    0.1$_{- 0.28}^{+ 0.19}$ &  1868.00$_{-  78.82}^{+  84.51}$ &     3.70$\pm$   0.022 &     4.31$\pm$   0.143 & \nodata               &     0.67$\pm$   0.070 &    1.8$_{- 0.59}^{+ 0.22}$ &    1.2$_{- 0.57}^{+ 0.72}$ \\
2M02342952+6147403 &    38.623001 &    61.794498 &    0.0$_{- 0.25}^{+ 0.23}$ &  1975.95$_{-  73.15}^{+  77.94}$ &     3.72$\pm$   0.027 &     4.11$\pm$   0.126 & \nodata               &     0.90$\pm$   0.069 &    2.3$_{- 0.34}^{+ 0.12}$ &    1.6$_{- 0.71}^{+ 0.88}$ \\
2M02343170+6116271 &    38.632099 &    61.274200 &    1.5$_{- 0.12}^{+ 0.10}$ &  2040.95$_{- 108.24}^{+ 104.60}$ &     3.73$\pm$   0.014 &     3.84$\pm$   0.201 & \nodata               &     1.01$\pm$   0.053 &    2.5$_{- 0.15}^{+ 0.11}$ &    1.2$_{- 0.25}^{+ 0.41}$ \\
2M02343670+6146583 &    38.652901 &    61.782902 &    1.7$_{- 0.09}^{+ 0.12}$ &  1938.33$_{-  76.38}^{+  88.28}$ &     3.76$\pm$   0.014 &     3.80$\pm$   0.121 & \nodata               &     1.23$\pm$   0.060 &    2.7$_{- 0.16}^{+ 0.14}$ &    1.6$_{- 0.41}^{+ 0.59}$ \\
2M02343995+6145231 &    38.666500 &    61.756401 &    1.2$_{- 0.18}^{+ 0.25}$ &  2031.93$_{-  59.21}^{+  56.31}$ &     3.74$\pm$   0.022 &     3.92$\pm$   0.165 & \nodata               &     1.20$\pm$   0.066 &    2.8$_{- 0.23}^{+ 0.10}$ &    1.2$_{- 0.54}^{+ 0.99}$ \\
2M02345682+6142075 &    38.736801 &    61.702099 &    1.9$_{- 0.04}^{+ 0.04}$ &  1851.73$_{-  72.71}^{+  76.49}$ &     3.69$\pm$   0.003 &     3.12$\pm$   0.084 & \nodata               &     1.29$\pm$   0.033 &    2.1$_{- 0.09}^{+ 0.15}$ &\nodata                     \\
2M02345825+6121161 &    38.742699 &    61.354500 &    3.0$_{- 0.23}^{+ 0.16}$ &  2041.91$_{- 127.78}^{+ 164.83}$ &     3.73$\pm$   0.022 &     4.16$\pm$   0.156 & \nodata               &     1.10$\pm$   0.067 &    2.6$_{- 0.32}^{+ 0.13}$ &    1.1$_{- 0.46}^{+ 0.78}$ \\
2M02350481+6110320 &    38.770100 &    61.175598 &    1.9$_{- 0.22}^{+ 0.10}$ &  2015.09$_{-  87.29}^{+  69.36}$ &     3.79$\pm$   0.010 &     4.38$\pm$   0.084 & \nodata               &     1.34$\pm$   0.056 &    2.5$_{- 0.08}^{+ 0.07}$ &    2.7$_{- 0.23}^{+ 0.30}$ \\
\enddata
\end{deluxetable*}
\end{longrotatetable}

%% file: figset1.tex
\figsetstart
\figsetnum{2}
\label{FS1}
\figsettitle{APOGEE-2 Target Position Maps}

\figsetend

%% file: figset2.tex
\figsetstart
\figsetnum{5}
\label{FS2}
\figsettitle{Spatial Distribution of Main Parameters}

\figsetend

%% file: figset3.tex
\figsetstart
\figsetnum{6}
\label{FS3}
\figsettitle{Hertzprung-Rusell Diagrams}

\figsetend

%% file: draft_03.bbl
\begin{thebibliography}{}
\expandafter\ifx\csname natexlab\endcsname\relax\def\natexlab#1{#1}\fi

\bibitem[{{Abdurro'uf} {et~al.}(2021){Abdurro'uf}, {Accetta}, {Aerts}, {Silva
  Aguirre}, {Ahumada}, {Ajgaonkar}, {Filiz Ak}, {Alam}, {Allende Prieto},
  {Almeida}, {Anders}, {Anderson}, {Andrews}, {Anguiano}, {Aquino-Ortiz},
  {Aragon-Salamanca}, {Argudo-Fernandez}, {Ata}, {Aubert}, {Avila-Reese},
  {Badenes}, {Barba}, {Barger}, {Barrera-Ballesteros}, {Beaton}, {Beers},
  {Belfiore}, {Bender}, {Bernardi}, {Bershady}, {Beutler}, {Moni Bidin},
  {Bird}, {Bizyaev}, {Blanc}, {Blanton}, {Boardman}, {Bolton}, {Boquien},
  {Borissova}, {Bovy}, {Brandt}, {Brown}, {Brownstein}, {Brusa}, {Buchner},
  {Bundy}, {Burchett}, {Bureau}, {Burgasser}, {Cabang}, {Campbell},
  {Cappellari}, {Carlberg}, {Carneiro Wanderley}, {Carrera}, {Cash}, {Chen},
  {Chen}, {Cherinka}, {Chiappini}, {Choi}, {Chojnowski}, {Chung}, {Clerc},
  {Cohen}, {Comerford}, {Comparat}, {da Costa}, {Covey}, {Crane},
  {Cruz-Gonzalez}, {Culhane}, {Cunha}, {Dai}, {Damke}, {Darling}, {Davidson},
  {Davies}, {Dawson}, {De Lee}, {Diamond-Stanic}, {Cano-Diaz}, {Dominguez
  Sanchez}, {Donor}, {Duckworth}, {Dwelly}, {Eisenstein}, {Elsworth},
  {Emsellem}, {Eracleous}, {Escoffier}, {Fan}, {Farr}, {Feng},
  {Fernandez-Trincado}, {Feuillet}, {Filipp}, {Fillingham}, {Frinchaboy},
  {Fromenteau}, {Galbany}, {Garcia}, {Garcia-Hernandez}, {Ge}, {Geisler},
  {Gelfand}, {Geron}, {Gibson}, {Goddy}, {Godoy-Rivera}, {Grabowski}, {Green},
  {Greener}, {Grier}, {Griffith}, {Guo}, {Guy}, {Hadjara}, {Harding},
  {Hasselquist}, {Hayes}, {Hearty}, {Hernndez}, {Hill}, {Hogg}, {Holtzman},
  {Horta}, {Hsieh}, {Hsu}, {Hsu}, {Huber}, {Huertas-Company}, {Hutchinson},
  {Hwang}, {Ibarra-Medel}, {Ider Chitham}, {Ilha}, {Imig}, {Jaekle},
  {Jayasinghe}, {Ji}, {Johnson}, {Jones}, {Jonsson}, {Katkov}, {Khalatyan},
  {Kinemuchi}, {Kisku}, {Knapen}, {Kneib}, {Kollmeier}, {Kong}, {Kounkel},
  {Kreckel}, {Krishnarao}, {Lacerna}, {Lane}, {Langgin}, {Lavender}, {Law},
  {Lazarz}, {Leung}, {Leung}, {Lewis}, {Li}, {Li}, {Lian}, {Liang}, {Lin},
  {Lin}, {Lin}, {Lintott}, {Long}, {Longa-Pena}, {Lopez-Coba}, {Lu},
  {Lundgren}, {Luo}, {Mackereth}, {de la Macorra}, {Mahadevan}, {Majewski},
  {Manchado}, {Mandeville}, {Maraston}, {Margalef-Bentabol}, {Masseron},
  {Masters}, {Mathur}, {McDermid}, {Mckay}, {Merloni}, {Merrifield},
  {Meszaros}, {Miglio}, {Di Mille}, {Minniti}, {Minsley}, {Monachesi}, {Moon},
  {Mosser}, {Mulchaey}, {Muna}, {Munoz}, {Myers}, {Myers}, {Nadathur}, {Nair},
  {Nandra}, {Neumann}, {Newman}, {Nidever}, {Nikakhtar}, {Nitschelm},
  {O'Connell}, {Garma-Oehmichen}, {Luan Souza de Oliveira}, {Olney}, {Oravetz},
  {Ortigoza-Urdaneta}, {Osorio}, {Otter}, {Pace}, {Padilla}, {Pan}, {Pan},
  {Parikh}, {Parker}, {Peirani}, {Pena Ramirez}, {Penny}, {Percival},
  {Perez-Fournon}, {Pinsonneault}, {Poidevin}, {Poovelil}, {Price-Whelan},
  {Queiroz}, {Raddick}, {Ray}, {Barboza Rembold}, {Riddle}, {Riffel}, {Riffel},
  {Rix}, {Robin}, {Rodriguez-Puebla}, {Santana Rojas}, {Roman-Lopes},
  {Roman-Zuniga}, {Rose}, {Ross}, {Rossi}, {Rubin}, {Salvato},
  {Sanchez-Gallego}, {Sanderson}, {Sarceno}, {Sarmiento}, {Sayres}, {Sazonova},
  {Schaefer}, {Schlegel}, {Schneider}, {Schultheis}, {Schwope}, {Serenelli},
  {Serna}, {Shao}, {Shapiro}, {Sharma}, {Shen}, {Shetrone}, {Shu}, {Simon},
  {Skrutskie}, {Smethurst}, {Smith}, {Sobeck}, {Spoo}, {Sprague}, {Stark},
  {Stassun}, {Steinmetz}, {Stello}, {Stone-Martinez}, {Storchi-Bergmann},
  {Stringfellow}, {Stutz}, {Su}, {Taghizadeh-Popp}, {Talbot}, {Tayar},
  {Telles}, {Teske}, {Thakar}, {Theissen}, {Thomas}, {Tkachenko}, {Tojeiro},
  {Hernandez Toledo}, {Troup}, {Trump}, {Trussler}, {Turner}, {Tuttle},
  {Unda-Sanzana}, {Vazquez-Mata}, {Valentini}, {Valenzuela}, {Vargas-Gonzalez},
  {Vargas-Magana}, {Alfaro}, {Villanova}, {Vincenzo}, {Wake}, {Warfield},
  {Washington}, {Weaver}, {Weijmans}, {Weinberg}, {Weiss}, {Westfall}, {Wild},
  {Wilde}, {Wilson}, {Wilson}, {Wilson}, {Wolf}, {Wood-Vasey}, {Yan}, {Zamora},
  {Zasowski}, {Zhang}, {Zhao}, {Zheng}, {Zheng}, \& {Zhu}}]{DR17}
{Abdurro'uf}, {Accetta}, K., {Aerts}, C., {et~al.} 2021, arXiv e-prints,
  arXiv:2112.02026

\bibitem[{{Abdurro'uf} {et~al.}(2022){Abdurro'uf}, {Accetta}, {Aerts}, {Silva
  Aguirre}, {Ahumada}, {Ajgaonkar}, {Filiz Ak}, {Alam}, {Allende Prieto},
  {Almeida}, {Anders}, {Anderson}, {Andrews}, {Anguiano}, {Aquino-Ort{\'\i}z},
  {Arag{\'o}n-Salamanca}, {Argudo-Fern{\'a}ndez}, {Ata}, {Aubert},
  {Avila-Reese}, {Badenes}, {Barb{\'a}}, {Barger}, {Barrera-Ballesteros},
  {Beaton}, {Beers}, {Belfiore}, {Bender}, {Bernardi}, {Bershady}, {Beutler},
  {Bidin}, {Bird}, {Bizyaev}, {Blanc}, {Blanton}, {Boardman}, {Bolton},
  {Boquien}, {Borissova}, {Bovy}, {Brandt}, {Brown}, {Brownstein}, {Brusa},
  {Buchner}, {Bundy}, {Burchett}, {Bureau}, {Burgasser}, {Cabang}, {Campbell},
  {Cappellari}, {Carlberg}, {Wanderley}, {Carrera}, {Cash}, {Chen}, {Chen},
  {Cherinka}, {Chiappini}, {Choi}, {Chojnowski}, {Chung}, {Clerc}, {Cohen},
  {Comerford}, {Comparat}, {da Costa}, {Covey}, {Crane}, {Cruz-Gonzalez},
  {Culhane}, {Cunha}, {Dai}, {Damke}, {Darling}, {Davidson}, {Davies},
  {Dawson}, {De Lee}, {Diamond-Stanic}, {Cano-D{\'\i}az}, {S{\'a}nchez},
  {Donor}, {Duckworth}, {Dwelly}, {Eisenstein}, {Elsworth}, {Emsellem},
  {Eracleous}, {Escoffier}, {Fan}, {Farr}, {Feng}, {Fern{\'a}ndez-Trincado},
  {Feuillet}, {Filipp}, {Fillingham}, {Frinchaboy}, {Fromenteau}, {Galbany},
  {Garc{\'\i}a}, {Garc{\'\i}a-Hern{\'a}ndez}, {Ge}, {Geisler}, {Gelfand},
  {G{\'e}ron}, {Gibson}, {Goddy}, {Godoy-Rivera}, {Grabowski}, {Green},
  {Greener}, {Grier}, {Griffith}, {Guo}, {Guy}, {Hadjara}, {Harding},
  {Hasselquist}, {Hayes}, {Hearty}, {Hern{\'a}ndez}, {Hill}, {Hogg},
  {Holtzman}, {Horta}, {Hsieh}, {Hsu}, {Hsu}, {Huber}, {Huertas-Company},
  {Hutchinson}, {Hwang}, {Ibarra-Medel}, {Chitham}, {Ilha}, {Imig}, {Jaekle},
  {Jayasinghe}, {Ji}, {Johnson}, {Jones}, {J{\"o}nsson}, {Katkov}, {Khalatyan},
  {Kinemuchi}, {Kisku}, {Knapen}, {Kneib}, {Kollmeier}, {Kong}, {Kounkel},
  {Kreckel}, {Krishnarao}, {Lacerna}, {Lane}, {Langgin}, {Lavender}, {Law},
  {Lazarz}, {Leung}, {Leung}, {Lewis}, {Li}, {Li}, {Lian}, {Liang}, {Lin},
  {Lin}, {Lin}, {Lintott}, {Long}, {Longa-Pe{\~n}a}, {L{\'o}pez-Cob{\'a}},
  {Lu}, {Lundgren}, {Luo}, {Mackereth}, {de la Macorra}, {Mahadevan},
  {Majewski}, {Manchado}, {Mandeville}, {Maraston}, {Margalef-Bentabol},
  {Masseron}, {Masters}, {Mathur}, {McDermid}, {Mckay}, {Merloni},
  {Merrifield}, {Meszaros}, {Miglio}, {Di Mille}, {Minniti}, {Minsley},
  {Monachesi}, {Moon}, {Mosser}, {Mulchaey}, {Muna}, {Mu{\~n}oz}, {Myers},
  {Myers}, {Nadathur}, {Nair}, {Nandra}, {Neumann}, {Newman}, {Nidever},
  {Nikakhtar}, {Nitschelm}, {O'Connell}, {Garma-Oehmichen}, {Luan Souza de
  Oliveira}, {Olney}, {Oravetz}, {Ortigoza-Urdaneta}, {Osorio}, {Otter},
  {Pace}, {Padilla}, {Pan}, {Pan}, {Parikh}, {Parker}, {Peirani}, {Pe{\~n}a
  Ram{\'\i}rez}, {Penny}, {Percival}, {Perez-Fournon}, {Pinsonneault},
  {Poidevin}, {Poovelil}, {Price-Whelan}, {B{\'a}rbara de Andrade Queiroz},
  {Raddick}, {Ray}, {Rembold}, {Riddle}, {Riffel}, {Riffel}, {Rix}, {Robin},
  {Rodr{\'\i}guez-Puebla}, {Roman-Lopes}, {Rom{\'a}n-Z{\'u}{\~n}iga}, {Rose},
  {Ross}, {Rossi}, {Rubin}, {Salvato}, {S{\'a}nchez}, {S{\'a}nchez-Gallego},
  {Sanderson}, {Santana Rojas}, {Sarceno}, {Sarmiento}, {Sayres}, {Sazonova},
  {Schaefer}, {Schiavon}, {Schlegel}, {Schneider}, {Schultheis}, {Schwope},
  {Serenelli}, {Serna}, {Shao}, {Shapiro}, {Sharma}, {Shen}, {Shetrone}, {Shu},
  {Simon}, {Skrutskie}, {Smethurst}, {Smith}, {Sobeck}, {Spoo}, {Sprague},
  {Stark}, {Stassun}, {Steinmetz}, {Stello}, {Stone-Martinez},
  {Storchi-Bergmann}, {Stringfellow}, {Stutz}, {Su}, {Taghizadeh-Popp},
  {Talbot}, {Tayar}, {Telles}, {Teske}, {Thakar}, {Theissen}, {Tkachenko},
  {Thomas}, {Tojeiro}, {Hernandez Toledo}, {Troup}, {Trump}, {Trussler},
  {Turner}, {Tuttle}, {Unda-Sanzana}, {V{\'a}zquez-Mata}, {Valentini},
  {Valenzuela}, {Vargas-Gonz{\'a}lez}, {Vargas-Maga{\~n}a}, {Alfaro},
  {Villanova}, {Vincenzo}, {Wake}, {Warfield}, {Washington}, {Weaver},
  {Weijmans}, {Weinberg}, {Weiss}, {Westfall}, {Wild}, {Wilde}, {Wilson},
  {Wilson}, {Wilson}, {Wolf}, {Wood-Vasey}, {Yan}, {Zamora}, {Zasowski},
  {Zhang}, {Zhao}, {Zheng}, {Zheng}, \& {Zhu}}]{DR172022}
---. 2022, \apjs, 259, 35

\bibitem[{{Ahumada} {et~al.}(2020){Ahumada}, {Prieto}, {Almeida}, {Anders},
  {Anderson}, {Andrews}, {Anguiano}, {Arcodia}, {Armengaud}, {Aubert}, {Avila},
  {Avila-Reese}, {Badenes}, {Balland}, {Barger}, {Barrera-Ballesteros}, {Basu},
  {Bautista}, {Beaton}, {Beers}, {Benavides}, {Bender}, {Bernardi}, {Bershady},
  {Beutler}, {Bidin}, {Bird}, {Bizyaev}, {Blanc}, {Blanton}, {Boquien},
  {Borissova}, {Bovy}, {Brandt}, {Brinkmann}, {Brownstein}, {Bundy}, {Bureau},
  {Burgasser}, {Burtin}, {Cano-D{\'\i}az}, {Capasso}, {Cappellari}, {Carrera},
  {Chabanier}, {Chaplin}, {Chapman}, {Cherinka}, {Chiappini}, {Doohyun Choi},
  {Chojnowski}, {Chung}, {Clerc}, {Coffey}, {Comerford}, {Comparat}, {da
  Costa}, {Cousinou}, {Covey}, {Crane}, {Cunha}, {Ilha}, {Dai}, {Damsted},
  {Darling}, {Davidson}, {Davies}, {Dawson}, {De}, {de la Macorra}, {De Lee},
  {Queiroz}, {Deconto Machado}, {de la Torre}, {Dell'Agli}, {du Mas des
  Bourboux}, {Diamond-Stanic}, {Dillon}, {Donor}, {Drory}, {Duckworth},
  {Dwelly}, {Ebelke}, {Eftekharzadeh}, {Davis Eigenbrot}, {Elsworth},
  {Eracleous}, {Erfanianfar}, {Escoffier}, {Fan}, {Farr},
  {Fern{\'a}ndez-Trincado}, {Feuillet}, {Finoguenov}, {Fofie},
  {Fraser-McKelvie}, {Frinchaboy}, {Fromenteau}, {Fu}, {Galbany}, {Garcia},
  {Garc{\'\i}a-Hern{\'a}ndez}, {Oehmichen}, {Ge}, {Maia}, {Geisler}, {Gelfand},
  {Goddy}, {Gonzalez-Perez}, {Grabowski}, {Green}, {Grier}, {Guo}, {Guy},
  {Harding}, {Hasselquist}, {Hawken}, {Hayes}, {Hearty}, {Hekker}, {Hogg},
  {Holtzman}, {Horta}, {Hou}, {Hsieh}, {Huber}, {Hunt}, {Chitham}, {Imig},
  {Jaber}, {Angel}, {Johnson}, {Jones}, {J{\"o}nsson}, {Jullo}, {Kim},
  {Kinemuchi}, {Kirkpatrick}, {Kite}, {Klaene}, {Kneib}, {Kollmeier}, {Kong},
  {Kounkel}, {Krishnarao}, {Lacerna}, {Lan}, {Lane}, {Law}, {Le Goff}, {Leung},
  {Lewis}, {Li}, {Lian}, {Lin}, {Long}, {Longa-Pe{\~n}a}, {Lundgren}, {Lyke},
  {Ted Mackereth}, {MacLeod}, {Majewski}, {Manchado}, {Maraston}, {Martini},
  {Masseron}, {Masters}, {Mathur}, {McDermid}, {Merloni}, {Merrifield},
  {M{\'e}sz{\'a}ros}, {Miglio}, {Minniti}, {Minsley}, {Miyaji}, {Mohammad},
  {Mosser}, {Mueller}, {Muna}, {Mu{\~n}oz-Guti{\'e}rrez}, {Myers}, {Nadathur},
  {Nair}, {Nandra}, {do Nascimento}, {Nevin}, {Newman}, {Nidever}, {Nitschelm},
  {Noterdaeme}, {O'Connell}, {Olmstead}, {Oravetz}, {Oravetz}, {Osorio},
  {Pace}, {Padilla}, {Palanque-Delabrouille}, {Palicio}, {Pan}, {Pan},
  {Parker}, {Paviot}, {Peirani}, {Ram{\'r}ez}, {Penny}, {Percival},
  {Perez-Fournon}, {P{\'e}rez-R{\`a}fols}, {Petitjean}, {Pieri},
  {Pinsonneault}, {Poovelil}, {Povick}, {Prakash}, {Price-Whelan}, {Raddick},
  {Raichoor}, {Ray}, {Rembold}, {Rezaie}, {Riffel}, {Riffel}, {Rix}, {Robin},
  {Roman-Lopes}, {Rom{\'a}n-Z{\'u}{\~n}iga}, {Rose}, {Ross}, {Rossi},
  {Rowlands}, {Rubin}, {Salvato}, {S{\'a}nchez}, {S{\'a}nchez-Menguiano},
  {S{\'a}nchez-Gallego}, {Sayres}, {Schaefer}, {Schiavon}, {Schimoia},
  {Schlafly}, {Schlegel}, {Schneider}, {Schultheis}, {Schwope}, {Seo},
  {Serenelli}, {Shafieloo}, {Shamsi}, {Shao}, {Shen}, {Shetrone}, {Shirley},
  {Aguirre}, {Simon}, {Skrutskie}, {Slosar}, {Smethurst}, {Sobeck}, {Sodi},
  {Souto}, {Stark}, {Stassun}, {Steinmetz}, {Stello}, {Stermer},
  {Storchi-Bergmann}, {Streblyanska}, {Stringfellow}, {Stutz}, {Su{\'a}rez},
  {Sun}, {Taghizadeh-Popp}, {Talbot}, {Tayar}, {Thakar}, {Theriault}, {Thomas},
  {Thomas}, {Tinker}, {Tojeiro}, {Toledo}, {Tremonti}, {Troup}, {Tuttle},
  {Unda-Sanzana}, {Valentini}, {Vargas-Gonz{\'a}lez}, {Vargas-Maga{\~n}a},
  {V{\'a}zquez-Mata}, {Vivek}, {Wake}, {Wang}, {Weaver}, {Weijmans}, {Wild},
  {Wilson}, {Wilson}, {Wolthuis}, {Wood-Vasey}, {Yan}, {Yang}, {Y{\`e}che},
  {Zamora}, {Zarrouk}, {Zasowski}, {Zhang}, {Zhao}, {Zhao}, {Zheng}, {Zheng},
  {Zhu}, \& {Zou}}]{DR162020}
{Ahumada}, R., {Prieto}, C.~A., {Almeida}, A., {et~al.} 2020, \apjs, 249, 3

\bibitem[{{Anderson}(1976)}]{Anderson76}
{Anderson}, G.~M. 1976, \gca, 40, 1533

\bibitem[{{Astropy Collaboration} {et~al.}(2013){Astropy Collaboration},
  {Robitaille}, {Tollerud}, {Greenfield}, {Droettboom}, {Bray}, {Aldcroft},
  {Davis}, {Ginsburg}, {Price-Whelan}, {Kerzendorf}, {Conley}, {Crighton},
  {Barbary}, {Muna}, {Ferguson}, {Grollier}, {Parikh}, {Nair}, {Unther},
  {Deil}, {Woillez}, {Conseil}, {Kramer}, {Turner}, {Singer}, {Fox}, {Weaver},
  {Zabalza}, {Edwards}, {Azalee Bostroem}, {Burke}, {Casey}, {Crawford},
  {Dencheva}, {Ely}, {Jenness}, {Labrie}, {Lim}, {Pierfederici}, {Pontzen},
  {Ptak}, {Refsdal}, {Servillat}, \& {Streicher}}]{Astropy13}
{Astropy Collaboration}, {Robitaille}, T.~P., {Tollerud}, E.~J., {et~al.} 2013,
  \aap, 558, A33

\bibitem[{{Astropy Collaboration} {et~al.}(2018){Astropy Collaboration},
  {Price-Whelan}, {Sip{\H{o}}cz}, {G{\"u}nther}, {Lim}, {Crawford}, {Conseil},
  {Shupe}, {Craig}, {Dencheva}, {Ginsburg}, {VanderPlas}, {Bradley},
  {P{\'e}rez-Su{\'a}rez}, {de Val-Borro}, {Aldcroft}, {Cruz}, {Robitaille},
  {Tollerud}, {Ardelean}, {Babej}, {Bach}, {Bachetti}, {Bakanov}, {Bamford},
  {Barentsen}, {Barmby}, {Baumbach}, {Berry}, {Biscani}, {Boquien}, {Bostroem},
  {Bouma}, {Brammer}, {Bray}, {Breytenbach}, {Buddelmeijer}, {Burke},
  {Calderone}, {Cano Rodr{\'\i}guez}, {Cara}, {Cardoso}, {Cheedella}, {Copin},
  {Corrales}, {Crichton}, {D'Avella}, {Deil}, {Depagne}, {Dietrich}, {Donath},
  {Droettboom}, {Earl}, {Erben}, {Fabbro}, {Ferreira}, {Finethy}, {Fox},
  {Garrison}, {Gibbons}, {Goldstein}, {Gommers}, {Greco}, {Greenfield},
  {Groener}, {Grollier}, {Hagen}, {Hirst}, {Homeier}, {Horton}, {Hosseinzadeh},
  {Hu}, {Hunkeler}, {Ivezi{\'c}}, {Jain}, {Jenness}, {Kanarek}, {Kendrew},
  {Kern}, {Kerzendorf}, {Khvalko}, {King}, {Kirkby}, {Kulkarni}, {Kumar},
  {Lee}, {Lenz}, {Littlefair}, {Ma}, {Macleod}, {Mastropietro}, {McCully},
  {Montagnac}, {Morris}, {Mueller}, {Mumford}, {Muna}, {Murphy}, {Nelson},
  {Nguyen}, {Ninan}, {N{\"o}the}, {Ogaz}, {Oh}, {Parejko}, {Parley}, {Pascual},
  {Patil}, {Patil}, {Plunkett}, {Prochaska}, {Rastogi}, {Reddy Janga},
  {Sabater}, {Sakurikar}, {Seifert}, {Sherbert}, {Sherwood-Taylor}, {Shih},
  {Sick}, {Silbiger}, {Singanamalla}, {Singer}, {Sladen}, {Sooley},
  {Sornarajah}, {Streicher}, {Teuben}, {Thomas}, {Tremblay}, {Turner},
  {Terr{\'o}n}, {van Kerkwijk}, {de la Vega}, {Watkins}, {Weaver}, {Whitmore},
  {Woillez}, {Zabalza}, \& {Astropy Contributors}}]{Astropy18}
{Astropy Collaboration}, {Price-Whelan}, A.~M., {Sip{\H{o}}cz}, B.~M., {et~al.}
  2018, \aj, 156, 123

\bibitem[{{Avedisova}(2002)}]{Avedisova02}
{Avedisova}, V.~S. 2002, Astronomy Reports, 46, 193

\bibitem[{{Bailer-Jones} {et~al.}(2021){Bailer-Jones}, {Rybizki}, {Fouesneau},
  {Demleitner}, \& {Andrae}}]{BJEDR3}
{Bailer-Jones}, C.~A.~L., {Rybizki}, J., {Fouesneau}, M., {Demleitner}, M., \&
  {Andrae}, R. 2021, \aj, 161, 147

\bibitem[{{Baraffe} {et~al.}(2015){Baraffe}, {Homeier}, {Allard}, \&
  {Chabrier}}]{Baraffe15}
{Baraffe}, I., {Homeier}, D., {Allard}, F., \& {Chabrier}, G. 2015, \aap, 577,
  A42

\bibitem[{{Baratella} {et~al.}(2020){Baratella}, {D'Orazi}, {Carraro},
  {Desidera}, {Randich}, {Magrini}, {Adibekyan}, {Smiljanic}, {Spina},
  {Tsantaki}, {Tautvai{\v{s}}ien{\.{e}}}, {Sousa}, {Jofr{\'e}},
  {Jim{\'e}nez-Esteban}, {Delgado-Mena}, {Martell}, {Van der Swaelmen},
  {Roccatagliata}, {Gilmore}, {Alfaro}, {Bayo}, {Bensby}, {Bragaglia},
  {Franciosini}, {Gonneau}, {Heiter}, {Hourihane}, {Jeffries}, {Koposov},
  {Morbidelli}, {Prisinzano}, {Sacco}, {Sbordone}, {Worley}, {Zaggia}, \&
  {Lewis}}]{Baratella20}
{Baratella}, M., {D'Orazi}, V., {Carraro}, G., {et~al.} 2020, \aap, 634, A34

\bibitem[{{Baratella} {et~al.}(2021){Baratella}, {D'Orazi}, {Sheminova},
  {Spina}, {Carraro}, {Gratton}, {Magrini}, {Randich}, {Lugaro}, {Pignatari},
  {Romano}, {Biazzo}, {Bragaglia}, {Casali}, {Desidera}, {Frasca}, {de Silva},
  {Melo}, {Van der Swaelmen}, {Tautvai{\v{s}}ien{\.{e}}},
  {Jim{\'e}nez-Esteban}, {Gilmore}, {Bensby}, {Smiljanic}, {Bayo},
  {Franciosini}, {Gonneau}, {Hourihane}, {Jofr{\'e}}, {Monaco}, {Morbidelli},
  {Sacco}, {Sbordone}, {Worley}, \& {Zaggia}}]{Baratella21}
{Baratella}, M., {D'Orazi}, V., {Sheminova}, V., {et~al.} 2021, arXiv e-prints,
  arXiv:2107.12381

\bibitem[{{Basri} \& {Mart{\'\i}n}(1999)}]{Basri99}
{Basri}, G., \& {Mart{\'\i}n}, E.~L. 1999, \apj, 510, 266

\bibitem[{{Beaton} {et~al.}(2021){Beaton}, {Oelkers}, {Hayes}, {Covey},
  {Chojnowski}, {De Lee}, {Sobeck}, {Majewski}, {Cohen}, {Fernandez-Trincado},
  {Longa-Pena}, {O'Connell}, {Santana}, {Stringfellow}, {Zasowski}, {Aerts},
  {Anguiano}, {Bender}, {Canas}, {Cunha}, {Fleming}, {Frinchaboy}, {Feuillet},
  {Harding}, {Hasselquist}, {Holtzman}, {Johnson}, {Kollmeier}, {Kounkel},
  {Mahadevan}, {Price-Whelan}, {Rojas-Arriagada}, {Roman-Zuniga}, {Schlafly},
  {Schultheis}, {Shetrone}, {Simon}, {Stassun}, {Stutz}, {Tayar}, {Teske},
  {Tkachenko}, {Troup}, {Albareti}, {Bizyaev}, {Bovy}, {Burgasser}, {Comparat},
  {Downes}, {Geisler}, {Inno}, {Manchado}, {Ness}, {Pinsonneault}, {Prada},
  {Roman-Lopes}, {Simonian}, {Smith}, {Yan}, \& {Zamora}}]{Beaton21N}
{Beaton}, R.~L., {Oelkers}, R.~J., {Hayes}, C.~R., {et~al.} 2021, arXiv
  e-prints, arXiv:2108.11907

\bibitem[{{Beccari} {et~al.}(2017){Beccari}, {Petr-Gotzens}, {Boffin},
  {Romaniello}, {Fedele}, {Carraro}, {De Marchi}, {de Wit}, {Drew}, {Kalari},
  {Manara}, {Martin}, {Mieske}, {Panagia}, {Testi}, {Vink}, {Walsh}, \&
  {Wright}}]{Beccari17}
{Beccari}, G., {Petr-Gotzens}, M.~G., {Boffin}, H.~M.~J., {et~al.} 2017, \aap,
  604, A22

\bibitem[{{Bica} {et~al.}(2003){Bica}, {Dutra}, \& {Barbuy}}]{Bica2003}
{Bica}, E., {Dutra}, C.~M., \& {Barbuy}, B. 2003, \aap, 397, 177

\bibitem[{{Binks} {et~al.}(2021){Binks}, {Jeffries}, {Jackson}, {Franciosini},
  {Sacco}, {Bayo}, {Magrini}, {Randich}, {Arancibia-Silva}, {Bergemann},
  {Bragaglia}, {Gilmore}, {Gonneau}, {Hourihane}, {Jofr{\'e}}, {Korn},
  {Morbidelli}, {Prisinzano}, {Worley}, \& {Zaggia}}]{Binks21}
{Binks}, A.~S., {Jeffries}, R.~D., {Jackson}, R.~J., {et~al.} 2021, \mnras,
  505, 1280

\bibitem[{{Blanton} {et~al.}(2017){Blanton}, {Bershady}, {Abolfathi},
  {Albareti}, {Allende Prieto}, {Almeida}, {Alonso-Garc{\'{\i}}a}, {Anders},
  {Anderson}, {Andrews}, \& et~al.}]{Blanton17}
{Blanton}, M.~R., {Bershady}, M.~A., {Abolfathi}, B., {et~al.} 2017, \aj, 154,
  28

\bibitem[{{Bonito} {et~al.}(2020){Bonito}, {Prisinzano}, {Venuti}, {Damiani},
  {Micela}, {Sacco}, {Traven}, {Biazzo}, {Sbordone}, {Masseron}, {Zwitter},
  {Gonneau}, {Bayo}, {Roccatagliata}, {Randich}, {Vink}, {Jofre}, {Flaccomio},
  {Magrini}, {Carraro}, {Morbidelli}, {Frasca}, {Monaco}, {Rigliaco}, {Worley},
  {Hourihane}, {Gilmore}, {Franciosini}, {Lewis}, \& {Koposov}}]{Bonito20}
{Bonito}, R., {Prisinzano}, L., {Venuti}, L., {et~al.} 2020, \aap, 642, A56

\bibitem[{{Borissova} {et~al.}(2019){Borissova}, {Roman-Lopes}, {Covey},
  {Medina}, {Kurtev}, {Roman-Zuniga}, {Kuhn}, {Contreras Pe{\~n}a}, {Lucas},
  {Ramirez Alegria}, {Minniti}, {Kounkel}, {Stringfellow}, {Barb{\'a}}, \&
  {Su{\'a}rez}}]{Borissova19}
{Borissova}, J., {Roman-Lopes}, A., {Covey}, K., {et~al.} 2019, \aj, 158, 46

\bibitem[{Bowen \& Vaughan(1973)}]{Bowen1973}
Bowen, I.~S., \& Vaughan, A.~H. 1973, Appl. Opt., 12, 1430

\bibitem[{{Bravi} {et~al.}(2018){Bravi}, {Zari}, {Sacco}, {Randich},
  {Jeffries}, {Jackson}, {Franciosini}, {Moraux}, {L{\'o}pez-Santiago},
  {Pancino}, {Spina}, {Wright}, {Jim{\'e}nez-Esteban}, {Klutsch},
  {Roccatagliata}, {Gilmore}, {Bragaglia}, {Flaccomio}, {Francois}, {Koposov},
  {Bayo}, {Carraro}, {Costado}, {Damiani}, {Frasca}, {Hourihane}, {Jofr{\'e}},
  {Lardo}, {Lewis}, {Magrini}, {Morbidelli}, {Prisinzano}, {Sousa}, {Worley},
  \& {Zaggia}}]{Bravi18}
{Bravi}, L., {Zari}, E., {Sacco}, G.~G., {et~al.} 2018, \aap, 615, A37

\bibitem[{{Bressan} {et~al.}(2012){Bressan}, {Marigo}, {Girardi}, {Salasnich},
  {Dal Cero}, {Rubele}, \& {Nanni}}]{Bressan12}
{Bressan}, A., {Marigo}, P., {Girardi}, L., {et~al.} 2012, \mnras, 427, 127

\bibitem[{{Brice{\~n}o} {et~al.}(2019){Brice{\~n}o}, {Calvet}, {Hern{\'a}ndez},
  {Vivas}, {Mateu}, {Downes}, {Loerincs}, {P{\'e}rez-Blanco}, {Berlind},
  {Espaillat}, {Allen}, {Hartmann}, {Mateo}, \& {Bailey}}]{Briceno19}
{Brice{\~n}o}, C., {Calvet}, N., {Hern{\'a}ndez}, J., {et~al.} 2019, \aj, 157,
  85

\bibitem[{{Caiazzo} {et~al.}(2020){Caiazzo}, {Heyl}, {Richer}, {Cummings},
  {Fleury}, {Hegarty}, {Kalirai}, {Kerr}, {Thiele}, {Tremblay}, \&
  {Villanueva}}]{Caiazzo20}
{Caiazzo}, I., {Heyl}, J., {Richer}, H., {et~al.} 2020, \apjl, 901, L14

\bibitem[{{Cambr{\'e}sy} {et~al.}(2013){Cambr{\'e}sy}, {Marton}, {Feher},
  {T{\'o}th}, \& {Schneider}}]{Cambresy13}
{Cambr{\'e}sy}, L., {Marton}, G., {Feher}, O., {T{\'o}th}, L.~V., \&
  {Schneider}, N. 2013, \aap, 557, A29

\bibitem[{Campello {et~al.}(2013)Campello, Moulavi, \& Sander}]{hdbscan_ref}
Campello, R. J. G.~B., Moulavi, D., \& Sander, J. 2013, in Advances in
  Knowledge Discovery and Data Mining, ed. J.~Pei, V.~S. Tseng, L.~Cao,
  H.~Motoda, \& G.~Xu (Berlin, Heidelberg: Springer Berlin Heidelberg),
  160--172

\bibitem[{{Cardelli} {et~al.}(1989){Cardelli}, {Clayton}, \& {Mathis}}]{CCM89}
{Cardelli}, J.~A., {Clayton}, G.~C., \& {Mathis}, J.~S. 1989, \apj, 345, 245

\bibitem[{{Choi} {et~al.}(2016){Choi}, {Dotter}, {Conroy}, {Cantiello},
  {Paxton}, \& {Johnson}}]{Choi16}
{Choi}, J., {Dotter}, A., {Conroy}, C., {et~al.} 2016, \apj, 823, 102

\bibitem[{{Cottaar} {et~al.}(2014){Cottaar}, {Covey}, {Meyer}, {Nidever},
  {Stassun}, {Foster}, {Tan}, {Chojnowski}, {da Rio}, {Flaherty}, {Frinchaboy},
  {Skrutskie}, {Majewski}, {Wilson}, \& {Zasowski}}]{Cotaar14}
{Cottaar}, M., {Covey}, K.~R., {Meyer}, M.~R., {et~al.} 2014, \apj, 794, 125

\bibitem[{{Cottaar} {et~al.}(2015){Cottaar}, {Covey}, {Foster}, {Meyer}, {Tan},
  {Nidever}, {Chojnowski}, {da Rio}, {Flaherty}, {Frinchaboy}, {Majewski},
  {Skrutskie}, {Wilson}, \& {Zasowski}}]{Cotaar15}
{Cottaar}, M., {Covey}, K.~R., {Foster}, J.~B., {et~al.} 2015, \apj, 807, 27

\bibitem[{{Cottle} {et~al.}(2018){Cottle}, {Covey}, {Su{\'a}rez},
  {Rom{\'a}n-Z{\'u}{\~n}iga}, {Schlafly}, {Downes}, {Ybarra}, {Hernandez},
  {Stassun}, {Stringfellow}, {Getman}, {Feigelson}, {Borissova}, {Kim},
  {Roman-Lopes}, {Da Rio}, {De Lee}, {Frinchaboy}, {Kounkel}, {Majewski},
  {Mennickent}, {Nidever}, {Nitschelm}, {Pan}, {Shetrone}, {Zasowski},
  {Chambers}, {Magnier}, \& {Valenti}}]{Cottle18}
{Cottle}, J.~N., {Covey}, K.~R., {Su{\'a}rez}, G., {et~al.} 2018, \apjs, 236,
  27

\bibitem[{{Da Rio} {et~al.}(2016){Da Rio}, {Tan}, {Covey}, {Cottaar}, {Foster},
  {Cullen}, {Tobin}, {Kim}, {Meyer}, {Nidever}, {Stassun}, {Chojnowski},
  {Flaherty}, {Majewski}, {Skrutskie}, {Zasowski}, \& {Pan}}]{daRio16}
{Da Rio}, N., {Tan}, J.~C., {Covey}, K.~R., {et~al.} 2016, \apj, 818, 59

\bibitem[{{Da Rio} {et~al.}(2017){Da Rio}, {Tan}, {Covey}, {Cottaar}, {Foster},
  {Cullen}, {Tobin}, {Kim}, {Meyer}, {Nidever}, {Stassun}, {Chojnowski},
  {Flaherty}, {Majewski}, {Skrutskie}, {Zasowski}, \& {Pan}}]{daRio17}
---. 2017, \apj, 845, 105

\bibitem[{{Dolan} \& {Mathieu}(2001)}]{Dolan01}
{Dolan}, C.~J., \& {Mathieu}, R.~D. 2001, \aj, 121, 2124

\bibitem[{{Dolan} \& {Mathieu}(2002)}]{Dolan02}
---. 2002, \aj, 123, 387

\bibitem[{{Dotter}(2016)}]{Dotter16}
{Dotter}, A. 2016, \apjs, 222, 8

\bibitem[{{Eisenstein} {et~al.}(2011){Eisenstein}, {Weinberg}, {Agol},
  {Aihara}, {Allende Prieto}, {Anderson}, {Arns}, {Aubourg}, {Bailey},
  {Balbinot}, {Barkhouser}, {Beers}, {Berlind}, {Bickerton}, {Bizyaev},
  {Blanton}, {Bochanski}, {Bolton}, {Bosman}, {Bovy}, {Brandt}, {Breslauer},
  {Brewington}, {Brinkmann}, {Brown}, {Brownstein}, {Burger}, {Busca},
  {Campbell}, {Cargile}, {Carithers}, {Carlberg}, {Carr}, {Chang}, {Chen},
  {Chiappini}, {Comparat}, {Connolly}, {Cortes}, {Croft}, {Cunha}, {da Costa},
  {Davenport}, {Dawson}, {De Lee}, {Porto de Mello}, {de Simoni}, {Dean},
  {Dhital}, {Ealet}, {Ebelke}, {Edmondson}, {Eiting}, {Escoffier}, {Esposito},
  {Evans}, {Fan}, {Femen{\'\i}a Castell{\'a}}, {Dutra Ferreira}, {Fitzgerald},
  {Fleming}, {Font-Ribera}, {Ford}, {Frinchaboy}, {Garc{\'\i}a P{\'e}rez},
  {Gaudi}, {Ge}, {Ghezzi}, {Gillespie}, {Gilmore}, {Girardi}, {Gott}, {Gould},
  {Grebel}, {Gunn}, {Hamilton}, {Harding}, {Harris}, {Hawley}, {Hearty},
  {Hennawi}, {Gonz{\'a}lez Hern{\'a}ndez}, {Ho}, {Hogg}, {Holtzman},
  {Honscheid}, {Inada}, {Ivans}, {Jiang}, {Jiang}, {Johnson}, {Jordan},
  {Jordan}, {Kauffmann}, {Kazin}, {Kirkby}, {Klaene}, {Knapp}, {Kneib},
  {Kochanek}, {Koesterke}, {Kollmeier}, {Kron}, {Lampeitl}, {Lang}, {Lawler},
  {Le Goff}, {Lee}, {Lee}, {Leisenring}, {Lin}, {Liu}, {Long}, {Loomis},
  {Lucatello}, {Lundgren}, {Lupton}, {Ma}, {Ma}, {MacDonald}, {Mack},
  {Mahadevan}, {Maia}, {Majewski}, {Makler}, {Malanushenko}, {Malanushenko},
  {Mandelbaum}, {Maraston}, {Margala}, {Maseman}, {Masters}, {McBride},
  {McDonald}, {McGreer}, {McMahon}, {Mena Requejo}, {M{\'e}nard},
  {Miralda-Escud{\'e}}, {Morrison}, {Mullally}, {Muna}, {Murayama}, {Myers},
  {Naugle}, {Neto}, {Nguyen}, {Nichol}, {Nidever}, {O'Connell}, {Ogando},
  {Olmstead}, {Oravetz}, {Padmanabhan}, {Paegert}, {Palanque-Delabrouille},
  {Pan}, {Pandey}, {Parejko}, {P{\^a}ris}, {Pellegrini}, {Pepper}, {Percival},
  {Petitjean}, {Pfaffenberger}, {Pforr}, {Phleps}, {Pichon}, {Pieri}, {Prada},
  {Price-Whelan}, {Raddick}, {Ramos}, {Reid}, {Reyle}, {Rich}, {Richards},
  {Rieke}, {Rieke}, {Rix}, {Robin}, {Rocha-Pinto}, {Rockosi}, {Roe},
  {Rollinde}, {Ross}, {Ross}, {Rossetto}, {S{\'a}nchez}, {Santiago}, {Sayres},
  {Schiavon}, {Schlegel}, {Schlesinger}, {Schmidt}, {Schneider}, {Sellgren},
  {Shelden}, {Sheldon}, {Shetrone}, {Shu}, {Silverman}, {Simmerer}, {Simmons},
  {Sivarani}, {Skrutskie}, {Slosar}, {Smee}, {Smith}, {Snedden}, {Stassun},
  {Steele}, {Steinmetz}, {Stockett}, {Stollberg}, {Strauss}, {Szalay},
  {Tanaka}, {Thakar}, {Thomas}, {Tinker}, {Tofflemire}, {Tojeiro}, {Tremonti},
  {Vargas Maga{\~n}a}, {Verde}, {Vogt}, {Wake}, {Wan}, {Wang}, {Weaver},
  {White}, {White}, {Wilson}, {Wisniewski}, {Wood-Vasey}, {Yanny}, {Yasuda},
  {Y{\`e}che}, {York}, {Young}, {Zasowski}, {Zehavi}, \&
  {Zhao}}]{Eisenstein2011}
{Eisenstein}, D.~J., {Weinberg}, D.~H., {Agol}, E., {et~al.} 2011, \aj, 142, 72

\bibitem[{{Esplin} {et~al.}(2018){Esplin}, {Luhman}, {Miller}, \&
  {Mamajek}}]{Esplin18}
{Esplin}, T.~L., {Luhman}, K.~L., {Miller}, E.~B., \& {Mamajek}, E.~E. 2018,
  \aj, 156, 75

\bibitem[{{Fang} {et~al.}(2020){Fang}, {Bidin}, {Zhao}, {Zhang}, \& {Bharat
  Kumar}}]{FangX20}
{Fang}, X.-S., {Bidin}, C.~M., {Zhao}, G., {Zhang}, L.-Y., \& {Bharat Kumar},
  Y. 2020, \mnras, 495, 2949

\bibitem[{{Fang} {et~al.}(2018){Fang}, {Zhao}, {Zhao}, \& {Bharat
  Kumar}}]{FangX18}
{Fang}, X.-S., {Zhao}, G., {Zhao}, J.-K., \& {Bharat Kumar}, Y. 2018, \mnras,
  476, 908

\bibitem[{{Fang} {et~al.}(2016){Fang}, {Zhao}, {Zhao}, {Chen}, \& {Bharat
  Kumar}}]{FangX16}
{Fang}, X.-S., {Zhao}, G., {Zhao}, J.-K., {Chen}, Y.-Q., \& {Bharat Kumar}, Y.
  2016, \mnras, 463, 2494

\bibitem[{{Fernandez} {et~al.}(2017){Fernandez}, {Covey}, {De Lee},
  {Chojnowski}, {Nidever}, {Ballantyne}, {Cottaar}, {Da Rio}, {Foster},
  {Majewski}, {Meyer}, {Reyna}, {Roberts}, {Skinner}, {Stassun}, {Tan},
  {Troup}, \& {Zasowski}}]{Fernandez17}
{Fernandez}, M.~A., {Covey}, K.~R., {De Lee}, N., {et~al.} 2017, \pasp, 129,
  084201

\bibitem[{{Foster} {et~al.}(2015){Foster}, {Cottaar}, {Covey}, {Arce}, {Meyer},
  {Nidever}, {Stassun}, {Tan}, {Chojnowski}, {da Rio}, {Flaherty}, {Rebull},
  {Frinchaboy}, {Majewski}, {Skrutskie}, {Wilson}, \& {Zasowski}}]{Foster15}
{Foster}, J.~B., {Cottaar}, M., {Covey}, K.~R., {et~al.} 2015, \apj, 799, 136

\bibitem[{{Gaia Collaboration} {et~al.}(2016){Gaia Collaboration}, {Prusti},
  {de Bruijne}, {Brown}, {Vallenari}, {Babusiaux}, {Bailer-Jones}, {Bastian},
  {Biermann}, {Evans}, {Eyer}, {Jansen}, {Jordi}, {Klioner}, {Lammers},
  {Lindegren}, {Luri}, {Mignard}, {Milligan}, {Panem}, {Poinsignon},
  {Pourbaix}, {Randich}, {Sarri}, {Sartoretti}, {Siddiqui}, {Soubiran},
  {Valette}, {van Leeuwen}, {Walton}, {Aerts}, {Arenou}, {Cropper}, {Drimmel},
  {H{\o}g}, {Katz}, {Lattanzi}, {O'Mullane}, {Grebel}, {Holland}, {Huc},
  {Passot}, {Bramante}, {Cacciari}, {Casta{\~n}eda}, {Chaoul}, {Cheek}, {De
  Angeli}, {Fabricius}, {Guerra}, {Hern{\'a}ndez}, {Jean-Antoine-Piccolo},
  {Masana}, {Messineo}, {Mowlavi}, {Nienartowicz}, {Ord{\'o}{\~n}ez-Blanco},
  {Panuzzo}, {Portell}, {Richards}, {Riello}, {Seabroke}, {Tanga},
  {Th{\'e}venin}, {Torra}, {Els}, {Gracia-Abril}, {Comoretto},
  {Garcia-Reinaldos}, {Lock}, {Mercier}, {Altmann}, {Andrae}, {Astraatmadja},
  {Bellas-Velidis}, {Benson}, {Berthier}, {Blomme}, {Busso}, {Carry},
  {Cellino}, {Clementini}, {Cowell}, {Creevey}, {Cuypers}, {Davidson}, {De
  Ridder}, {de Torres}, {Delchambre}, {Dell'Oro}, {Ducourant}, {Fr{\'e}mat},
  {Garc{\'\i}a-Torres}, {Gosset}, {Halbwachs}, {Hambly}, {Harrison}, {Hauser},
  {Hestroffer}, {Hodgkin}, {Huckle}, {Hutton}, {Jasniewicz}, {Jordan},
  {Kontizas}, {Korn}, {Lanzafame}, {Manteiga}, {Moitinho}, {Muinonen},
  {Osinde}, {Pancino}, {Pauwels}, {Petit}, {Recio-Blanco}, {Robin}, {Sarro},
  {Siopis}, {Smith}, {Smith}, {Sozzetti}, {Thuillot}, {van Reeven}, {Viala},
  {Abbas}, {Abreu Aramburu}, {Accart}, {Aguado}, {Allan}, {Allasia},
  {Altavilla}, {{\'A}lvarez}, {Alves}, {Anderson}, {Andrei}, {Anglada Varela},
  {Antiche}, {Antoja}, {Ant{\'o}n}, {Arcay}, {Atzei}, {Ayache}, {Bach},
  {Baker}, {Balaguer-N{\'u}{\~n}ez}, {Barache}, {Barata}, {Barbier}, {Barblan},
  {Baroni}, {Barrado y Navascu{\'e}s}, {Barros}, {Barstow}, {Becciani},
  {Bellazzini}, {Bellei}, {Bello Garc{\'\i}a}, {Belokurov}, {Bendjoya},
  {Berihuete}, {Bianchi}, {Bienaym{\'e}}, {Billebaud}, {Blagorodnova},
  {Blanco-Cuaresma}, {Boch}, {Bombrun}, {Borrachero}, {Bouquillon}, {Bourda},
  {Bouy}, {Bragaglia}, {Breddels}, {Brouillet}, {Br{\"u}semeister},
  {Bucciarelli}, {Budnik}, {Burgess}, {Burgon}, {Burlacu}, {Busonero}, {Buzzi},
  {Caffau}, {Cambras}, {Campbell}, {Cancelliere}, {Cantat-Gaudin}, {Carlucci},
  {Carrasco}, {Castellani}, {Charlot}, {Charnas}, {Charvet}, {Chassat},
  {Chiavassa}, {Clotet}, {Cocozza}, {Collins}, {Collins}, {Costigan}, {Crifo},
  {Cross}, {Crosta}, {Crowley}, {Dafonte}, {Damerdji}, {Dapergolas}, {David},
  {David}, {De Cat}, {de Felice}, {de Laverny}, {De Luise}, {De March}, {de
  Martino}, {de Souza}, {Debosscher}, {del Pozo}, {Delbo}, {Delgado},
  {Delgado}, {di Marco}, {Di Matteo}, {Diakite}, {Distefano}, {Dolding}, {Dos
  Anjos}, {Drazinos}, {Dur{\'a}n}, {Dzigan}, {Ecale}, {Edvardsson}, {Enke},
  {Erdmann}, {Escolar}, {Espina}, {Evans}, {Eynard Bontemps}, {Fabre},
  {Fabrizio}, {Faigler}, {Falc{\~a}o}, {Farr{\`a}s Casas}, {Faye}, {Federici},
  {Fedorets}, {Fern{\'a}ndez-Hern{\'a}ndez}, {Fernique}, {Fienga}, {Figueras},
  {Filippi}, {Findeisen}, {Fonti}, {Fouesneau}, {Fraile}, {Fraser}, {Fuchs},
  {Furnell}, {Gai}, {Galleti}, {Galluccio}, {Garabato}, {Garc{\'\i}a-Sedano},
  {Gar{\'e}}, {Garofalo}, {Garralda}, {Gavras}, {Gerssen}, {Geyer}, {Gilmore},
  {Girona}, {Giuffrida}, {Gomes}, {Gonz{\'a}lez-Marcos},
  {Gonz{\'a}lez-N{\'u}{\~n}ez}, {Gonz{\'a}lez-Vidal}, {Granvik}, {Guerrier},
  {Guillout}, {Guiraud}, {G{\'u}rpide}, {Guti{\'e}rrez-S{\'a}nchez}, {Guy},
  {Haigron}, {Hatzidimitriou}, {Haywood}, {Heiter}, {Helmi}, {Hobbs},
  {Hofmann}, {Holl}, {Holland}, {Hunt}, {Hypki}, {Icardi}, {Irwin}, {Jevardat
  de Fombelle}, {Jofr{\'e}}, {Jonker}, {Jorissen}, {Julbe}, {Karampelas},
  {Kochoska}, {Kohley}, {Kolenberg}, {Kontizas}, {Koposov}, {Kordopatis},
  {Koubsky}, {Kowalczyk}, {Krone-Martins}, {Kudryashova}, {Kull}, {Bachchan},
  {Lacoste-Seris}, {Lanza}, {Lavigne}, {Le Poncin-Lafitte}, {Lebreton},
  {Lebzelter}, {Leccia}, {Leclerc}, {Lecoeur-Taibi}, {Lemaitre}, {Lenhardt},
  {Leroux}, {Liao}, {Licata}, {Lindstr{\o}m}, {Lister}, {Livanou}, {Lobel},
  {L{\"o}ffler}, {L{\'o}pez}, {Lopez-Lozano}, {Lorenz}, {Loureiro},
  {MacDonald}, {Magalh{\~a}es Fernandes}, {Managau}, {Mann}, {Mantelet},
  {Marchal}, {Marchant}, {Marconi}, {Marie}, {Marinoni}, {Marrese},
  {Marschalk{\'o}}, {Marshall}, {Mart{\'\i}n-Fleitas}, {Martino}, {Mary},
  {Matijevi{\v{c}}}, {Mazeh}, {McMillan}, {Messina}, {Mestre}, {Michalik},
  {Millar}, {Miranda}, {Molina}, {Molinaro}, {Molinaro}, {Moln{\'a}r},
  {Moniez}, {Montegriffo}, {Monteiro}, {Mor}, {Mora}, {Morbidelli}, {Morel},
  {Morgenthaler}, {Morley}, {Morris}, {Mulone}, {Muraveva}, {Musella},
  {Narbonne}, {Nelemans}, {Nicastro}, {Noval}, {Ord{\'e}novic},
  {Ordieres-Mer{\'e}}, {Osborne}, {Pagani}, {Pagano}, {Pailler}, {Palacin},
  {Palaversa}, {Parsons}, {Paulsen}, {Pecoraro}, {Pedrosa}, {Pentik{\"a}inen},
  {Pereira}, {Pichon}, {Piersimoni}, {Pineau}, {Plachy}, {Plum}, {Poujoulet},
  {Pr{\v{s}}a}, {Pulone}, {Ragaini}, {Rago}, {Rambaux}, {Ramos-Lerate},
  {Ranalli}, {Rauw}, {Read}, {Regibo}, {Renk}, {Reyl{\'e}}, {Ribeiro},
  {Rimoldini}, {Ripepi}, {Riva}, {Rixon}, {Roelens}, {Romero-G{\'o}mez},
  {Rowell}, {Royer}, {Rudolph}, {Ruiz-Dern}, {Sadowski}, {Sagrist{\`a}
  Sell{\'e}s}, {Sahlmann}, {Salgado}, {Salguero}, {Sarasso}, {Savietto},
  {Schnorhk}, {Schultheis}, {Sciacca}, {Segol}, {Segovia}, {Segransan},
  {Serpell}, {Shih}, {Smareglia}, {Smart}, {Smith}, {Solano}, {Solitro},
  {Sordo}, {Soria Nieto}, {Souchay}, {Spagna}, {Spoto}, {Stampa}, {Steele},
  {Steidelm{\"u}ller}, {Stephenson}, {Stoev}, {Suess}, {S{\"u}veges}, {Surdej},
  {Szabados}, {Szegedi-Elek}, {Tapiador}, {Taris}, {Tauran}, {Taylor},
  {Teixeira}, {Terrett}, {Tingley}, {Trager}, {Turon}, {Ulla}, {Utrilla},
  {Valentini}, {van Elteren}, {Van Hemelryck}, {van Leeuwen}, {Varadi},
  {Vecchiato}, {Veljanoski}, {Via}, {Vicente}, {Vogt}, {Voss}, {Votruba},
  {Voutsinas}, {Walmsley}, {Weiler}, {Weingrill}, {Werner}, {Wevers},
  {Whitehead}, {Wyrzykowski}, {Yoldas}, {{\v{Z}}erjal}, {Zucker}, {Zurbach},
  {Zwitter}, {Alecu}, {Allen}, {Allende Prieto}, {Amorim},
  {Anglada-Escud{\'e}}, {Arsenijevic}, {Azaz}, {Balm}, {Beck}, {Bernstein},
  {Bigot}, {Bijaoui}, {Blasco}, {Bonfigli}, {Bono}, {Boudreault}, {Bressan},
  {Brown}, {Brunet}, {Bunclark}, {Buonanno}, {Butkevich}, {Carret}, {Carrion},
  {Chemin}, {Ch{\'e}reau}, {Corcione}, {Darmigny}, {de Boer}, {de Teodoro}, {de
  Zeeuw}, {Delle Luche}, {Domingues}, {Dubath}, {Fodor}, {Fr{\'e}zouls},
  {Fries}, {Fustes}, {Fyfe}, {Gallardo}, {Gallegos}, {Gardiol}, {Gebran},
  {Gomboc}, {G{\'o}mez}, {Grux}, {Gueguen}, {Heyrovsky}, {Hoar}, {Iannicola},
  {Isasi Parache}, {Janotto}, {Joliet}, {Jonckheere}, {Keil}, {Kim},
  {Klagyivik}, {Klar}, {Knude}, {Kochukhov}, {Kolka}, {Kos}, {Kutka}, {Lainey},
  {LeBouquin}, {Liu}, {Loreggia}, {Makarov}, {Marseille}, {Martayan},
  {Martinez-Rubi}, {Massart}, {Meynadier}, {Mignot}, {Munari}, {Nguyen},
  {Nordlander}, {Ocvirk}, {O'Flaherty}, {Olias Sanz}, {Ortiz}, {Osorio},
  {Oszkiewicz}, {Ouzounis}, {Palmer}, {Park}, {Pasquato}, {Peltzer}, {Peralta},
  {P{\'e}turaud}, {Pieniluoma}, {Pigozzi}, {Poels}, {Prat}, {Prod'homme},
  {Raison}, {Rebordao}, {Risquez}, {Rocca-Volmerange}, {Rosen}, {Ruiz-Fuertes},
  {Russo}, {Sembay}, {Serraller Vizcaino}, {Short}, {Siebert}, {Silva},
  {Sinachopoulos}, {Slezak}, {Soffel}, {Sosnowska}, {Strai{\v{z}}ys}, {ter
  Linden}, {Terrell}, {Theil}, {Tiede}, {Troisi}, {Tsalmantza}, {Tur},
  {Vaccari}, {Vachier}, {Valles}, {Van Hamme}, {Veltz}, {Virtanen}, {Wallut},
  {Wichmann}, {Wilkinson}, {Ziaeepour}, \& {Zschocke}}]{Gaia16b}
{Gaia Collaboration}, {Prusti}, T., {de Bruijne}, J.~H.~J., {et~al.} 2016,
  \aap, 595, A1

\bibitem[{{Galli} {et~al.}(2019){Galli}, {Loinard}, {Bouy}, {Sarro},
  {Ortiz-Le{\'o}n}, {Dzib}, {Olivares}, {Heyer}, {Hernandez},
  {Rom{\'a}n-Z{\'u}{\~n}iga}, {Kounkel}, \& {Covey}}]{Galli19}
{Galli}, P.~A.~B., {Loinard}, L., {Bouy}, H., {et~al.} 2019, \aap, 630, A137

\bibitem[{{Garc{\'\i}a P{\'e}rez} {et~al.}(2016){Garc{\'\i}a P{\'e}rez},
  {Allende Prieto}, {Holtzman}, {Shetrone}, {M{\'e}sz{\'a}ros}, {Bizyaev},
  {Carrera}, {Cunha}, {Garc{\'\i}a-Hern{\'a}ndez}, {Johnson}, {Majewski},
  {Nidever}, {Schiavon}, {Shane}, {Smith}, {Sobeck}, {Troup}, {Zamora},
  {Weinberg}, {Bovy}, {Eisenstein}, {Feuillet}, {Frinchaboy}, {Hayden},
  {Hearty}, {Nguyen}, {O'Connell}, {Pinsonneault}, {Wilson}, \&
  {Zasowski}}]{GarciaP16}
{Garc{\'\i}a P{\'e}rez}, A.~E., {Allende Prieto}, C., {Holtzman}, J.~A.,
  {et~al.} 2016, \aj, 151, 144

\bibitem[{{Gilmore} {et~al.}(2012){Gilmore}, {Randich}, {Asplund}, {Binney},
  {Bonifacio}, {Drew}, {Feltzing}, {Ferguson}, {Jeffries}, {Micela},
  {Negueruela}, {Prusti}, {Rix}, {Vallenari}, {Alfaro}, {Allende-Prieto},
  {Babusiaux}, {Bensby}, {Blomme}, {Bragaglia}, {Flaccomio}, {Fran{\c{c}}ois},
  {Irwin}, {Koposov}, {Korn}, {Lanzafame}, {Pancino}, {Paunzen},
  {Recio-Blanco}, {Sacco}, {Smiljanic}, {Van Eck}, {Walton}, {Aden}, {Aerts},
  {Affer}, {Alcala}, {Altavilla}, {Alves}, {Antoja}, {Arenou}, {Argiroffi},
  {Asensio Ramos}, {Bailer-Jones}, {Balaguer-Nunez}, {Bayo}, {Barbuy},
  {Barisevicius}, {Barrado y Navascues}, {Battistini}, {Bellas Velidis},
  {Bellazzini}, {Belokurov}, {Bergemann}, {Bertelli}, {Biazzo}, {Bienayme},
  {Bland-Hawthorn}, {Boeche}, {Bonito}, {Boudreault}, {Bouvier}, {Brandao},
  {Brown}, {de Bruijne}, {Burleigh}, {Caballero}, {Caffau}, {Calura},
  {Capuzzo-Dolcetta}, {Caramazza}, {Carraro}, {Casagrande}, {Casewell},
  {Chapman}, {Chiappini}, {Chorniy}, {Christlieb}, {Cignoni}, {Cocozza},
  {Colless}, {Collet}, {Collins}, {Correnti}, {Covino}, {Crnojevic}, {Cropper},
  {Cunha}, {Damiani}, {David}, {Delgado}, {Duffau}, {Edvardsson}, {Eldridge},
  {Enke}, {Eriksson}, {Evans}, {Eyer}, {Famaey}, {Fellhauer}, {Ferreras},
  {Figueras}, {Fiorentino}, {Flynn}, {Folha}, {Franciosini}, {Frasca},
  {Freeman}, {Fremat}, {Friel}, {Gaensicke}, {Gameiro}, {Garzon}, {Geier},
  {Geisler}, {Gerhard}, {Gibson}, {Gomboc}, {Gomez}, {Gonzalez-Fernandez},
  {Gonzalez Hernandez}, {Gosset}, {Grebel}, {Greimel}, {Groenewegen},
  {Grundahl}, {Guarcello}, {Gustafsson}, {Hadrava}, {Hatzidimitriou}, {Hambly},
  {Hammersley}, {Hansen}, {Haywood}, {Heber}, {Heiter}, {Held}, {Helmi},
  {Hensler}, {Herrero}, {Hill}, {Hodgkin}, {Huelamo}, {Huxor}, {Ibata},
  {Jackson}, {de Jong}, {Jonker}, {Jordan}, {Jordi}, {Jorissen}, {Katz},
  {Kawata}, {Keller}, {Kharchenko}, {Klement}, {Klutsch}, {Knude}, {Koch},
  {Kochukhov}, {Kontizas}, {Koubsky}, {Lallement}, {de Laverny}, {van Leeuwen},
  {Lemasle}, {Lewis}, {Lind}, {Lindstrom}, {Lobel}, {Lopez Santiago}, {Lucas},
  {Ludwig}, {Lueftinger}, {Magrini}, {Maiz Apellaniz}, {Maldonado}, {Marconi},
  {Marino}, {Martayan}, {Martinez-Valpuesta}, {Matijevic}, {McMahon},
  {Messina}, {Meyer}, {Miglio}, {Mikolaitis}, {Minchev}, {Minniti}, {Moitinho},
  {Momany}, {Monaco}, {Montalto}, {Monteiro}, {Monier}, {Montes}, {Mora},
  {Moraux}, {Morel}, {Mowlavi}, {Mucciarelli}, {Munari}, {Napiwotzki},
  {Nardetto}, {Naylor}, {Naze}, {Nelemans}, {Okamoto}, {Ortolani}, {Pace},
  {Palla}, {Palous}, {Parker}, {Penarrubia}, {Pillitteri}, {Piotto}, {Posbic},
  {Prisinzano}, {Puzeras}, {Quirrenbach}, {Ragaini}, {Read}, {Read}, {Reyle},
  {De Ridder}, {Robichon}, {Robin}, {Roeser}, {Romano}, {Royer}, {Ruchti},
  {Ruzicka}, {Ryan}, {Ryde}, {Santos}, {Sanz Forcada}, {Sarro Baro},
  {Sbordone}, {Schilbach}, {Schmeja}, {Schnurr}, {Schoenrich}, {Scholz},
  {Seabroke}, {Sharma}, {De Silva}, {Smith}, {Solano}, {Sordo}, {Soubiran},
  {Sousa}, {Spagna}, {Steffen}, {Steinmetz}, {Stelzer}, {Stempels},
  {Tabernero}, {Tautvaisiene}, {Thevenin}, {Torra}, {Tosi}, {Tolstoy}, {Turon},
  {Walker}, {Wambsganss}, {Worley}, {Venn}, {Vink}, {Wyse}, {Zaggia},
  {Zeilinger}, {Zoccali}, {Zorec}, {Zucker}, {Zwitter}, \& {Gaia-ESO Survey
  Team}}]{GaiaESO12}
{Gilmore}, G., {Randich}, S., {Asplund}, M., {et~al.} 2012, The Messenger, 147,
  25

\bibitem[{{Gossage} {et~al.}(2018){Gossage}, {Conroy}, {Dotter}, {Choi},
  {Rosenfield}, {Cargile}, \& {Dolphin}}]{Gossage18}
{Gossage}, S., {Conroy}, C., {Dotter}, A., {et~al.} 2018, \apj, 863, 67

\bibitem[{{Gunn} {et~al.}(2006){Gunn}, {Siegmund}, {Mannery}, {Owen}, {Hull},
  {Leger}, {Carey}, {Knapp}, {York}, {Boroski}, {Kent}, {Lupton}, {Rockosi},
  {Evans}, {Waddell}, {Anderson}, {Annis}, {Barentine}, {Bartoszek}, {Bastian},
  {Bracker}, {Brewington}, {Briegel}, {Brinkmann}, {Brown}, {Carr},
  {Czarapata}, {Drennan}, {Dombeck}, {Federwitz}, {Gillespie}, {Gonzales},
  {Hansen}, {Harvanek}, {Hayes}, {Jordan}, {Kinney}, {Klaene}, {Kleinman},
  {Kron}, {Kresinski}, {Lee}, {Limmongkol}, {Lindenmeyer}, {Long}, {Loomis},
  {McGehee}, {Mantsch}, {Neilsen}, {Neswold}, {Newman}, {Nitta}, {Peoples},
  {Pier}, {Prieto}, {Prosapio}, {Rivetta}, {Schneider}, {Snedden}, \&
  {Wang}}]{Gunn06}
{Gunn}, J.~E., {Siegmund}, W.~A., {Mannery}, E.~J., {et~al.} 2006, \aj, 131,
  2332

\bibitem[{Hahsler {et~al.}(2019)Hahsler, Piekenbrock, \& Doran}]{fast_hdbscan}
Hahsler, M., Piekenbrock, M., \& Doran, D. 2019, Journal of Statistical
  Software, Articles, 91, 1

\bibitem[{{Holtzman} {et~al.}(2015){Holtzman}, {Shetrone}, {Johnson}, {Allende
  Prieto}, {Anders}, {Andrews}, {Beers}, {Bizyaev}, {Blanton}, {Bovy},
  {Carrera}, {Chojnowski}, {Cunha}, {Eisenstein}, {Feuillet}, {Frinchaboy},
  {Galbraith-Frew}, {Garc{\'\i}a P{\'e}rez}, {Garc{\'\i}a-Hern{\'a}ndez},
  {Hasselquist}, {Hayden}, {Hearty}, {Ivans}, {Majewski}, {Martell},
  {Meszaros}, {Muna}, {Nidever}, {Nguyen}, {O'Connell}, {Pan}, {Pinsonneault},
  {Robin}, {Schiavon}, {Shane}, {Sobeck}, {Smith}, {Troup}, {Weinberg},
  {Wilson}, {Wood-Vasey}, {Zamora}, \& {Zasowski}}]{Holtzman15}
{Holtzman}, J.~A., {Shetrone}, M., {Johnson}, J.~A., {et~al.} 2015, \aj, 150,
  148

\bibitem[{{Hunt} \& {Reffert}(2021)}]{hunt21}
{Hunt}, E.~L., \& {Reffert}, S. 2021, \aap, 646, A104

\bibitem[{{Kollmeier} {et~al.}(2017){Kollmeier}, {Zasowski}, {Rix}, {Johns},
  {Anderson}, {Drory}, {Johnson}, {Pogge}, {Bird}, {Blanc}, {Brownstein},
  {Crane}, {De Lee}, {Klaene}, {Kreckel}, {MacDonald}, {Merloni}, {Ness},
  {O'Brien}, {Sanchez-Gallego}, {Sayres}, {Shen}, {Thakar}, {Tkachenko},
  {Aerts}, {Blanton}, {Eisenstein}, {Holtzman}, {Maoz}, {Nandra}, {Rockosi},
  {Weinberg}, {Bovy}, {Casey}, {Chaname}, {Clerc}, {Conroy}, {Eracleous},
  {G{\"a}nsicke}, {Hekker}, {Horne}, {Kauffmann}, {McQuinn}, {Pellegrini},
  {Schinnerer}, {Schlafly}, {Schwope}, {Seibert}, {Teske}, \& {van
  Saders}}]{Kollmeier17}
{Kollmeier}, J.~A., {Zasowski}, G., {Rix}, H.-W., {et~al.} 2017, arXiv
  e-prints, arXiv:1711.03234

\bibitem[{{Kos} {et~al.}(2021){Kos}, {Bland-Hawthorn}, {Buder}, {Nordlander},
  {Spina}, {Beeson}, {Lind}, {Asplund}, {Freeman}, {Hayden}, {Lewis},
  {Martell}, {Sharma}, {De Silva}, {Simpson}, {Zucker}, {Zwitter},
  {{\v{C}}otar}, {Horner}, {Ting}, \& {Traven}}]{Kos21}
{Kos}, J., {Bland-Hawthorn}, J., {Buder}, S., {et~al.} 2021, \mnras, 506, 4232

\bibitem[{{Kounkel} {et~al.}(2018){Kounkel}, {Covey}, {Su{\'a}rez},
  {Rom{\'a}n-Z{\'u}{\~n}iga}, {Hernandez}, {Stassun}, {Jaehnig}, {Feigelson},
  {Pe{\~n}a Ram{\'\i}rez}, {Roman-Lopes}, {Da Rio}, {Stringfellow}, {Kim},
  {Borissova}, {Fern{\'a}ndez-Trincado}, {Burgasser},
  {Garc{\'\i}a-Hern{\'a}ndez}, {Zamora}, {Pan}, \& {Nitschelm}}]{Kounkel18}
{Kounkel}, M., {Covey}, K., {Su{\'a}rez}, G., {et~al.} 2018, \aj, 156, 84

\bibitem[{{Lada} {et~al.}(2009){Lada}, {Lombardi}, \& {Alves}}]{Lada09}
{Lada}, C.~J., {Lombardi}, M., \& {Alves}, J.~F. 2009, \apj, 703, 52

\bibitem[{{Lindegren} {et~al.}(2018){Lindegren}, {Hern{\'a}ndez}, {Bombrun},
  {Klioner}, {Bastian}, {Ramos-Lerate}, {de Torres}, {Steidelm{\"u}ller},
  {Stephenson}, {Hobbs}, {Lammers}, {Biermann}, {Geyer}, {Hilger}, {Michalik},
  {Stampa}, {McMillan}, {Casta{\~n}eda}, {Clotet}, {Comoretto}, {Davidson},
  {Fabricius}, {Gracia}, {Hambly}, {Hutton}, {Mora}, {Portell}, {van Leeuwen},
  {Abbas}, {Abreu}, {Altmann}, {Andrei}, {Anglada}, {Balaguer-N{\'u}{\~n}ez},
  {Barache}, {Becciani}, {Bertone}, {Bianchi}, {Bouquillon}, {Bourda},
  {Br{\"u}semeister}, {Bucciarelli}, {Busonero}, {Buzzi}, {Cancelliere},
  {Carlucci}, {Charlot}, {Cheek}, {Crosta}, {Crowley}, {de Bruijne}, {de
  Felice}, {Drimmel}, {Esquej}, {Fienga}, {Fraile}, {Gai}, {Garralda},
  {Gonz{\'a}lez-Vidal}, {Guerra}, {Hauser}, {Hofmann}, {Holl}, {Jordan},
  {Lattanzi}, {Lenhardt}, {Liao}, {Licata}, {Lister}, {L{\"o}ffler},
  {Marchant}, {Martin-Fleitas}, {Messineo}, {Mignard}, {Morbidelli}, {Poggio},
  {Riva}, {Rowell}, {Salguero}, {Sarasso}, {Sciacca}, {Siddiqui}, {Smart},
  {Spagna}, {Steele}, {Taris}, {Torra}, {van Elteren}, {van Reeven}, \&
  {Vecchiato}}]{Lindegren18}
{Lindegren}, L., {Hern{\'a}ndez}, J., {Bombrun}, A., {et~al.} 2018, \aap, 616,
  A2

\bibitem[{{Luhman}(2020)}]{Luhman20}
{Luhman}, K.~L. 2020, \aj, 160, 186

\bibitem[{{Majewski} {et~al.}(2017){Majewski}, {Schiavon}, {Frinchaboy},
  {Allende Prieto}, {Barkhouser}, {Bizyaev}, {Blank}, {Brunner}, {Burton},
  {Carrera}, {Chojnowski}, {Cunha}, {Epstein}, {Fitzgerald}, {Garc{\'\i}a
  P{\'e}rez}, {Hearty}, {Henderson}, {Holtzman}, {Johnson}, {Lam}, {Lawler},
  {Maseman}, {M{\'e}sz{\'a}ros}, {Nelson}, {Nguyen}, {Nidever}, {Pinsonneault},
  {Shetrone}, {Smee}, {Smith}, {Stolberg}, {Skrutskie}, {Walker}, {Wilson},
  {Zasowski}, {Anders}, {Basu}, {Beland}, {Blanton}, {Bovy}, {Brownstein},
  {Carlberg}, {Chaplin}, {Chiappini}, {Eisenstein}, {Elsworth}, {Feuillet},
  {Fleming}, {Galbraith-Frew}, {Garc{\'\i}a}, {Garc{\'\i}a-Hern{\'a}ndez},
  {Gillespie}, {Girardi}, {Gunn}, {Hasselquist}, {Hayden}, {Hekker}, {Ivans},
  {Kinemuchi}, {Klaene}, {Mahadevan}, {Mathur}, {Mosser}, {Muna}, {Munn},
  {Nichol}, {O'Connell}, {Parejko}, {Robin}, {Rocha-Pinto}, {Schultheis},
  {Serenelli}, {Shane}, {Silva Aguirre}, {Sobeck}, {Thompson}, {Troup},
  {Weinberg}, \& {Zamora}}]{APOGEE17}
{Majewski}, S.~R., {Schiavon}, R.~P., {Frinchaboy}, P.~M., {et~al.} 2017, \aj,
  154, 94

\bibitem[{{Massi} {et~al.}(2019){Massi}, {Weiss}, {Elia}, {Csengeri},
  {Schisano}, {Giannini}, {Hill}, {Lorenzetti}, {Menten}, {Olmi}, {Schuller},
  {Strafella}, {De Luca}, {Motte}, \& {Wyrowski}}]{Massi2019}
{Massi}, F., {Weiss}, A., {Elia}, D., {et~al.} 2019, \aap, 628, A110

\bibitem[{{Medina} {et~al.}(2021){Medina}, {Borissova}, {Kurtev},
  {Alonso-Garc{\'\i}a}, {Rom{\'a}n-Z{\'u}{\~n}iga}, {Bayo}, {Kounkel},
  {Roman-Lopes}, {Lucas}, {Covey}, {F{\'o}rster}, {Minniti}, {Adame}, \&
  {Hern{\'a}ndez}}]{Medina21}
{Medina}, N., {Borissova}, J., {Kurtev}, R., {et~al.} 2021, \apj, 914, 28

\bibitem[{{Meisner} \& {Finkbeiner}(2014)}]{WSSA14}
{Meisner}, A.~M., \& {Finkbeiner}, D.~P. 2014, \apj, 781, 5

\bibitem[{{Mejia-Narvaez} {et~al.}(2021){Mejia-Narvaez}, {Bruzual}, {Sanchez},
  {Carigi}, {Barrera-Ballesteros}, {Valerdi}, {Yan}, \& {Drory}}]{Mejia21}
{Mejia-Narvaez}, A., {Bruzual}, G., {Sanchez}, S.~F., {et~al.} 2021, arXiv
  e-prints, arXiv:2108.01697

\bibitem[{{Nidever} {et~al.}(2015){Nidever}, {Holtzman}, {Allende Prieto},
  {Beland}, {Bender}, {Bizyaev}, {Burton}, {Desphande}, {Fleming}, {Garc{\'\i}a
  P{\'e}rez}, {Hearty}, {Majewski}, {M{\'e}sz{\'a}ros}, {Muna}, {Nguyen},
  {Schiavon}, {Shetrone}, {Skrutskie}, {Sobeck}, \& {Wilson}}]{Nidever15}
{Nidever}, D.~L., {Holtzman}, J.~A., {Allende Prieto}, C., {et~al.} 2015, \aj,
  150, 173

\bibitem[{{Olney} {et~al.}(2020){Olney}, {Kounkel}, {Schillinger}, {Scoggins},
  {Yin}, {Howard}, {Covey}, {Hutchinson}, \& {Stassun}}]{Olney20}
{Olney}, R., {Kounkel}, M., {Schillinger}, C., {et~al.} 2020, \aj, 159, 182

\bibitem[{{Pecaut} \& {Mamajek}(2013)}]{Pecaut13}
{Pecaut}, M.~J., \& {Mamajek}, E.~E. 2013, \apjs, 208, 9

\bibitem[{{Percival} {et~al.}(2005){Percival}, {Salaris}, \&
  {Groenewegen}}]{Percival05}
{Percival}, S.~M., {Salaris}, M., \& {Groenewegen}, M.~A.~T. 2005, \aap, 429,
  887

\bibitem[{{Porras} {et~al.}(2003){Porras}, {Christopher}, {Allen}, {Di
  Francesco}, {Megeath}, \& {Myers}}]{Porras03}
{Porras}, A., {Christopher}, M., {Allen}, L., {et~al.} 2003, \aj, 126, 1916

\bibitem[{{R Core Team}(2018)}]{R_ref}
{R Core Team}. 2018, R: A Language and Environment for Statistical Computing, R
  Foundation for Statistical Computing, Vienna, Austria

\bibitem[{{Ram{\'\i}rez-Preciado} {et~al.}(2020){Ram{\'\i}rez-Preciado},
  {Roman-Lopes}, {Rom{\'a}n-Z{\'u}{\~n}iga}, {Hern{\'a}ndez},
  {Garc{\'\i}a-Hern{\'a}ndez}, {Stassun}, {Stringfellow}, \& {Kim}}]{Ramirez20}
{Ram{\'\i}rez-Preciado}, V.~G., {Roman-Lopes}, A., {Rom{\'a}n-Z{\'u}{\~n}iga},
  C.~G., {et~al.} 2020, \apj, 894, 5

\bibitem[{{Roman-Lopes} {et~al.}(2020{\natexlab{a}}){Roman-Lopes},
  {Rom{\'a}n-Z{\'u}{\~n}iga}, {Borissova}, {Ram{\'\i}rez-Preciado},
  {Hern{\'a}ndez}, \& {Minniti}}]{RomanLopes20b}
{Roman-Lopes}, A., {Rom{\'a}n-Z{\'u}{\~n}iga}, C.~G., {Borissova}, J., {et~al.}
  2020{\natexlab{a}}, \apj, 891, 107

\bibitem[{{Roman-Lopes} {et~al.}(2020{\natexlab{b}}){Roman-Lopes},
  {Rom{\'a}n-Z{\'u}{\~n}iga}, {Tapia}, {Minniti}, \&
  {Borissova}}]{RomanLopes20}
{Roman-Lopes}, A., {Rom{\'a}n-Z{\'u}{\~n}iga}, C.~G., {Tapia}, M., {Minniti},
  D., \& {Borissova}, J. 2020{\natexlab{b}}, \apjs, 247, 17

\bibitem[{{Roman-Lopes} {et~al.}(2018){Roman-Lopes},
  {Rom{\'a}n-Z{\'u}{\~n}iga}, {Tapia}, {Chojnowski}, {G{\'o}mez Maqueo Chew},
  {Garc{\'\i}a-Hern{\'a}ndez}, {Borissova}, {Minniti}, {Covey},
  {Longa-Pe{\~n}a}, {Fernandez-Trincado}, {Zamora}, \&
  {Nitschelm}}]{RomanLopes18}
{Roman-Lopes}, A., {Rom{\'a}n-Z{\'u}{\~n}iga}, C., {Tapia}, M., {et~al.} 2018,
  \apj, 855, 68

\bibitem[{{Santana} {et~al.}(2021){Santana}, {Beaton}, {Covey}, {O'Connell},
  {Longa-Pe{\~n}a}, {Cohen}, {Fern{\'a}ndez-Trincado}, {Hayes}, {Zasowski},
  {Sobeck}, {Majewski}, {Chojnowski}, {De Lee}, {Oelkers}, {Stringfellow},
  {Almeida}, {Anguiano}, {Donor}, {Frinchaboy}, {Hasselquist}, {Johnson},
  {Kollmeier}, {Nidever}, {Price-Whelan}, {Rojas-Arriagada}, {Schultheis},
  {Shetrone}, {Simon}, {Aerts}, {Borissova}, {Drout}, {Geisler}, {Law},
  {Medina}, {Minniti}, {Monachesi}, {Mu{\~n}oz}, {Poleski}, {Roman-Lopes},
  {Schlaufman}, {Stutz}, {Teske}, {Tkachenko}, {Van Saders}, {Weinberger}, \&
  {Zoccali}}]{Santana21S}
{Santana}, F.~A., {Beaton}, R.~L., {Covey}, K.~R., {et~al.} 2021, arXiv
  e-prints, arXiv:2108.11908

\bibitem[{{Spina} {et~al.}(2014){Spina}, {Randich}, {Palla}, {Biazzo}, {Sacco},
  {Alfaro}, {Franciosini}, {Magrini}, {Morbidelli}, {Frasca}, {Adibekyan},
  {Delgado-Mena}, {Sousa}, {Gonz{\'a}lez Hern{\'a}ndez}, {Montes}, {Tabernero},
  {Tautvai{\v{s}}ien{\.{e}}}, {Bonito}, {Lanzafame}, {Gilmore}, {Jeffries},
  {Vallenari}, {Bensby}, {Bragaglia}, {Flaccomio}, {Korn}, {Pancino},
  {Recio-Blanco}, {Smiljanic}, {Bergemann}, {Costado}, {Damiani}, {Hill},
  {Hourihane}, {Jofr{\'e}}, {de Laverny}, {Lardo}, {Masseron}, {Prisinzano}, \&
  {Worley}}]{Spina14}
{Spina}, L., {Randich}, S., {Palla}, F., {et~al.} 2014, \aap, 568, A2

\bibitem[{{Spina} {et~al.}(2017){Spina}, {Randich}, {Magrini}, {Jeffries},
  {Friel}, {Sacco}, {Pancino}, {Bonito}, {Bravi}, {Franciosini}, {Klutsch},
  {Montes}, {Gilmore}, {Vallenari}, {Bensby}, {Bragaglia}, {Flaccomio},
  {Koposov}, {Korn}, {Lanzafame}, {Smiljanic}, {Bayo}, {Carraro}, {Casey},
  {Costado}, {Damiani}, {Donati}, {Frasca}, {Hourihane}, {Jofr{\'e}}, {Lewis},
  {Lind}, {Monaco}, {Morbidelli}, {Prisinzano}, {Sousa}, {Worley}, \&
  {Zaggia}}]{Spina17}
{Spina}, L., {Randich}, S., {Magrini}, L., {et~al.} 2017, \aap, 601, A70

\bibitem[{{Sprague} {et~al.}(2022){Sprague}, {Culhane}, {Kounkel}, {Olney},
  {Covey}, {Hutchinson}, {Lingg}, {Stassun}, {Rom{\'a}n-Z{\'u}{\~n}iga},
  {Roman-Lopes}, {Nidever}, {Beaton}, {Borissova}, {Stutz}, {Stringfellow},
  {Ram{\'\i}rez}, {Ram{\'\i}rez-Preciado}, {Hern{\'a}ndez}, {Kim}, \&
  {Lane}}]{Sprague22}
{Sprague}, D., {Culhane}, C., {Kounkel}, M., {et~al.} 2022, \aj, 163, 152

\bibitem[{{Stutz}(2018)}]{Stutz18}
{Stutz}, A.~M. 2018, \mnras, 473, 4890

\bibitem[{{Stutz} \& {Gould}(2016)}]{Stutz16}
{Stutz}, A.~M., \& {Gould}, A. 2016, \aap, 590, A2

\bibitem[{{Taylor}(2005)}]{TOPCAT05}
{Taylor}, M.~B. 2005, in Astronomical Society of the Pacific Conference Series,
  Vol. 347, Astronomical Data Analysis Software and Systems XIV, ed.
  P.~{Shopbell}, M.~{Britton}, \& R.~{Ebert}, 29

\bibitem[{{Ting} {et~al.}(2019){Ting}, {Conroy}, {Rix}, \& {Cargile}}]{Ting19}
{Ting}, Y.-S., {Conroy}, C., {Rix}, H.-W., \& {Cargile}, P. 2019, \apj, 879, 69

\bibitem[{{Tognelli} {et~al.}(2011){Tognelli}, {Prada Moroni}, \&
  {Degl'Innocenti}}]{Tognelli11}
{Tognelli}, E., {Prada Moroni}, P.~G., \& {Degl'Innocenti}, S. 2011, \aap, 533,
  A109

\bibitem[{{Wenger} {et~al.}(2000){Wenger}, {Ochsenbein}, {Egret}, {Dubois},
  {Bonnarel}, {Borde}, {Genova}, {Jasniewicz}, {Lalo{\"e}}, {Lesteven}, \&
  {Monier}}]{Wenger2000}
{Wenger}, M., {Ochsenbein}, F., {Egret}, D., {et~al.} 2000, \aaps, 143, 9

\bibitem[{{Wilson} {et~al.}(2019){Wilson}, {Hearty}, {Skrutskie}, {Majewski},
  {Holtzman}, {Eisenstein}, {Gunn}, {Blank}, {Henderson}, {Smee}, {Nelson},
  {Nidever}, {Arns}, {Barkhouser}, {Barr}, {Beland}, {Bershady}, {Blanton},
  {Brunner}, {Burton}, {Carey}, {Carr}, {Colque}, {Crane}, {Damke}, {Davidson},
  {Dean}, {Di Mille}, {Don}, {Ebelke}, {Evans}, {Fitzgerald}, {Gillespie},
  {Hall}, {Harding}, {Harding}, {Hammond}, {Hancock}, {Harrison}, {Hope},
  {Horne}, {Karakla}, {Lam}, {Leger}, {MacDonald}, {Maseman}, {Matsunari},
  {Melton}, {Mitcheltree}, {O'Brien}, {O'Connell}, {Patten}, {Richardson},
  {Rieke}, {Rieke}, {Roman-Lopes}, {Schiavon}, {Sobeck}, {Stolberg}, {Stoll},
  {Tembe}, {Trujillo}, {Uomoto}, {Vernieri}, {Walker}, {Weinberg}, {Young},
  {Anthony-Brumfield}, {Bizyaev}, {Breslauer}, {De Lee}, {Downey}, {Halverson},
  {Huehnerhoff}, {Klaene}, {Leon}, {Long}, {Mahadevan}, {Malanushenko},
  {Nguyen}, {Owen}, {S{\'a}nchez-Gallego}, {Sayres}, {Shane}, {Shectman},
  {Shetrone}, {Skinner}, {Stauffer}, \& {Zhao}}]{Wilson2019}
{Wilson}, J.~C., {Hearty}, F.~R., {Skrutskie}, M.~F., {et~al.} 2019, \pasp,
  131, 055001

\bibitem[{{Wright} {et~al.}(2019){Wright}, {Jeffries}, {Jackson}, {Bayo},
  {Bonito}, {Damiani}, {Kalari}, {Lanzafame}, {Pancino}, {Parker},
  {Prisinzano}, {Randich}, {Vink}, {Alfaro}, {Bergemann}, {Franciosini},
  {Gilmore}, {Gonneau}, {Hourihane}, {Jofr{\'e}}, {Koposov}, {Lewis},
  {Magrini}, {Micela}, {Morbidelli}, {Sacco}, {Worley}, \& {Zaggia}}]{Wright19}
{Wright}, N.~J., {Jeffries}, R.~D., {Jackson}, R.~J., {et~al.} 2019, \mnras,
  486, 2477

\bibitem[{{Yao} {et~al.}(2018){Yao}, {Meyer}, {Covey}, {Tan}, \& {Da
  Rio}}]{Yao18}
{Yao}, Y., {Meyer}, M.~R., {Covey}, K.~R., {Tan}, J.~C., \& {Da Rio}, N. 2018,
  \apj, 869, 72

\bibitem[{{Zasowski} {et~al.}(2017){Zasowski}, {Cohen}, {Chojnowski},
  {Santana}, {Oelkers}, {Andrews}, {Beaton}, {Bender}, {Bird}, {Bovy},
  {Carlberg}, {Covey}, {Cunha}, {Dell'Agli}, {Fleming}, {Frinchaboy},
  {Garc{\'\i}a-Hern{\'a}ndez}, {Harding}, {Holtzman}, {Johnson}, {Kollmeier},
  {Majewski}, {M{\'e}sz{\'a}ros}, {Munn}, {Mu{\~n}oz}, {Ness}, {Nidever},
  {Poleski}, {Rom{\'a}n-Z{\'u}{\~n}iga}, {Shetrone}, {Simon}, {Smith},
  {Sobeck}, {Stringfellow}, {Szigeti{\'a}ros}, {Tayar}, \&
  {Troup}}]{Zasowski17}
{Zasowski}, G., {Cohen}, R.~E., {Chojnowski}, S.~D., {et~al.} 2017, \aj, 154,
  198

\bibitem[{{Zhao} {et~al.}(2012){Zhao}, {Zhao}, {Chu}, {Jing}, \&
  {Deng}}]{LAMOST12}
{Zhao}, G., {Zhao}, Y., {Chu}, Y., {Jing}, Y., \& {Deng}, L. 2012, arXiv
  e-prints, arXiv:1206.3569

\end{thebibliography}
